\documentclass[12pt,a4paper]{article}

\usepackage{amsmath}
\usepackage{mathtools}
\usepackage{amsfonts}
\usepackage{amssymb}
\usepackage{graphicx}
\usepackage{color}
\usepackage{booktabs}
\usepackage{inputenc}
\usepackage[T1]{fontenc}
\usepackage{mathrsfs}
\usepackage{enumerate}
\usepackage{xcolor,url}
\usepackage{dsfont} 
\usepackage{multirow}
\usepackage[normalem]{ulem} 
\usepackage{makecell}
\usepackage{rotating}
\usepackage{cancel}
\usepackage{braket}
\usepackage{cite}

\newcommand{\eea}{\end{eqnarray}}
\newcommand{\bea}{\begin{eqnarray}}
\newcommand{\be}{\begin{equation}}
\newcommand{\ee}{\end{equation}}

\newcommand{\rmd}{\mathrm{d}}

\def\be{\begin{equation}}
\def\ee{\end{equation}}

\newcommand{\de}{\partial}
\def\knl{k_{\rm NL}}
\renewcommand{\(}{\left(}
\renewcommand{\)}{\right)}
\renewcommand{\[}{\left[}
\renewcommand{\]}{\right]}

\newcommand{\gam}{\gamma}
\newcommand{\del}{\delta}
\newcommand{\eps}{\epsilon}

\newcommand{\Om}{\Omega}

\newcommand{\tchi}{\tilde{\chi}}
\newcommand{\hn}{\hat{n}}
\newcommand{\hk}{\hat{k}}
\newcommand{\vk}{\vec k}
\newcommand{\vq}{\vec q}
\newcommand{\vkmq}{\vec k-\vec q}

\newcommand{\km}{k_{\rm M}}

\def\hinvMpc{h\,{\rm Mpc}^{-1}}
\def\Mpcinvh{{\rm Mpc}/h}

\newcommand{\code}[1]{\texttt{#1}}

\usepackage[colorlinks,bookmarks]{hyperref}
\definecolor{linkblue}{rgb}{0,0,0.8}
\definecolor{linkgreen}{rgb}{0,0.5,0}

\hypersetup{pdfpagemode=UseNone, pdfstartview=FitH, linkcolor=linkblue,
            citecolor=linkgreen, urlcolor=linkblue}

\begin{document}

\vspace*{-25mm}

\begin{center}

{\Large \bf The cosmological analysis of DES 3$\times$2pt data\\[0.3cm] from the Effective Field Theory of Large-Scale Structure}  \\[0.7cm]
{\large   Guido D'Amico${}^{1,2}$,  Alexandre Refregier${}^{3}$, \\[0.3cm] Leonardo Senatore${}^{4}$, and  Pierre Zhang${}^{3,4,5}$ \\[0.7cm]}

\end{center}

\begin{center}

\vspace{.0cm}

\begin{small}

{ { \sl $^{1}$ Department of Mathematical, Physical and Computer Sciences,\\ University of Parma, 43124 Parma, Italy}}
\vspace{.05in}

{ { \sl $^{2}$ INFN Gruppo Collegato di Parma, 43124 Parma, Italy}}
\vspace{.05in}

{ { \sl $^{3}$ Institute for Particle Physics and Astrophysics, ETH Zürich, 8093 Zürich, Switzerland}}
\vspace{.05in}

{ { \sl $^{4}$ Institute for Theoretical Physics, ETH Zürich, 8093 Zürich, Switzerland}}
\vspace{.05in}

{ { \sl $^{5}$ Dipartimento di Fisica “Aldo Pontremoli”, Universit\`a degli Studi di Milano, 20133 Milan, Italy}}
\vspace{.05in}

\end{small}
\end{center}

\hrule \vspace{0.3cm}
{\small  \noindent \textbf{Abstract} 
\noindent  
We analyze the Dark Energy Survey (DES) Year 3 data using predictions from the Effective Field Theory of Large-Scale Structure (EFTofLSS). 
Specifically, we fit three two-point observables (3$\times$2pt), galaxy clustering, galaxy-galaxy lensing, and cosmic shear, using the one-loop expressions for the projected angular correlation functions. 
We validate our pipeline against numerical simulations and we check for several internal consistencies before applying it to the observational data. 
Fixing the spectral tilt and the baryons abundance, we measure $S_8=0.833\pm 0.032$, $\Omega_m = 0.272\pm 0.022$, and $h = 0.773\pm 0.049$, to about $3.8\%$, $8.1\%$, and $6.3\%$, at $68\%$CL, respectively. 
Our results are consistent at the $\sim 1.5-2\sigma$ level with those from Planck and the BOSS full-shape analyses, as well as with those from DES collaboration 3$\times$2pt analysis combined with a Big-Bang Nucleosynthesis prior and a Planck prior on $n_s$.
The shift in the posterior compared to DES collaboration results highlights the impact of modeling, scale cuts, and choice of prior.
The theory code and likelihood used for our analyses, \texttt{PyFowl}, is made publicly available. 
\vspace{0.3cm}}
\hrule

\vspace{0.3cm}
\newpage

\tableofcontents

\section{Introduction}  \label{sec:intro}

The last few decades have seen the establishment of the $\Lambda$CDM model as the standard model for cosmology. This was achieved through the combination of different cosmological probes, which provide high-precision constraints and tests of this concordance model. In particular, recent stringent constraints have been derived by combining weak gravitational lensing, galaxy clustering, and galaxy-galaxy lensing. These 3$\times$2pt measurements make use of the direct measurement of the density field via weak lensing, with the higher signal-to-noise ratio afforded by the galaxy number field. It also allows for the cross-calibration of systematics effects such as galaxy bias and intrinsic alignments.

Recent data releases from DESI~\cite{DESI:2025zgx}, ACT~\cite{ACT:2025tim} and SPT~\cite{SPT-3G:2025bzu} have shown some tensions (or quasi-tensions) with the standard $\Lambda$CDM model: in particular, we refer to the hint for a time-dependent dark energy and to the fact that the upper bound on neutrino masses is dangerously close to the lower bound from neutrino oscillations.
It is therefore important to further increase the constraints on the cosmological model to test these deviations and possible new physics. For this purpose, the inclusion of smaller scales (or higher $k$-modes) is both crucial to reduce statistical error and also challenging as it requires the accurate modeling of the non-linear evolution of cosmic structures. 
In this paper, we apply the effective field theory of large-scale structure (EFTofLSS) to robustly model the observables of the 3$\times$2pt analysis, and we apply our model to the Year 3 data release from the Dark Energy Survey collaboration (DES)~\cite{DES:2018gui}.

The EFTofLSS is a field-theory based approach that aims at accurately describing the long-wavelength dynamics of the universe.
On scales longer than the non-linear scale, at around 10 Mpc, the predictions of the EFTofLSS are supposed to approach the exact dynamics, up to tiny non-perturbative eﬀects.
Such an accuracy comes at the cost that, on scales shorter than the non-linear scale, the theory ceases to be trustworthy.\footnote{The EFTofLSS is expected to be able to describe every aspect of the long-wavelength dynamics of the universe: dark matter~\cite{Baumann:2010tm,Carrasco:2012cv,Carrasco:2013sva,Carrasco:2013mua,Pajer:2013jj,Porto:2013qua,Carroll:2013oxa,Mercolli:2013bsa,Senatore:2014via,Senatore:2014vja,Angulo:2014tfa,Baldauf:2014qfa,Baldauf:2015zga,Baldauf:2015xfa,Foreman:2015lca,Vlah:2015sea,Baldauf:2015tla,Abolhasani:2015mra,Baldauf:2015aha,McQuinn:2015tva,Bertolini:2016bmt,Cataneo:2016suz,Blas:2016sfa,Senatore:2017pbn,Lewandowski:2018ywf,Lewandowski:2017kes,Konstandin:2019bay},
baryons (see e.g.~\cite{Lewandowski:2014rca,Braganca:2020nhv}), neutrinos (see e.g.~\cite{Senatore:2017hyk,deBelsunce:2018xtd}), halos, galaxies, and
any collapsed objects~\cite{Senatore:2014eva,Mirbabayi:2014zca,Angulo:2015eqa,Fujita:2016dne,Perko:2016puo,Nadler:2017qto,Donath:2020abv} (see also~\cite{McDonald:2009dh}), time-dependent and smooth or clustering dark energy (see e.g.~\cite{Lewandowski:2016yce,Cusin:2017wjg,Bose:2018orj,DAmico:2020tty,Lu:2025gki}.
This includes predictions for both density and velocity fields, in real and redshift space (see e.g.~\cite{Senatore:2014vja,Perko:2016puo,Lewandowski:2015ziq}).}
It took several essential steps to develop the EFTofLSS to such a level that it could be reliably applied to data,\footnote{A thorough analysis of comparison of the various predictions against simulations was carried out in many papers (see e.g.~\cite{Carrasco:2012cv,Angulo:2015eqa,Foreman:2015lca,Angulo:2014tfa,DAmico:2019fhj,Colas:2019ret,Nishimichi:2020tvu,Chen:2020zjt}). After these comparisons, it was still unclear if the EFTofLSS was a useful tool to actually extract cosmological information from observations. This required the embedding of the EFTofLSS in Monte Carlo sampling tools, and the development of techniques for the fast calculation of the EFTofLSS predictions (see e.g.~\cite{Simonovic:2017mhp,Anastasiou:2022udy,Bakx:2024zgu}).
The outcome was the first direct application of the EFTofLSS to observational data, in particular the BOSS galaxy clustering data~\cite{DAmico:2019fhj,Ivanov:2019pdj,Colas:2019ret,Chen:2021wdi,DAmico:2022osl}. With the EFTofLSS, one was able to analyze the power spectrum and bispectrum and reliably extract cosmological parameters with error bars, at least for some of the parameters, competitive with cosmic microwave background (CMB) experiments. Once the technology has been developed, the final data analysis is conceptually relatively simple: one uses all the data below a certain wavenumber, and for this reason the method has been dubbed ``full shape''. It is now a standard tool also for official collaborations such as DESI~\cite{DESI:2024hhd}.} and, in this regard, the application to lensing data and the full 3$\times$2pt analysis represents a very important step further.

To date, the application of EFTofLSS to lensing observables has not been fully developed.
In the context of the EFTofLSS, lensing predictions were first explored in the context of CMB by~\cite{Braganca:2020nhv,Foreman:2015uva}, and a hybrid approach, in combination with numerical simulations, was applied to DES Year 1 data in~\cite{Hadzhiyska:2021xbv}. 
Ref.~\cite{Chen:2024vvk} (see also ref.~\cite{Verdiani:2025jgz}) recently applied the EFTofLSS to analyze DESI galaxy clustering in harmonic space, combined with galaxy-galaxy lensing using DES Year 3 source galaxies. 
Here we perform an analysis exclusively based on the prediction of the EFTofLSS, to be applied to current and future lensing surveys.
Our first step is to develop the formalism needed to apply EFTofLSS predictions to the observables of the 3$\times$2pt analysis. 
In contrast to ref.~\cite{Chen:2024vvk}, we work in angular space, rely solely on DES Year 3 data, and additionally model the shear within the EFTofLSS, enabling the first EFTofLSS-based 3$\times$2pt analysis using purely photometric data. 
We then introduce a method to determine the scale cuts up to which the data can be reliably analyzed.
We will also include several systematic effects: some, such as the redshift uncertainties, are of observational origin, and we adopt the collaboration prescription for them; others, as the baryonic effects, are physical effects which we will model with the EFTofLSS. 

\begin{figure}[ht!]
 \centering
 \includegraphics[width=0.49\textwidth]{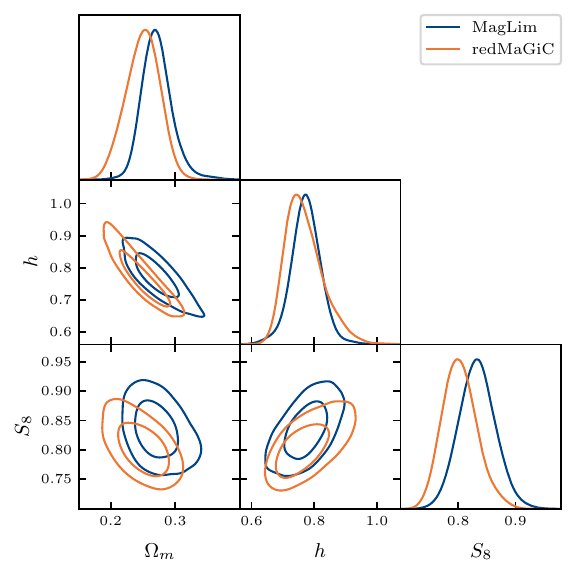}
  \includegraphics[width=0.49\textwidth]{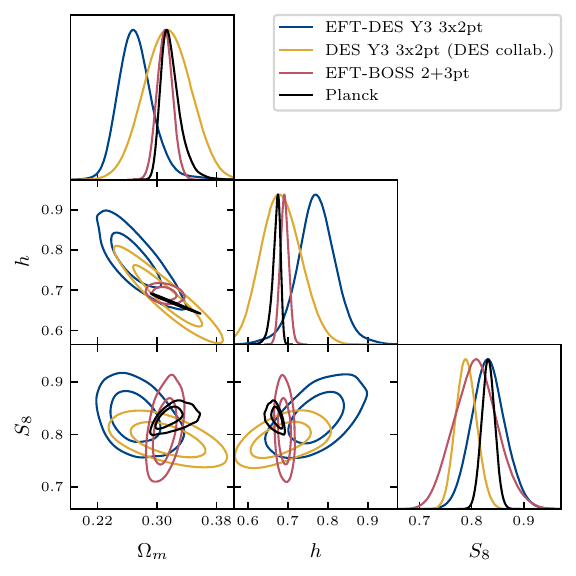}
  \caption{\footnotesize  
   Triangle plots of $\Lambda$CDM cosmological parameters from the EFTofLSS analysis of DES Y3 3$\times$2pt data, with $\omega_b$ and $n_{s}$ set to BBN and Planck preferred values, respectively. 
   \emph{Left panel}: Results using either the \texttt{MagLim} or \texttt{redMaGiC} sample. 
   \emph{Right panel}: In comparison to the \texttt{MagLim} results are shown constraints from \textit{i)} DES Y3 3$\times$2pt analysis by the DES collaboration~\cite{DES:2021wwk} where the publicly-released MCMC chain products are post-processed with a BBN prior on $\omega_b$ and a Planck prior on $n_s$; \textit{ii)} BOSS galaxy clustering power spectrum and bispectrum analysis from the EFTofLSS at one loop~\cite{DAmico:2022osl}; and \textit{iii)} CMB data from Planck with free neutrino mass~\cite{Planck:2018vyg}. 
   }
  \label{fig:main}
  \end{figure}

We believe our analysis is important for three main reasons.
First, it is essential to analyze all the data produced by the observational efforts using the best theoretical models available.
Second, lensing observables exhibit degeneracies that differ from those in clustering or CMB data, making their combination particularly powerful. While we do not perform a joint analysis in this paper, we lay the groundwork for a combined analysis by establishing a reliable EFTofLSS-based pipeline for lensing data.
Third, given the current tensions among different cosmological datasets, it is valuable to obtain parameter constraints that are subject to different degeneracies and systematics than other observables. 
Our main results are presented in fig.~\ref{fig:main}, and we publicly release our EFTofLSS 3$\times$2pt analysis code, \texttt{PyFowl}\footnote{\url{https://github.com/pierrexyz/pyfowl}} (Python code for Fast Observables in Weak Lensing).

The paper is organized as follows. 
In sec.~\ref{sec:data}, we describe the data products and the measurements we use. 
The theory model, including observational aspects, are described in sec.~\ref{sec:theory}. 
The likelihood, prior, and inference setup are presented in sec.~\ref{sec:setup}.
We validate our pipeline in sec.~\ref{sec:validation} through various checks and tests against simulations. 
Finally, our cosmological results are presented in sec.~\ref{sec:results}, where we also provide comparison with other cosmological probes. 
We conclude and outline future directions in sec.~\ref{sec:conclusion}. 
Supplementary materials and technical aspects are provided in the appendices.


\section{Data}  \label{sec:data}

\begin{figure}[h]
 \centering
 \includegraphics[width=0.66\textwidth]{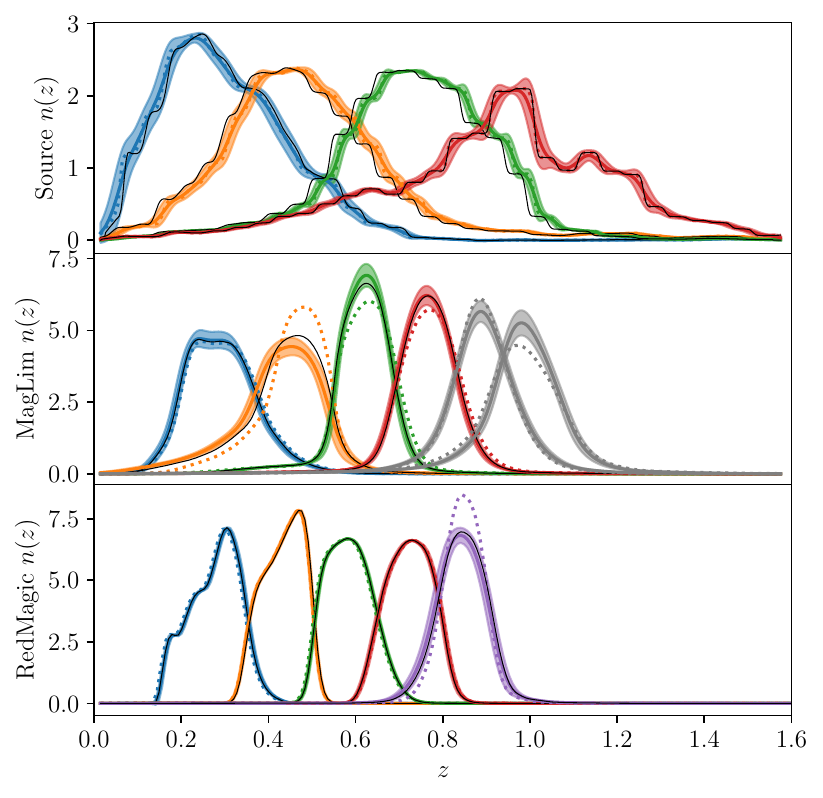}
  \caption{\footnotesize  
   DES Y3 redshift distributions of source galaxies (upper panel) and lens galaxies selected with \texttt{MagLim} or \texttt{redMaGiC} (lower panels). 
   Distributions are normalised such their integral over $z$ is one.
   As explained in the main text, when analysing the \texttt{MagLim} sample only the first four redshift bins are used. 
   The dotted lines are the redshifts determined from the photometric data, whereas the continuous lines together with the shaded regions represent the $1\sigma$-confidence intervals obtained from calibrating the redshifts to spectroscopic data as explained in~\cite{DES:2020sjz} and sec.~\ref{sec:observational}. 
   Those are used as prior on photo-$z$ uncertainties in the DES collaboration analysis~\cite{DES:2021wwk} and the present analysis.  
The thin black lines are obtained using the photo-$z$ parameters from the best-fits in our cosmological analysis. 
The main cosmological results presented in this analysis, when not explicitly specified, are obtained using the first four redshift bins of \texttt{MagLim} sample, while the \texttt{redMaGiC} sample is mainly used for comparison purposes. }
  \label{fig:nz}
  \end{figure}

The Dark Energy Survey (DES)~\cite{DES:2016jjg}, operating the Blanco 4m telescope with the Dark Energy Camera~\cite{Flaugher_2015} at Cerro Tololo in Chile, images galaxies through $grizY$-band photometry over a $\sim 5000 \ \textrm{deg}^2$ sky. 
After processing the first three years of observations, the survey released the wide-field `Gold' sample, which is the dataset for cosmological use, consisting of $\sim 319$ million galaxies~\cite{DES:2020aks}. 
The catalog is calibrated with the use of a deep-field survey~\cite{DES:2020drs}, where in particular selection functions are tested through the \texttt{Balrog} framework~\cite{DES:2020jnm}.
In the end, DES Y3 data are gathered into four redshift bins of source galaxies, with uncertainties summarised in ref.~\cite{DES:2020sjz}, and five, or six, redshift bins of lens galaxies either selected through the \texttt{redMaGiC}~\cite{DES:2015pcw,DES:2021zxv} or \texttt{MagLim}~\cite{DES:2020ajx,DES:2021bpo} pipelines, respectively. 
Those final selection functions are shown in fig.~\ref{fig:nz}. 
After unblinding, the last two redshift bins in \texttt{MagLim} were removed from the cosmological analysis for issues discussed in ref.~\cite{DES:2021wwk}. 
We therefore also use only the first four redshift bins when analysing the \texttt{MagLim} sample. 
The shear catalog~\cite{DES:2020ekd} consisting of $100$ millions galaxies is obtained by inferring the ellipticity and other relevant quantities using the \texttt{METACALIBRATION} pipeline~\cite{Huff:2017qxu,Sheldon:2017szh}.

\begin{figure}[ht!]
\centering
\includegraphics[width=0.99\textwidth]{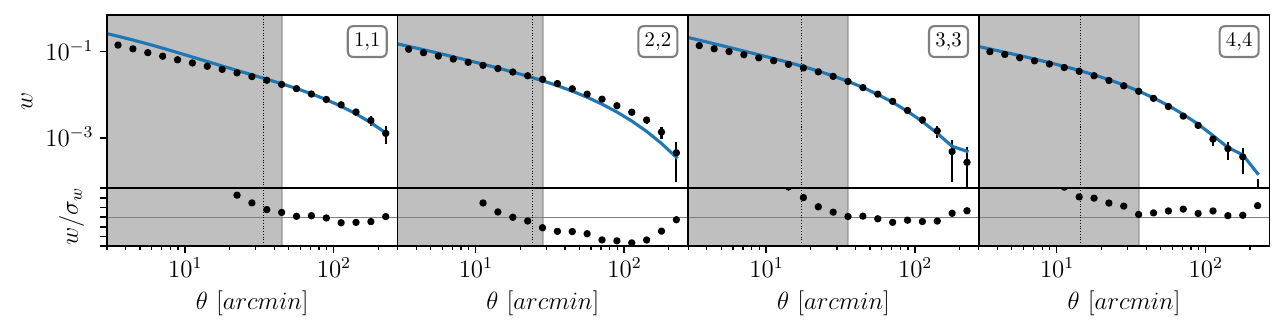}
\includegraphics[width=0.99\textwidth]{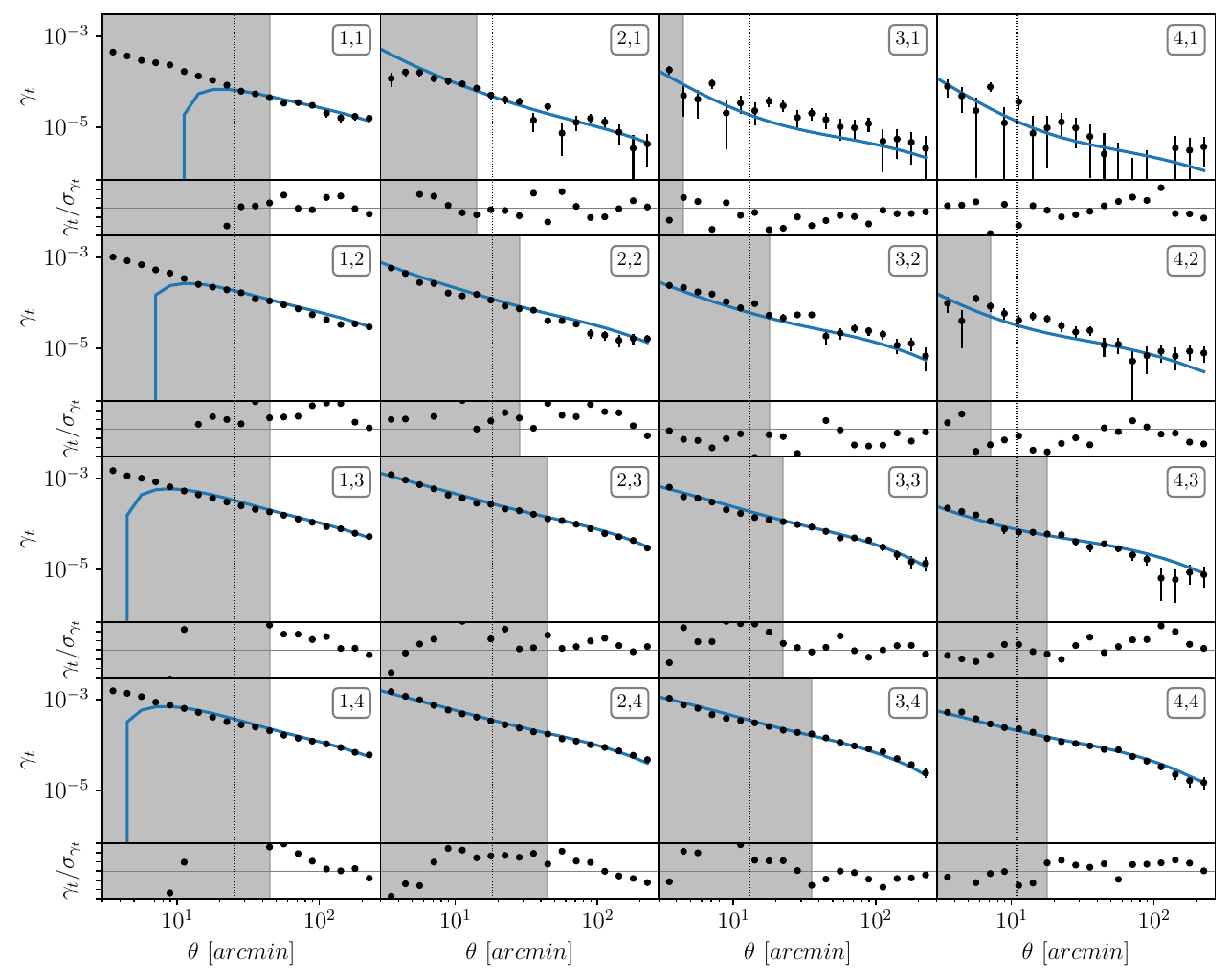}
\caption{\footnotesize  
DES Y3 two-point angular correlation functions: galaxy clustering $w$ and galaxy-galaxy lensing $\gamma_t$.
In the upper part of each plot, the black dots are the data points with their error bars, and the blue lines are the best-fit predictions from the EFTofLSS presented in this work.
The lower part of each plot shows the residuals of the best-fit curves relative to the data diagonal errors (with $y$-axis corresponding to $\pm 3\sigma$).
The shaded regions are excluded by the scale cuts used in this analysis; for reference, the DES collaboration scale cut choice~\cite{DES:2021wwk} is shown in dotted vertical lines.
For the $\gamma_t$ correlations, the rows (first indices) correspond to the lenses, while the columns (second indices) corresponds to the sources.
}
\label{fig:bestfit_gal}
\end{figure}

\begin{figure}[ht!]
\centering
\includegraphics[width=0.99\textwidth]{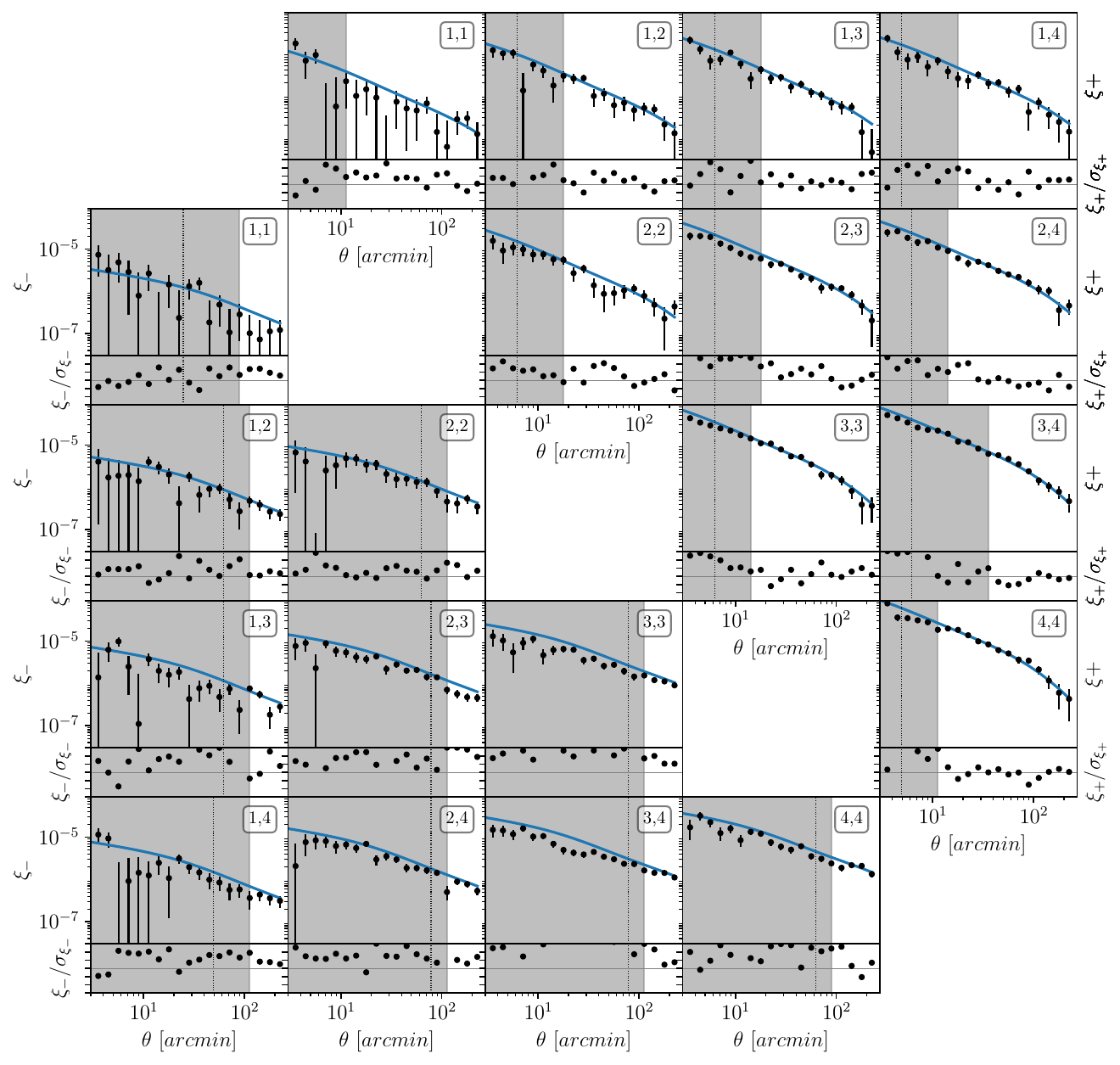}
\caption{\footnotesize  
DES Y3 two-point angular correlation functions: cosmic shear $\xi_\pm$.
In the upper part of each plot, the black dots are the data points with their error bars, and the blue lines are the best-fit predictions from the EFTofLSS presented in this work.
The lower part of each plot shows the residuals of the best-fit curves relative to the data diagonal errors (with $y$-axis corresponding to $\pm 3\sigma$).
The shaded regions are excluded by the scale cuts used in this analysis; for reference, the DES collaboration scale cut choice~\cite{DES:2021wwk} is shown in dotted vertical lines.
}
\label{fig:bestfit_shear}
\end{figure}

In this work, we use the final data products from DES Year 3 3$\times$2pt cosmological analysis~\cite{DES:2021wwk}.\footnote{Publicly available at \url{https://des.ncsa.illinois.edu/releases/y3a2/Y3key-products}}
It consists of three sets of two-point projected correlation functions over angular separation $\theta$:
\begin{itemize}
\item \textit{Galaxy clustering}: the auto-correlation of lens galaxy positions $w^i(\theta)$ within lens redshift bin $i = 1, \dots, N_{lens}$, estimated in ref.~\cite{DES:2021bat}.
\item \textit{Galaxy-galaxy lensing}: The correlation between lens galaxy positions and source galaxy tangential shear in redshift bins $i = 1, \dots, N_{lens}$ and $j=1, \dots, N_{source}$, respectively, $\gamma_t^{ij}(\theta)$, estimated in ref.~\cite{DES:2021qnp}. 
\item \textit{Cosmic shear}: The correlation functions between source galaxy shears in redshift bins $i$ and $j$, $\xi_{\pm}^{ij}(\theta)$, estimated in ref.~\cite{DES:2021bvc,DES:2021vln}. 
\end{itemize}
The full data vector, split into $20$ logarithmic angular bins from $2.5$ to $250$ arcmin, from the \texttt{MagLim} sample is shown in figs.~\ref{fig:bestfit_gal}~and~\ref{fig:bestfit_shear} while the one from the \texttt{redMaGiC} sample is shown in app.~\ref{app:redmag}. 
With the scale cuts determined using the procedure described in sec.~\ref{sec:scalecut}, there is a total of respectively 382 and 494 data points that we include our analysis. 
As for the data covariance, we use the analytic estimate of ref.~\cite{DES:2020ypx}. 

We also make use of the \texttt{Buzzard} simulation suites~\cite{DES:2019jmj,DES:2021bwg}. 
The \texttt{Buzzard v2.0}~\cite{DES:2021bwg} are realistic full-survey simulations of DES Y3 3$\times$2pt data that are used for validation of cosmological analysis pipelines. 
In total, there are $18$ independent realisations whose data vectors we average such that the simulation data has small enough noise to gauge the level of systematics in our cosmological measurements.

%
%

%
%

\section{Theory}
\label{sec:theory} 

After reviewing the underlying fields probed in photometric surveys in sec.~\ref{sec:shear} and sec.~\ref{sec:gal_delta}, we describe their angular two-point functions in sec.~\ref{sec:2pt} and connect them with predictions in the EFTofLSS in sec.~\ref{sec:eftoflss}. 
Additional observational effects are discussed in sec.~\ref{sec:observational}. 
Some details are relegated to appendices.
We provide a re-derivation of the magnification bias in app.~\ref{app:mag}, and of the Limber approximation in app.~\ref{app:limber}. 
Baryonic effects predicted in the EFTofLSS and how they enter our observables are discussed in sec.~\ref{app:baryons}.

\subsection{Lensing potential, convergence, and shear}
\label{sec:shear}
In a weak lensing survey, we measure correlations of the distortions of the images of distant galaxies (the sources) due to the intervening gravitational potential along the line of sight.
In a spherical coordinate system centered on the observer, we observe galaxies in the angular direction $\hn = (\sin \vartheta \, \cos \varphi, \sin \vartheta \, \sin \varphi, \cos \vartheta)$, which is our line of sight, at a redshift $z$, related to their comoving distance $\chi$ in an FLRW universe by $\chi = \int_0^z \frac{\rmd z'}{H(z')}$, $H(z)$ being the Hubble function.
The observed shape for a galaxy which is at angular position $\hn$ can be described by (2-dimensional) vectors in the plane orthogonal to $\hat{n}$, i.e. the plane tangent to the celestial sphere.
As a basis for this tangent space we choose two unit vectors $\hat{e}_1$, $\hat{e}_2$, with coordinates $\hat{e}_1 = (\cos \vartheta \, \cos \varphi, \cos \vartheta \, \sin \varphi, - \sin \vartheta)$, $\hat{e}_2 = (- \sin \varphi, \cos \varphi, 0)$.\footnote{To help the intuition, we note that for $\vartheta = 0$, $\varphi=0$, we have $\hn = (0, 0, 1)$, $\hat{e}_1 = (1, 0, 0)$, $\hat{e}_2 = (0, 1, 0)$.}
Because of lensing, a light ray that would be observed at a 2D position $\vec{\alpha}$ on the tangent plane is instead observed at $\vec{\beta}$.
To lowest order in the gravitational potential, this map can be written as $\beta_i = A_{ij} \alpha_j = \alpha_i - M_{ij} \alpha_j$, where we introduce the $2 \times 2$ distortion matrix $M_{ij}$:
\begin{equation} 
  M = 
  \begin{pmatrix}
    \kappa + \gam_1 & \gam_2 \\
    \gam_2 & \kappa - \gam_1
  \end{pmatrix} \, ,
  \label{eq:Mmatrix}
\end{equation}
parametrized by the convergence $\kappa$ and shear components $\gam_1$, $\gam_2$ (for a derivation, see e.g.~\cite{Dodelson:2003ft,Weinberg:2008zzc}).
Physically, if we consider the lensed image of a circular shape, the convergence quantifies the increase in the surface area and the shear component describes the elliptical (area-preserving) distortion of the image.
The distortion matrix is given by the second derivatives (on the plane tangent to the sphere) of a \emph{lensing potential}, $M_{ij} = \frac{\de}{\de \alpha_i}  \frac{\de}{\de \alpha_j} \psi(\hn)$.

The lensing potential is a projection of the gravitational potential $\phi$.
For a distribution of sources with normalized number density\footnote{The two densities, $N_{source}$ and $n_{source}$, are both normalized in the $z$ variable. The definition stems from the fact that $\rmd \chi \, N_{source}(\chi) = \rmd z \, n_{source}(z)$.} $N_{source}(\chi) \equiv H(z(\chi)) \, n_{source}(z(\chi))$, $\psi$ can be written as the following integral:
\begin{equation}
  \psi(\hn) = 2 \int_0^\infty \rmd \chi  \, N_{source}(\chi) \int_0^\chi \rmd \tchi \frac{\chi - \tchi}{\chi \tchi} \phi\(\tchi \hn, z(\tchi) \)
  = 2 \int_0^\chi \rmd \chi \, \frac{g(\chi)}{\chi^2} \phi\(\chi \hn, z(\chi) \) \, ,
  \label{eq:psi}
\end{equation}
where\footnote{To derive this form of the equation, we first notice that the domain of integration in $\tchi$, $\chi$, written as $0 < \tchi < \chi$, $0 < \chi < \infty$, can be equivalently expressed as $0 < \tchi < \infty$, $\tchi < \chi < \infty$:
\begin{equation}
  \psi(\hn) = 2 \int_0^\infty \rmd \chi  \int_0^\chi \rmd \tchi \, N_{source}(\chi) \frac{\chi - \tchi}{\chi \tchi} \phi\(\tchi \hn, z(\tchi) \)
  = 2 \int_0^\infty \rmd \tchi \int_{\tchi}^\infty \rmd \chi \, N_{source}(\chi)  \frac{\chi - \tchi}{\chi \tchi} \phi\(\tchi \hn, z(\tchi)\) \, .
\end{equation}
We finally rename the integration variables $\chi \leftrightarrow \tchi$ to arrive at Eq.~\eqref{eq:gdef}.}
\begin{equation}
  \label{eq:gdef}
  g(\chi) = \chi \int_\chi^\infty \rmd \tchi \, N_{source}(\tchi) \frac{\tchi - \chi}{\tchi} \, .
\end{equation}
To calculate correlation functions, we start with the spherical harmonic expansion $\psi(\hn) = \sum_{\ell m} \psi_{\ell m} Y_{\ell m}(\hn)$.
To calculate the $\psi_{\ell m}$, it is convenient to Fourier transform $\phi$:
\begin{equation}
  \psi(\hn) = 2 \int_0^\infty \rmd \chi \, \frac{g(\chi)}{\chi^2} \int_{\vk} e^{i \vk \cdot \hn \chi} \, \phi(\vk, z(\chi) ) \, ,
\end{equation}
and using the spherical wave expansion of the exponential\footnote{
We remind the reader of the expression
$
e^{i \vk \cdot \hn \chi} = 4 \pi \sum_{\ell m} i^\ell j_\ell(k \chi) Y_{\ell m}(\hn) Y^\ast_{\ell m}(\hk) \, ,
$ where $j_\ell$ is the spherical Bessel function of order $\ell$.},
we can read the multipole coefficients
\begin{equation}
  \psi_{\ell m} = i^\ell \, 8 \pi \int_{\vk} \,  Y^\ast_{\ell m}(\hk) \int_0^\infty \rmd \chi \,  j_\ell(k \chi) \, \frac{g(\chi)}{\chi^2} \,\phi(\vk, z(\chi) ) \, .
  \label{eq:psilm}
\end{equation}
We can finally use the Poisson equation $k^2 \phi(\vk, a) = - \frac{3}{2 a} \Om_m H_0^2 \del_m(\vk, a)$ to rewrite:
\begin{equation}
  \psi_{\ell m} = - i^\ell \, 8 \pi  \int_{\vk} \,  Y^\ast_{\ell m}(\hk) \int_0^\infty \rmd \chi \,  j_\ell(k \chi)  \, \frac{f_\kappa(\chi)}{k^2 \chi^2} \,  \del_m(\vk, z(\chi) ) \, ,
\end{equation}
where we define the \emph{lensing efficiency}:
\begin{equation}\label{eq:lensing_efficiency}
  f_\kappa(\chi) = \frac{3}{2} \Om_m H_0^2 (1+z(\chi)) \,  \chi \int_\chi^\infty \rmd \tchi \, N_{source}(\tchi) \frac{\tchi - \chi}{\tchi} \, .
\end{equation}

Given this expression for $\psi$, we can use the definition of $M_{ij}$ to get an expression for the shear.
First, it is convenient to define a complex shear $\gamma$ by the relation $\gam = \gam_1 + i \gam_2 = e_+^i e_+^j M_{ij}$, where $\hat{e}_+ = (\hat{e}_1 + i \hat{e}_2)/\sqrt{2}$.
This definition makes manifest the fact that the shear is a spin-2 quantity, because, under a rotation of angle $\alpha$ around the axis $\hat{n}$, it transforms as $\gam \to \gam e^{2 i \alpha}$: intuitively, an ellipse goes into itself under a rotation of $\pi$~\footnote{Fixing coordinates such as $\hn=(0,0,1)$, a rotation of angle $\alpha$ around $\hn$ is described by the standard 2$\times$2 matrix $\begin{pmatrix}
  \cos \alpha & \sin \alpha \\ - \sin \alpha & \cos \alpha \end{pmatrix}$, which brings $\hat{e}_+ \to \hat{e}_+ e^{ i \alpha}$. }.
Therefore, $\gamma$ has to be expanded in spin-2 spherical harmonics, ${}_{2}Y_{\ell m}(\hn)$, as it is the case for the CMB polarization, for which the formalism is very similar~\cite{Weinberg:2008zzc,Dodelson:2003ft}.
Using the relation $M_{ij} = \de_i \de_j \psi(\hat{n})$, and the definition of the spin-2 spherical harmonics~\cite{Goldberg:1966uu}, we find
\begin{equation}
  \gam(\hn) =  \frac{1}{2} \sum_{\ell m} \sqrt{\frac{(\ell+2)!}{(\ell-2)!}} \, {}_{2}Y_{\ell m}(\hn) \, \psi_{\ell m}  \, .
\end{equation}

As defined, the components $\gam_1$ and $\gam_2$, hence also $\gam$, depend on the choice of coordinates.
There is, however, a natural reference direction to consider when measuring shapes, which is the one connecting the lens to the lensed image of the source, projected on the tangent plane.
To make contact with observations, it is then useful to define the \emph{tangential shear} $\gamma_{\parallel}$ and \emph{cross shear} $\gamma_{\times}$ components with respect to this reference direction:
\begin{equation}
  \gamma_{\parallel} = - \Re[\gam e^{-2 i \alpha}] \, , \quad
  \gamma_{\times} = - \Im[\gam e^{-2 i \alpha}] \, ,
\end{equation}
where $\alpha$ is the angle between the (arbitrary) coordinate axis $\hat{e}_1$ and the reference direction.
The $\gamma_{\parallel}$ is nonzero for distortions oriented along or orthogonal with respect to the line connecting galaxy and lens; $\gamma_{\times}$ is nonzero for distortions along axes rotated $\pm 45^\circ$ with respect to it.

We now make an important comment about parity. 
Under parity, we have $\hat{n} \to -\hat{n}$, and a right-handed coordinate system transforms into a left-handed coordinate system.
In particular, in the tangent plane we see a reflection with respect to the axis $\hat{e}_1$.
Under parity, $\gamma_1 \to \gamma_1$ and $\gamma_2 \to - \gamma_2$ (so $\gamma \to \gamma^\star$), and the angle $\alpha \to - \alpha$.
It follows that $\gamma_{\parallel}$ is parity-even while $\gamma_{\times}$ is parity-odd.
Thus, in the absence of systematics (and assuming a parity symmetric Universe), the two-point functions of $\gamma_\times$ with parity-even fields such as $\gamma_\parallel$ or the overdensity vanish.

\subsection{Projected galaxy density}
\label{sec:gal_delta}
In the 3$\times$2pt analysis, the lens galaxies are described by the angular projection of their density field.
In the direction $\hn$ and for the redshift bin $i$, the (projected) density contrast can be written as the sum of three contributions, that we describe explicitly in the following:
\begin{equation}
  \del_G^{i}(\hn) = \del_g^{i}(\hn) + \del_{RSD}^{i}(\hn) + \del_\mu^{i}(\hn) \, .
\end{equation}
Here $\del^{i}_g(\hn)$ is given by
\begin{equation}
  \del^{i}_g(\hn) = \int \rmd \chi \, N_{lens}^i(\chi) \del_g(\chi \hn, z(\chi))
  = \int \rmd \chi \, N_{lens}^i(\chi) \int_{\vk} e^{i \vk \cdot \hn \chi} \del_g(\vk, z(\chi)) \, ,
  \label{eq:deltan}
\end{equation}
where $N_{lens}^i(\chi) = H(z(\chi)) \, n_g^i(z(\chi))$ is the selection function, for bin $i$, of lenses with number density $n_g^i(z)$, normalized to $\int_0^{\infty} \rmd z \, n_g^i(z) = 1$.\footnote{As for the source galaxies, this expression comes from $n_{g}^i(z) \, \rmd z = N_{lens}^i(\chi) \, \rmd \chi$. From here and now on, we will sometimes drop from the notation of the redshift $z \equiv z(\chi)$, or the comoving distance $\chi \equiv \chi(z)$, their inter-dependence, as it is clear how to interchange them.}
In harmonic space, after expanding the exponential in spherical harmonics\footnote{We remind the reader of the expression
\begin{equation*}
  e^{i \vk \cdot \hn \chi} = 4 \pi \sum_{\ell m} i^\ell j_\ell(k \chi) Y^*_{\ell m}(\hat{k}) Y_{\ell m}(\hat{n}) \, .
\end{equation*}}, we read the coefficients:
\begin{equation}
  \del_{g, \ell m}^{i} = 4 \pi i^\ell \int_{\vk} Y^*_{\ell m}(\hk) \int_0^\infty \rmd \chi j_\ell(k \chi) N_{lens}^i(\chi) \del_g(\vk, z(\chi)) \, .
\end{equation}
In the integrand, assuming the background relation between $z$ and $\chi$, $\delta_g(\vk, z)$ is then the nonlinear density field in real space.

\begin{figure}[ht!]
\centering
\includegraphics[width=0.99\textwidth]{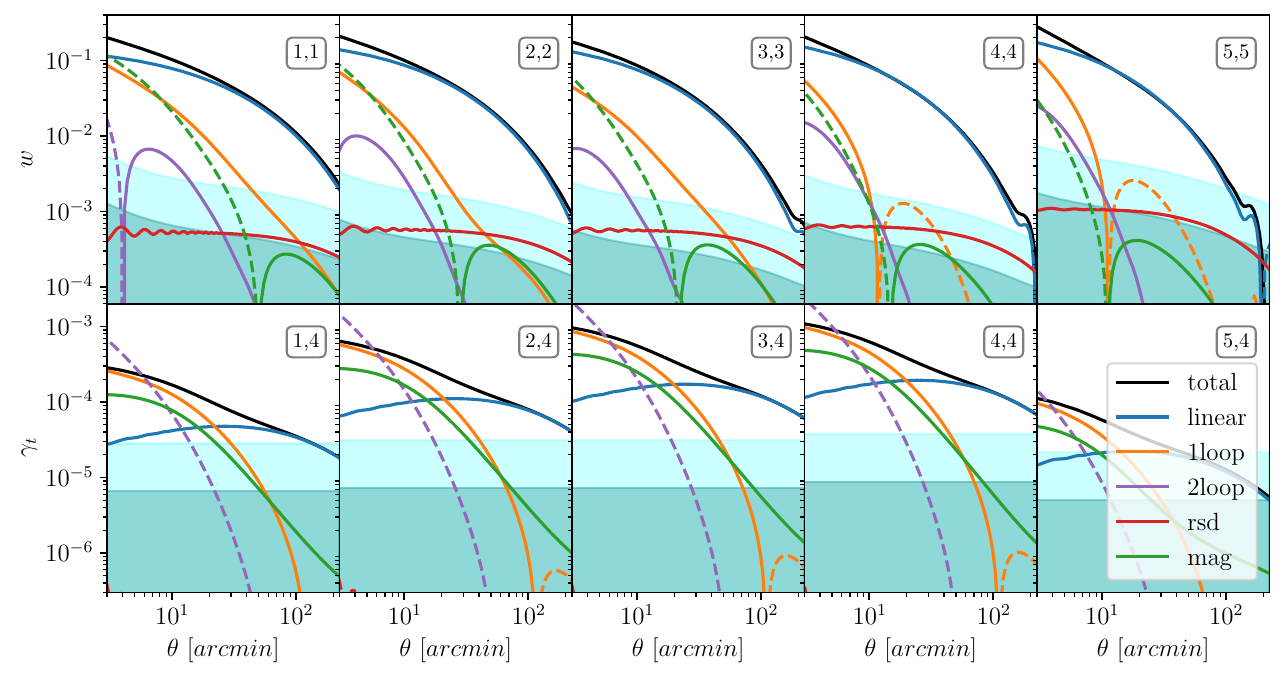}
\caption{\footnotesize  
Summary of the main contributions entering the predictions of galaxy clustering $w$ and galaxy-galaxy lensing $\gamma_t$ (shown for one source redshift bin only) considered in this work, from the best-fit obtained fitting the \texttt{Buzzard} simulations.
The total signal is compared to the individual contributions in the EFTofLSS: linear, one loop, and two loop. 
The latter, approximated by eq.~\eqref{eq:2loop}, does not enter in our baseline analysis setup and is used to calibrate our choice of scale-cuts.
The linear and 1-loop lines correspond to the real-space contribution. The redshift-space distortions and magnification bias, modelled at the linear level only, are explicitly shown.
When negative, contributions are shown in dashed lines.
The shaded regions are the $1-$ and $2-\sigma$ error bars from DES data, shown for reference.
}
\label{fig:contributions}
\end{figure}

In observations, the redshift is determined by the Hubble flow plus the effect of peculiar velocities, which is commonly referred to as redshift-space distortion (RSD).
For the angular correlation functions in DES redshift bins, the RSD are a small effect as shown in fig.~\ref{fig:contributions}. 
We therefore only consider the linear contribution
\begin{equation}
  \label{eq:del_RSD_hatn}
  \begin{split}
    \del_{RSD}^{i}(\hat{n}) &= - \int \rmd \chi \, N_{lens}^i(\chi) \frac{1+z}{H} \de_\chi (\vec{v} \cdot \hn)  \\
  &= \int \rmd \chi \, N_{lens}^i(\chi) \int_{\vk} e^{i \vk \cdot \hn \chi} f(z) \frac{(\vk \cdot \hat{n})^2}{k^2} D(z) \del_{\rm in}(\vk) \, ,
  \end{split}
\end{equation}
where, in the second line, we have substituted the linear solution for the velocity $\vec{v}$ in terms of the matter density, \textit{i.e.}, $v_i = - a H f D \frac{\partial_i}{\partial^2} \delta_{\rm in}$, with $f(z)$ the linear growth rate, $D(z)$ the linear growth factor normalized to $D(0)=1$, and $\del_{\rm in}(\vk)$ the linear matter density at $z=0$.
To compute the harmonic coefficients, we use the substitution:
\begin{equation}
  (\vk \cdot \hat{n})^2 \, e^{i \vk \cdot \hn \chi}= - \de_\chi^2 \, e^{i \vk \cdot \hn \chi}  = 
  - 4 \pi \sum_{\ell m} i^\ell k^2 j''_\ell(k \chi) Y^*_{\ell m}(\hat{k}) Y_{\ell m}(\hat{n}) \, .
\end{equation}
It follows that, in harmonic space,
\begin{equation}
  \del_{RSD,\ell m}^{i} = - 4 \pi i^\ell \int_{\vk} Y^*_{\ell m}(\hk) \int_0^\infty \rmd \chi N_{lens}^i(\chi) j''_\ell(k \chi) f(z) D(z) \del_{\rm in}(\vk) \, .
\end{equation}

An additional term to be considered for the projected lens galaxy density is the magnification bias, for which a detailed derivation is presented in App.~\ref{app:mag}.
Again, we can limit ourselves to linear theory; the result is the following term:
\begin{equation}
  \begin{split}
  \del_\mu^{i}(\hn) &= \int_0^{\infty} \rmd \chi N_{lens}^i(\chi) b_{\rm mag}(z) \kappa(\hn, z) \\
  &= \int_0^{\infty} \rmd \chi N_{lens}^i(\chi) b_{\rm mag}(z) \int_0^\chi \rmd \tchi \frac{\chi - \tchi}{\chi \tchi} \nabla_{\hat{n}}^2 \phi(\tchi, \tchi \hat{n}) \\
  &= \int_0^{\infty} \rmd \chi \nabla_{\hat{n}}^2 \phi(\chi, \chi \hat{n})  \int_\chi^\infty \rmd \tchi N_{lens}^i(\tchi) b_{\rm mag}(\tchi) \frac{\tchi - \chi}{\chi \tchi} \, ,
  \end{split}
\end{equation}
where we have used eq.~\eqref{eq:kappa} and, in the last line, we have rewritten the domain of integration as in eq.~\eqref{eq:psi}.
To deal with the $\nabla_{\hat{n}}^2 \phi(\chi, \chi \hat{n})$ term, as in eqs.~(\ref{eq:kappa}, \ref{eq:kappa_ylm}) we Fourier transform it and expand the exponential in spherical harmonics:
\begin{equation}
  \nabla_{\hat{n}}^2 \phi(\chi, \chi \hat{n}) = \nabla_{\hat{n}}^2 \int_{\vk} e^{i \chi \hn \cdot \vk} \phi(\chi, \vk) =
  - 4 \pi \sum_{\ell m} i^\ell \ell (\ell+1) Y_{\ell m}(\hn) \int_{\vk} j_\ell(k \chi) Y^*_{\ell m}(\hk) \phi(\chi, \vk) \, .
\end{equation}
Using the Poisson equation, we can read the harmonic coefficients:
\begin{equation}
  \begin{split}
    \del_{\mu,\ell m}^{i}
    =& 4 \pi i^\ell \ell (\ell+1) \frac{3}{2} \Omega_{m,0} H_0^2 \int_{\vk} Y^*_{\ell m}(\hk) \int_0^{\infty} \rmd \chi \, j_\ell(k \chi) \frac{1+z(\chi)}{k^2} D(z(\chi)) \delta_{\rm in}(\vk) \\
    &    \times \int_\chi^\infty \rmd \tchi N_{lens}^i(\tchi) b_{\rm mag}(\tchi)\frac{\tchi - \chi}{\chi \tilde{\chi}}  = \\
    =& 4 \pi i^\ell \ell (\ell+1) \int_{\vk} Y^*_{\ell m}(\hk) \int_0^{\infty} \rmd \chi \, j_\ell(k \chi) \frac{f_g^i(\chi)}{k^2 \chi^2} D(z) \delta_{\rm in}(\vk) \, ,
  \end{split}
\end{equation}
where we introduced the following kernel, similar to the lensing efficiency, but for the lens galaxies instead of the source galaxies:
\begin{equation}
  f_g^i(\chi) = \frac{3}{2} \Omega_{m,0} H_0^2 (1+z(\chi)) \chi 
  \int_{\chi}^\infty \rmd \tchi N_{lens}^i(\tchi) b_{\rm mag}(\tchi) \frac{\tchi - \chi}{\tilde{\chi}}  \, .
\end{equation}

\subsection{2-pt functions}\label{sec:2pt}

In the following, we will gather and summarize all the standard expressions for the projected angular 2-point functions. These can be found in the literature, e.g.~\cite{Krause:2017ekm}.

\paragraph{Fields} Let us summarize the fields out of which we compute the correlation functions measured by the survey.
The shear components, for the redshift bin $j$, are
\begin{equation}
  \gam^j(\hn) = \frac{1}{2} \sum_{\ell m} \sqrt{\frac{(\ell+2)!}{(\ell-2)!}} {}_{2} Y_{\ell m}(\hn) \psi_{\ell m}^j \, ,
  \label{eq:gampm}
\end{equation}
where
\begin{align}
  \psi_{\ell m}^j &= \int_{\vk} Y^\ast_{\ell m}(\hk) \int_0^\infty \rmd \chi W^j_{\psi, \ell}(k, \chi) \del_m(\vk, z(\chi) ) \, , \\
  W^j_{\psi, \ell}(k, \chi) &= - i^\ell \, 8 \pi \,  j_\ell(k \chi)  \, \frac{f^{j}_\kappa(\chi)}{k^2 \chi^2} \, .
\end{align}

The projected lens galaxy density field in redshift bin $i$ is the sum
\begin{equation}
  \del_G^{i}(\hn) = \del_g^{i}(\hn) + \del_{RSD}^{i}(\hn) + \del_\mu^{i}(\hn) \, .
\end{equation}
Each component is expanded in spherical harmonics, with coefficients
\begin{align}
  \del_{g,\ell m}^{i} &= \int_{\vk} Y^*_{\ell m}(\hk) \int_0^\infty \rmd \chi W^i_{g,\ell}(k, \chi) \del_g(\vk, z(\chi)) \, , \\
  W^i_{g,\ell}(k, \chi) &= 4 \pi i^\ell j_\ell(k \chi) N_{lens}^i(\chi) \, .
\end{align}
\begin{align}
  \del_{R,\ell m}^{i} &= \int_{\vk} Y^*_{\ell m}(\hk) \int_0^\infty \rmd \chi W^i_{R,\ell}(k, \chi) \del_{\rm in}(\vk) \, , \\
  W^i_{R, \ell}(k, \chi) &= - 4 \pi i^\ell j''_\ell(k \chi) N_{lens}^i(\chi) f(z) D(z) \, .
\end{align}
\begin{align}
  \del_{\mu,\ell m}^{i} &= \int_{\vk} Y^*_{\ell m}(\hk) \int_0^\infty \rmd \chi W^i_{\mu,\ell}(k, \chi) \del_{\rm in}(\vk) \, , \\
  W^i_{\mu,\ell}(k, \chi) &= 4 \pi i^\ell \ell (\ell+1) j_\ell(k \chi) \frac{f_g(\chi)}{k^2 \chi^2} D(z(\tchi)) \, .
\end{align}

\paragraph{Shear-shear} Let us start from the power spectrum of the lensing potential.
Considering sources at redshift bins $i$ and $j$, we compute $\braket{\psi^i_{\ell m} \psi^{j, \ast}_{l' m'}} = \del_{ll'} \del_{mm'} C^{ij}_{\psi \psi}(\ell)$, with
\begin{equation}
  \begin{split}
  C_{\psi \psi}^{ij}(\ell) &= \int_0^{\infty} \frac{\rmd k}{(2 \pi)^3} k^2 \int_0^\infty \rmd \chi \int_0^\infty \rmd \tchi W_{\psi,\ell}^i(k,\chi) W_{\psi,\ell}^j(k,\tchi) P_{mm}(k, z, \tilde{z}) = \\
  &= \frac{8}{\pi} \int_0^\infty \rmd \chi \int_0^\infty \rmd \tchi \int_0^\infty \rmd k \, k^2 \,  j_\ell(k \chi) j_\ell(k \tchi) \, \frac{f^i_\kappa(\chi)}{k^2 \chi^2} \frac{f^j_\kappa(\tchi)}{k^2 \tchi^2} \, P_{mm}(k, z, \tilde{z}) \, ,
  \label{eq:Cpsi}
  \end{split}
\end{equation}
where for simplicity of notation we use $z=z(\chi)$, $\tilde{z} = z(\tchi)$.
In this expression, $P_{mm}(k, z, \tilde{z})$ denotes the unequal-time power spectrum, which will be given by the EFTofLSS.
In the Limber approximation, derived in app.~\ref{app:limber}, this expression simplifies to the following:
\begin{equation}
  \begin{split}
    C^{ij}_{\psi \psi}(\ell) &= \, \frac{4}{(\ell+\frac{1}{2})^5} \int_0^\infty \rmd k \, f_{\kappa}^i\(\frac{\ell+\frac{1}{2}}{k}\) f_{\kappa}^j\(\frac{\ell+\frac{1}{2}}{k}\) P_{mm}(k, z) \\
    &= \frac{4}{(\ell+\frac{1}{2})^4} \int_0^\infty \frac{\rmd \chi}{\chi^2} \, f_{\kappa}^i(\chi) f_{\kappa}^j(\chi) P_{mm}\(\frac{\ell+\frac{1}{2}}{\chi}, z \) \, .
  \end{split}
  \label{eq:CpsiLimber}
\end{equation}
As we explain later in sec.~\ref{sec:eftoflss}, we use the Limber approximation only for the loop terms.

In terms of $C_{\psi \psi}$, it is now easy to compute the shear correlations.
Out of $\gamma$, we can construct two independent correlations, which DES measures:
\begin{align}
  \xi_+(\theta) &= \braket{\gamma_{\parallel}(\hn_1) \gamma_{\parallel}(\hn_2)} + \braket{\gamma_\times(\hn_1) \gamma_\times(\hn_2)}
  = \braket{\gamma^*(\hn_1) \gamma(\hn_2)} \, , \\
  \xi_-(\theta) &= \braket{\gamma_{\parallel}(\hn_1) \gamma_{\parallel}(\hn_2)} - \braket{\gamma_\times(\hn_1) \gamma_\times(\hn_2)}
  = \Re[\braket{\gamma(\hn_1) \gamma(\hn_2)} e^{-4 i \alpha}] \, ,
\end{align}
where $\cos \theta = \hn_1 \cdot \hn_2$.\footnote{As explained at the end of~\ref{sec:shear}, these are the two combinations one can construct that respect parity invariance.
The correlation function $\braket{\gamma_{\parallel}(\hn_1) \gamma_{\times}(\hn_2)} = 0$ in the absence of systematics.}
Using eq.~\eqref{eq:gampm}, it is easy to write down the following expressions in terms of $C_{\psi \psi}$,\footnote{Let us clarify how to obtain these expressions. For $\xi_+$,
\begin{equation}
  \begin{split}
    \xi_+(\theta) &= \frac{1}{4} \sum_{\ell m} \sum_{\ell' m'} \sqrt{\frac{(\ell+2)!}{(\ell-2)!}} \sqrt{\frac{(\ell'+2)!}{(\ell'-2)!}} \, {}_{2}Y^*_{\ell m}(\hn_1) \, {}_{2}Y_{\ell' m'}(\hn_2) \del_{\ell \ell'} \del_{m m'} \, C_{\psi \psi}(\ell) = \\
    &= \frac{1}{4} \sum_{\ell} \frac{(\ell+2)!}{(\ell-2)!} \, C_{\psi \psi}(\ell) \sum_m {}_{2}Y^*_{\ell m}(\hn_1) \, {}_{2}Y_{\ell m}(\hn_2)
    = \frac{2 \ell + 1}{4 \pi} \frac{1}{4} \sum_\ell \frac{(\ell+2)!}{(\ell-2)!} C_{\psi \psi}(\ell) \, d^\ell_{2,-2}(\theta) \, .
  \end{split}
\end{equation}
For $\xi_-$, the derivation is similar, but we have to use that
\begin{equation}
  \sum_m \Re[ {}_{2}Y_{\ell m}(\hn_1) \, {}_{2}Y_{\ell m}(\hn_2) e^{-4 i \alpha}] = \frac{2 \ell + 1}{4 \pi} d^\ell_{2,2}(\theta) \, .
\end{equation}
}
\begin{align}
  \xi_+(\theta) &= \frac{2 \ell + 1}{4 \pi} \frac{1}{4} \sum_\ell \frac{(\ell+2)!}{(\ell-2)!} C_{\psi \psi}(\ell) \, d^\ell_{2,-2}(\theta) \, , \label{eq:xi+_sum} \\
  \xi_-(\theta) &= \frac{2 \ell + 1}{4 \pi} \frac{1}{4} \sum_\ell \frac{(\ell+2)!}{(\ell-2)!} C_{\psi \psi}(\ell) \, d^\ell_{2,2}(\theta) \, , \label{eq:xi-_sum}
\end{align}
where $d_{m,m'}^\ell(\theta)$ are the Wigner $d$-matrices~\cite{Martin_1960}.

Since the angular separation of DES are very small, up to 250 arcmins, we will simplify the full sums to integrals, applying the flat-sky approximation.
This is valid for small angular separations, and it amounts to approximate the sphere with its tangent plane.
The flat-sky approximation works extremely well for DES, and for most near-future LSS surveys.\footnote{For example, we have checked that, on synthetic data with covariance of volume 18 times the one DES Y3 generated with~\eqref{eq:xi+_sum}~and~\eqref{eq:xi-_sum} but analyzed with the asymptotic expressions~\eqref{eq:xi+}~and~\eqref{eq:xi-}, the change in $\chi^2$  is less than $1$ on the DES angular separations. }
Mathematically, this amounts to use the asymptotic expression~\cite{Gradshteyn:1943cpj}:
\begin{equation}
  d^\ell_{m,m'}(\theta) \to J_{m-m'}(\ell \theta) \, ,
\end{equation}
valid in the limit $\ell \to \infty$, $\theta \to 0$, for $\ell \theta$ fixed.
Here $J_\alpha$ is the Bessel function of order $\alpha$.
Using this, approximating the sum with an integral and taking $\ell \gg 1$ in the prefactors, we get our final expressions
\begin{align}
  \xi^{ij}_+(\theta) &= \int_0^\infty \frac{\rmd \ell \, \ell}{2 \pi} \frac{\ell^4}{4} J_0(\ell \theta) C^{ij}_{\psi \psi}(\ell) \, , \label{eq:xi+} \\
  \xi^{ij}_-(\theta) &= \int_0^\infty \frac{\rmd \ell \, \ell}{2 \pi} \frac{\ell^4}{4} J_4(\ell \theta) C^{ij}_{\psi \psi}(\ell) \label{eq:xi-} \, ,
\end{align}
with $C^{ij}_{\psi \psi}$ given by eq.~\eqref{eq:Cpsi}, or eq.~\eqref{eq:CpsiLimber} in the Limber approximation.

\paragraph{Galaxy-shear} The second 2-pt function that is measured is the correlation between lens galaxies and the shear of the sources, commonly referred to as galaxy-galaxy lensing and denoted by $\gam_{t}(\theta)$.
Theoretically, it is the cross-correlation $\braket{\delta_g^i \gamma_{\parallel}^j}(\theta)$.\footnote{Because of parity invariance, in the absence of systematics $\braket{\delta_g^i \gamma_{\times}^j} = 0$.}
This is given by the following expression:
\begin{equation}
  \begin{split}
  \gam_t^{ij}(\theta) =& \braket{\delta_g^i(\hn_1) \gamma_{\parallel}^j(\hn_2)}
  = \frac{1}{2} \sum_{\ell m} \sqrt{\frac{(\ell+2)!}{(\ell-2)!}} C_{\psi g}^{ij}(\ell) Y_{\ell m}(\hn_1) \Re[{}_{+ 2}Y_{\ell m}(\hn_2) e^{-2 i \alpha}] \\
  =& \frac{1}{2} \sum_{\ell} \sqrt{\frac{(\ell+2)!}{(\ell-2)!}} \frac{2 \ell+1}{4 \pi} C_{\psi g}^{ij}(\ell) d^{\ell}_{20}(\theta)
  = \frac{1}{2} \sum_{\ell} \frac{2 \ell+1}{4 \pi} C_{\psi g}^{ij}(\ell) P_\ell^2(\cos(\theta)) \, ,
  \end{split}
\end{equation}
where $\cos \theta = \hn_1 \cdot \hn_2$, $\alpha$ is the angle of $\hn_2-\hn_1$ with the $\hat{e}_1$ vector, and $P_\ell^m$ denotes the associated Legendre function.
Using the asymptotic expression for $d^{\ell}_{20}(\theta)$, we arrive at the flat-sky approximation:
\begin{equation}
  \gam_{t}^{ij}(\theta) = \int_0^{\infty} \frac{\rmd \ell \, \ell}{2 \pi} \frac{\ell^2}{4} J_2(\ell \theta)  C_{g \psi}^{ij}(\ell) \, .
\label{eq:gammat}
\end{equation}
The harmonic cross-spectrum is given by
\begin{equation}
  C_{g \psi}^{ij}(\ell) = \int_0^{\infty} \frac{\rmd k}{(2 \pi)^3} k^2 \int_0^\infty \rmd \chi \int_0^\infty \rmd \tchi W_{g,l}^i(k,\chi) W_{\psi,l}^j(k,\tchi) P_{gm}(k, z, \tilde{z}) \, ,
  \label{eq:Cgpsi}
\end{equation}
where $P_{gm}$ is the unequal-time galaxy-matter power spectrum, given by the EFTofLSS.
In the Limber approximation, using the formulas from App.~\ref{app:limber}, the previous expression simplifies to:
\begin{equation}
    C^{ij}_{g \psi}(\ell)
    = 2 \int_0^\infty \frac{\rmd \chi}{\chi^4} \, N_{lens}^i(\chi) f_{\kappa}^j(\chi) P_{gm}\(\frac{\ell+\frac{1}{2}}{\chi}, z(\chi) \) \, .
  \label{eq:CgpsiLimber}
\end{equation}
To this, we should add the contributions $\braket{\delta_{RSD}^i \gamma_{\parallel}^j}$ and $\braket{\delta_{\mu}^i \gamma_{\parallel}^j}$, for which we have similar expressions.

\paragraph{Galaxy-galaxy} Finally, we compute the galaxy-galaxy autocorrelation for the lenses, $w^i(\theta)$.
Substituting the harmonic expansion, we find:
\begin{equation}
  w^{i}(\theta) = \sum_{\ell} \frac{2 \ell + 1}{4\pi} P_{\ell}(\cos \theta) C^{i}(\ell) \, .
  \label{eq:wi}
\end{equation}
In the flat-sky approximation, using that $P_\ell(\cos \theta) = d^{\ell}_{00}(\theta)$, we get:
\begin{equation}
  w^{i}(\theta) = \int_0^{\infty} \frac{\rmd \ell \ell}{2 \pi} J_0\(\ell \theta \) C^{i}(\ell) \, .
  \label{eq:wiflat}
\end{equation}

The expression for $C_{\ell}^{i}$ has now several terms, since we have to consider the correlations obtained by expanding $\braket{(\delta_g + \delta_{RSD} + \delta_\mu)^i(\hn_1) (\delta_g + \delta_{RSD} + \delta_\mu)^i(\hn_2)}$.
The $gg$ contribution gives
\begin{equation}
  C_{gg}^{i}(\ell) = \int_0^\infty \frac{\rmd k \, k^2}{(2 \pi)^3} \int_0^\infty \rmd \chi \int_0^\infty \rmd \tchi W_{g,\ell}(k, \chi) W_{g,\ell}(k, \tchi) P_{gg}(k, z, \tilde{z}) \, ,
  \label{eq:Cgg}
\end{equation}
with $P_{gg}$ the real-space galaxy-galaxy power spectrum at unequal times, given by the EFTofLSS.

\subsection{EFTofLSS}
\label{sec:eftoflss}

As we have shown, the observables for clustering and lensing are written as functions of the power spectra $P_{mm}(k)$, $P_{gm}(k)$ and $P_{gg}(k)$.
We model them in perturbation theory up to one-loop order using the EFTofLSS. 
We note that all terms, except the loop ones, are calculated beyond the Limber approximation.
This is justified from the size of the one-loop contributions that are shown in fig.~\ref{fig:contributions}, that starts to be important only at small angular separations where the Limber approximation works well.
In fig.~\ref{fig:contributions}, we provide a summary of the relevant contributions entering our predictions.

\paragraph{Matter-matter power spectrum}
At linear level, we use the unequal-time matter-matter power spectrum, given by
\begin{equation}
  P_{mm,\textrm{lin}}(k, z_1, z_2) = D(z_1) D(z_2) P_{11}(k) \, .
  \label{eq:Pgglin}
\end{equation}
Beyond linear level, because of the Limber approximation, we only need the equal-time power spectrum~\cite{Baumann:2010tm,Carrasco:2012cv}:
\begin{equation}
  P_{mm,\textrm{1L}}(k, z) = D^4(z) \( P_{22}(k) + P_{13}(k) \) + 2 c_{s}(z)^2 D^2(z) P_{11}(k) \frac{k^2}{\knl^2} \, ,
\end{equation}
where $P_{22}$ and $P_{13}$ are the standard contributions at one loop, $D(z)$ is the growth function, and $c_{s}(z)^2$ is the dark matter counterterm, whose time dependence is not determined.

The shear-shear correlations will then consider one effective counterterm for each combination of redshift bins, in the following way
\begin{equation}
  \xi^{ij}_{+/-}(\theta) \supset 2 \frac{c_{ss,+/-}^{ij}}{\knl^2} \int \frac{\rmd \ell \, \ell}{2 \pi} J_{0/4}(\ell \theta) \int \rmd \chi \frac{f_{\kappa}^i(\chi) f_{\kappa}^j(\chi)}{\chi^2} D^2(z) P_{11}\(\frac{\ell}{\chi}\) \frac{\ell^2}{\chi^2} \, .
\end{equation}
This approximation comes from the fact that the counterterms have an unkown, but mild, redshift dependence, and for each bin pair there would be a different redshift integral.
Because the counterterm values vary slowly over the kernel width, we approximate the integral by considering a constant effective counterterm for each bin pair, $c^{ij}_{ss,+/-}$.
Within DES error bars, this is a good approximation as we discuss in subsection~\ref{sec:prior}.
Notice that the linear matter-matter power spectrum at equal time contributes to the following correlation functions, all evaluated in the Limber approximation: $\braket{\delta_{RSD} \delta_{RSD}}$, $\braket{\delta_{RSD} \delta_{\mu}}$ and $\braket{\delta_{\mu} \delta_{\mu}}$, that are part of the galaxy-galaxy autocorrelation $w^i$, and  $\braket{\delta_{RSD} \gamma_{\parallel}}$, $\braket{\delta_{\mu} \gamma_{\parallel}}$, that are part of the galaxy-shear cross-correlation $\gamma_t^{ij}$.

\paragraph{Galaxy-galaxy power spectrum}
At linear level, we use the unequal-time galaxy-galaxy power spectrum, as we evaluate the full eq.~\eqref{eq:Cgg} without resorting to the Limber approximation.
This is given by
\begin{equation}
  P_{gg,\textrm{lin}}(k, z_1, z_2) = D(z_1) D(z_2) b_1(z_1) b_1(z_2) P_{11}(k) \, .
  \label{eq:Pgglin}
\end{equation}
Beyond linear level, as mentioned above, we evaluate the integrals in the Limber approximation, which puts the two different redshifts equal to their mean.
Therefore, at 1-loop order, we only need the equal-time power spectrum, which is
\begin{equation}
  P_{gg,\textrm{1L}}(k, z) = D^4(z) \( P_{gg,22}(k) + P_{gg,13}(k) \)
  + 2 D^2(z) b_1(z) c_{gg}(z) P_{11}(k) \frac{k^2}{\km^2} \, ,
  \label{eq:Pgg1L}
\end{equation}
where $c_{gg} = b_1 c_s^2 + c_{hd}$ with $c_{hd}$ a higher derivative bias,
and 
\begin{align}
  &P_{gg,22}(k) = 2 \int_{\vq} K_2(\vkmq, \vq)^2 P_{11}(q) P_{11}(|\vkmq|) \, , \\
  &P_{gg,13}(k) = 6 \, b_1 P_{11}(k) \int_{\vq} K_3(\vq, -\vq, \vk) P_{11}(q) \, .
\end{align} 
The $K_2$, $K_3$ are the following galaxy kernels in real space~\cite{Senatore:2014eva,Angulo:2015eqa,Fujita:2016dne}:
\begin{align}
  \label{eq:K2}
  &K_2(\vq_1, \vq_2) = b_1 \frac{(q_1^2 + q_2^2) \vq_1 \cdot \vq_2}{2 q_1^2 q_2^2} + \sqrt{2} \, c_2  \( F_2(\vq_1,\vq_2) - \frac{(q_1^2 + q_2^2) \vq_1 \cdot \vq_2}{2 q_1^2 q_2^2} + 1 \) \, , \\
  \label{eq:K3}
  &K_{3, \rm int}(\vq, -\vq, \vk) = \frac{b_1}{504 k^3 q^3} \[ - 38 k^5 q + 48 k^3 q^3 - 18 k q^5 + 9 (k^2-q^2)^3\, \ln \frac{k-q}{k+q} \] + \\
  & \qquad + \frac{b_3}{756 k^3 q^5} \[ 2 k q (k^2+q^2)(3 k^4 - 14 k^2 q^2 + 3 q^4) + 3 (k^2-q^2)^4 \, \ln \frac{k-q}{k+q} \] \nonumber \, ,
\end{align}
where for simplicity we show the $K_3$ integrated on $\hat k \cdot \hat q$.
The bias and counterterm coefficients have an unknown but mild time dependence.
Since the lens galaxy selection functions are narrow, we approximate them as constants in each bin.
Thus, for each redshift bin, we consider one counterterm, $c_{gg}^i$, and three bias coefficients, $b_1^i$, $b_{24}^i$, $b_3^i$.\footnote{For clarity in the presentation, we have introduced a single bias coefficient $c_2/\sqrt{2} = b_2 \simeq b_4$, as the combination $b_2 - b_4$ is found to be too weak to be detected in a two-point function analysis at one loop, see e.g.~\cite{DAmico:2019fhj}.}

\paragraph{Galaxy-matter power spectrum}
Also for the $P_{gm}$, we use the unequal-time expression only at linear level, evaluating the full eq.~\eqref{eq:Cgpsi}.
The linear expression is 
\begin{equation}
  P_{gm,\textrm{lin}}(k, z_1, z_2) = D(z_1) D(z_2) b_1(z_1) P_{11}(k) \, ,
  \label{eq:Pgmlin}
\end{equation}
where $z_1$ is the redshift of the lens galaxies.
At 1-loop order we instead use the Limber approximation, eq.~\eqref{eq:CgpsiLimber}, for which we need the equal-time expression
\begin{equation}
  P_{gm,\textrm{1L}}(k, z) = D^4(z) \( P_{gm,22}(k) + P_{gm,13}(k) \)
  + 2c_{gs}(z) D^2(z) P_{11}(k) \frac{k^2}{\km^2} \, ,
\end{equation}
where $2c_{gs} = 2b_1c_s^2 + c_{hd}$ and $P_{gm,22}$ and $P_{gm,13}$ are given by:
\begin{align}
  P_{gm,22}(k) =& 2 \int_{\vq} K_2(\vkmq, \vq) F_2(\vkmq, \vq) P_{11}(q) P_{11}(|\vkmq|) \, , \\
  P_{gm,13}(k) =& 3 \, P_{11}(k) \int_{\vq} \( b_1 F_3(\vq, -\vq, \vk) + K_3(\vq, -\vq, \vk) \) P_{11}(q) \, ,
\end{align}
where $F_2$ and $F_3$ are the usual standard perturbation theory kernels.
As discussed before, we consider the biases as constants in each lens redshift bin. However, since in $\gamma_{\rm t}^{ij}(\theta)$ the redshift integrals are different than for $w^i(\theta)$, we cannot use the same constants $b_1^i$, $b_{24}^i$, $b_3^i$ for the two correlation functions.
Thus, we choose to use different biases, for the same redshift bin, in $w^i(\theta)$ and in $\gamma_{\rm t}^{ij}(\theta)$, on which we impose a correlation at the level of the prior, as explained in sec.~\ref{sec:prior}.
As for the counterterms, we approximate the time integrals involving $c_{gs}$ by choosing a different effective coefficient $c_{gs}^{ij}$ for each bin combination.

We note that, in all the previous expressions, we have not accounted for the infrared resummation in the power spectra.
In principle, effects from long-wavelength modes should be resummed to faithfully describe the peak of the baryon acoustic oscillations (BAO)~\cite{Senatore:2014via}. 
However, on DES Y3 projected angles, the BAO peak is practically unseen.\footnote{See however ref.~\cite{DES:2024pwq} that selects a sub-sample of DES galaxies with particularly well resolved redshifts, allowing for a BAO angular distance determination. }
In principle, also, the resummation of unequal-time correlators damps their power at high-$k$'s~\cite{Zhang:2021uyp}. 
However, this competes with the fact that high modes get projected onto small separations where the correlations are effectively equal time. 
Therefore, for numerical simplicity, we neglect the IR-resummation in our analysis.

\subsection{Approximations and other effects}
\label{sec:observational}

In addition to the previous discussion, there are other effects that one should consider to properly model the data, some physical and other observational. 
Among the physical effects, we have already discussed the magnification bias. 
Details on prior used for the additional nuisance parameters introduced to model the effects presented in this section are given in sec.~\ref{sec:prior}. 

\paragraph{Baryons} One physical effect is that, on small scales, baryons give a contribution to clustering different than that of cold dark matter.
In the EFTofLSS, this can be taken into account in a systematic way, by going to the adiabatic/isocurvature basis~\cite{Lewandowski:2014rca}.
In our modeling, we include baryonic effects as explained in~\ref{app:baryons}, in which the main contributions are automatically captured by, and are degenerate with, the counterterms we have discussed previously.
Moreover, we add the leading correction, that is the linear adiabatic-isocurvature cross-correlation, to our power spectra.
This adds $N_{lens}$ bias parameters to marginalise over. 

\paragraph{Intrinsic alignments} Another important effect goes under the name of intrinsic alignments.
The shear modifies the shape of lensed galaxies; however, the same lensed galaxies have an intrinsic ellipticity, whose distribution is expected to be correlated with large-scale structure.
Therefore, since the shear is inferred by the observed ellipticity of background galaxies, one has to add an intrinsic alignment term to the shear from lensing: $\gam_{ab}(\hn) \to \gam_{ab}(\hn) + \gam^{IA}_{ab}(\hn)$.
A most recent modeling~\cite{Chen:2023yyb} makes use of the EFTofLSS.
DES-Y3~\cite{DES:2021rex} adopts the so-called TATT model~\cite{Blazek:2017wbz}, but we use the simpler NLA model~\cite{Hirata:2004gc} employed in DES-Y1~\cite{Krause:2017ekm}.
This prescription amounts to replace the lensing efficiency with the following function:
\begin{equation}
  f_{\kappa}^{j}(z) \rightarrow f_{\kappa}^{j}(z) - A \left( \frac{1+z}{1+z_0} \right)^{\alpha} \frac{C_{1} \rho_{m,0}}{D(z)} \frac{n_{source}^j(z)}{\bar{n}_s^j} \frac{dz}{d\chi} \, ,
\end{equation}
where $z_0 = 0.62$ is the pivot redshift and $C_{1} = 0.0134 / \rho_{\rm crit}$ is a normalization derived from SuperCOSMOS observations~\cite{Hambly:2001yj}.
The intrinsic alignment then add 2 parameters to the model.

\paragraph{Shear calibration} The shear is additionally affected by the observational uncertainty in the shear calibration.
We follow~\cite{DES:2021rex} and multiply the shear components in each of the $N_{source}$ redshift bins by a parameter $m^i$, such that
\begin{equation}
  \xi_\pm^{ij} \to (1 + m^i) (1 + m^j) \xi_\pm^{ij} \, , \qquad \gamma_t^{ij} \to (1 + m^j) \gamma_t^{ij} \ . 
\end{equation}
Finally, the most important observational uncertainty is the determination of the photometric redshifts.
In our analysis we correct the photo-$z$ error as eq.~(14) of ref.~\cite{DES:2020sjz}: for a given redshift bin, characterized by the selection function $n_{\rm pz}(z)$ with mean redshift $\bar z$, the selection function used in the analysis is $n(z) = n_{\rm pz}((z - \bar z - \Delta z)/\sigma_z + \bar z)$, where $\sigma_z$ and $\Delta z$ are respectively the stretch and the shift parameters (whose values are precised in sec.~\ref{sec:prior}).\footnote{Regardless of the typo in eq.~(15) of ref.~\cite{DES:2021wwk} concerning the parametrisation for the photo-$z$ errors, DES Y3 analysis~\cite{DES:2021rex} swaps the stretch and shift in practice (see \href{https://github.com/joezuntz/cosmosis-standard-library/blob/main/examples/des-y3.ini}{here}), adopting $n(z) = n_{\rm pz}((z - \bar z)/\sigma_z + \bar z  - \Delta z)$. 
This choice differs from the stated prior specification but is likely negligible in impact. 
We thank Ross Cawthon for private correspondence on this point.}
The effect of photo-$z$ uncertainties are illustrated in fig.~\ref{fig:nz}.

\paragraph{Galaxy-galaxy lensing point mass}
The galaxy-shear correlation $\gamma_t^{ij}$ presented in sec.~\ref{sec:2pt}, also referred as galaxy-galaxy lensing, is measured as the stacking of azimuthal angle-averages (circular scans) of the convergence field around lens galaxies.
This average therefore depends on the whole mass density around the lens, that contributes to the measured correlation function. We model this dependence as follow, first focusing on the situation for one source galaxy and one lens before generalising to the situation where we average populations of sources and lenses.
The galaxy-galaxy lensing signal of a source galaxy at radial (comoving) distance $\tilde \chi$ by the matter around a galaxy at radial distance $\chi$ and at a transverse distance $\rho$ (orthogonal to the line of sight to the source galaxy) can be written as
\begin{equation}
\gamma_t \left(\theta \simeq \frac{\rho}{\tilde \chi-\chi}\right) = \frac{\overline \Sigma_<(\rho,\chi)  - \Sigma(\rho, \chi)}{\Sigma_{\rm crit}(\chi, \tilde \chi)} \ .
\end{equation} 
It is the (anti-)excess in the projected (or surface) mass density $\Sigma(\rho, \chi)$ \emph{at} transverse distance $\rho$  with respect to the average $\overline \Sigma_<(\rho, \chi)$ \emph{within} $\rho$, relative to the critical surface mass density $\Sigma_{\rm crit}$ given by\footnote{Note that once averaged over a population of sources, $\Sigma_{\rm crit}^{-1}$ relates to the lensing efficiency~\eqref{eq:lensing_efficiency}. It is called the \emph{critical} surface mass density as it is the minimal surface density for which a source galaxy will be magnified.} 
\begin{equation}
\Sigma_{\rm crit}(\chi, \tilde \chi) = \frac{c^2}{4\pi G} \frac{\tilde\chi}{\chi (\tilde \chi - \chi)} \ ,
\end{equation}
where $c$ is the speed of light and $G$ the Newton constant.\footnote{Here we keep an explicit factor of $c$ instead of setting it to $1$ as in the rest of the paper, to make clear that  $B^{ij}$, defined below, has units of mass.}
Since 
\begin{equation}
\overline \Sigma_<(\rho, \chi) = \frac{2}{\rho^2} \int_0^\rho d\rho'  \, \rho' \, \Sigma(\rho', \chi) \ ,
\end{equation}
$\gamma_t$ gets contributions both from large scales, well described by the EFTofLSS ($\rho' \lesssim \rho$), but also from small scales that are not under perturbative control ($\rho' \ll \rho$). 
To model the dependence of $\gamma_t$ on these small scales, we introduce, following~\cite{Baldauf:2009vj,DES:2021zxv}, an effective ``point mass parameter'' $B^{ij}$ to model the averaged mass distribution within the non-perturbative range for each bin pair $ij$, 
\begin{equation}\label{eq:pm}
\gamma_t^{ij}(\theta) \rightarrow \gamma_t^{ij}(\theta) + \frac{B^{ij} \beta^{ij}}{\theta^2} \ , 
\end{equation}
where
\begin{equation}\label{eq:beta}
\beta^{ij} = \frac{4\pi G}{c^2} \int_0^{\infty}d\chi \int_0^{\infty} d\tilde \chi \, N_{l}^i(\chi) N_{s}^j(\tilde \chi)  \frac{\tilde \chi - \chi}{\tilde \chi \chi} \ .
\end{equation}
In the analysis, we marginalize over the $B^{ij}$ parameters, with a Gaussian prior of mean $0$ and standard deviation $\sigma_{B} = 10^{17} \, M_{\odot}$, as explained below in Sec.~\ref{sec:setup}.


\section{Inference setup}\label{sec:setup}

Here we describe the likelihood, prior, and posterior sampling used in the cosmological analysis that we test and validate in the next section. 
More details on the construction of credible intervals used to report the cosmological results in this paper are given in app.~\ref{app:CI}.

\subsection{Parametrisation and likelihood}\label{sec:likelihood}
To infer cosmological parameters $\pmb{\lambda_c}$, along with nuisance parameters $\pmb{\lambda_n}$ that are usually marginalized over, we make use of the following Gaussian likelihood function $\mathcal{L}(y | \pmb{\lambda})$, where $y$ is the data vector, i.e. the concatenated 3$\times$2pt measurements after scale cuts described in sec.~\ref{sec:data}, and $\pmb{\lambda} = \pmb{\lambda_c} \cup \pmb{\lambda_n}$:
\begin{equation}\label{eq:likelihood}
-2 \log \mathcal{L}(y | \pmb{\lambda}) = (T(\pmb{\lambda}) - y) \cdot C^{-1} \cdot (T(\pmb{\lambda}) - y) \ .
\end{equation}
Here $T(\pmb{\lambda})$ is the theory prediction for the data vector $y$, as described in sec.~\ref{sec:theory}, and $C$ is the data covariance, for which we use the estimate by the DES collaboration~\cite{DES:2020ypx}.
Posterior distributions $\mathcal{P}(\pmb \lambda|y)$ can then be sampled from the likelihood~\eqref{eq:likelihood} with a prior $\pi(\pmb{\lambda})$ we describe below. 
Posteriors of cosmological parameters $\mathcal{P}(\pmb \lambda_c|y)$ are then obtained upon marginalisation of nuisance parameters $\pmb{\lambda_n}$. 
 
\paragraph{Nuisance parameters}
The nuisances are the EFT parameters introduced in sec.~\ref{sec:eftoflss} to account for small-scale physics, galaxy biasing, magnification bias, or baryons, together with the additional parameters introduced in sec.~\ref{sec:observational} to account for further observational effects in 3$\times$2pt analyses.
In detail, for $N_{lens}$ and $N_{source}$ redshift bins of lens and source galaxies, respectively, we consider the following parameters:
\begin{itemize}
\item For the galaxy clustering correlations $w^i(\theta)$, $N_{lens} \times 4$ EFT parameters $\lbrace b_{1,i}^{w}, c_{2,i}^w, b_{3,i}^w , c_{ct,i}^w \rbrace$ describing lens galaxies in redshift bin $i$, where $i = 1, \dots, N_{lens}$;

\item For the galaxy-galaxy lensing $\gamma_t^{ij}(\theta)$, $N_{lens} \times 3$ EFT parameters $\lbrace b_{1,i}^{\gamma_t}, c_{2,i}^{\gamma_t}, b_{3,i}^{\gamma_t} \rbrace$ describing lens galaxies in redshift bin $i$, plus $N_{lens} \times N_{source}$ counterterms $c_{ct,(i,j)}^{\gamma_t}$ --- which are a linear combination of the lens galaxy counterterms and the source galaxy shear (matter) counterterms --- per bin $(i,j)$, where $i = 1, \dots, N_{lens} \ , \ j = 1, \dots, N_{source}$;

\item For both $w^i(\theta)$ and $\gamma_t^{ij}(\theta)$, one common set of $N_{lens}$ magnification biases $b_{mag,i}$, one per lens $i$;

\item Finally, for the cosmic shear $\xi^{ij}_{\pm}(\theta)$, $N_{source} \times (N_{source} + 1) / 2$ matter counterterms $c_{ct,(i,j)}^{\xi_\pm}$ per bin $(i,j)$ in each correlation function, where $i,j = 1, \dots, N_{source}$ with $i \leq j$. 
\end{itemize}
As explained in sec.~\ref{sec:eftoflss}, the rationale to consider a different set of EFT parameters for lens galaxies entering the $w^i$ or $\gamma_t^{ij}$ correlations, or different matter counterterms when considering $\xi_+^{ij}$ or $\xi_-^{ij}$, is that the integral kernels are in principle different for all these observational bins.
This parametrisation is thus so far as general as possible.
Once the selection functions are known, given that the time evolution of the EFT parameters is mild, correlations among them can be imposed as we do in sec.~\ref{sec:prior}. 
Note that for the magnification bias, as their contributions are small, we have already assumed that it is safe to let the ones in $w^i$ and $\gamma_t^{ij}$ to be the same. 
Following the list above, we have for $N_{lens} = 4 \ (5)$ and $N_{source} = 4$, for \texttt{MagLim} (\texttt{redMaGiC}) respectively, a total of $(4 N_{lens}) + (3  N_{lens} + N_{lens} \cdot N_{source}) + (N_{lens}) + (2 N_{source} \cdot (N_{source}+1)/2)  = 68 \  (80)$ EFT parameters to marginalise over. 

\paragraph{Analytic marginalisation} 
As some nuisance parameters $\pmb {\lambda_l} \subset \pmb{\lambda_n}$ appear only linearly in the theory model and therefore at most quadratically in the $\log$-posterior $\ln \mathcal{P}(\pmb{\lambda})$, we can use properties of Gaussian integrals to analytically marginalise over them. 
Assuming a Gaussian prior on $\pmb {\lambda_l}$ centred on $\pmb {\hat \lambda_l}$ with covariance $\mathcal{C}$, we are led to a simple partially-marginalised posterior that is function of the remaining parameters~\cite{DAmico:2019fhj}:
\begin{equation}\label{eq:marg}
\ln \mathcal{P}^{\rm marg}(\pmb{\lambda_c}) = \frac{1}{2}\mathcal{J}_\alpha \mathcal{F}^{-1}_{\alpha\beta} \mathcal{J}_\beta  -\frac{1}{2}\ln \det |\mathcal{F}| + \ln \mathcal{P}|_{\pmb{\lambda_l} = 0} \ .
\end{equation}
Here $\ln \mathcal{P}|_{\pmb{\lambda_l} = 0}$ is the original $\log$-posterior obtained from sampling~\eqref{eq:likelihood} but with Gaussian parameters $\pmb \lambda_l$ set to $0$ in the theory model, and
\begin{align}
\mathcal{F}_{\alpha\beta}(\pmb{\lambda_c})  & = \partial_\alpha T \cdot C^{-1} \cdot \partial_\beta T + \mathcal{C}^{-1}_{\alpha\beta} \ , \\
\mathcal{J}_\alpha(\pmb{\lambda_c})  & = -\partial_\alpha T \cdot C^{-1} \cdot (T|_{\pmb{\lambda_l} = 0} - y) + \mathcal{C}^{-1}_{\alpha\beta} \hat \lambda_{l,\beta} \ ,
\end{align}
where $\partial_\alpha \equiv \partial/ \partial \lambda_{l,\alpha}$.
Since only $b_1$'s, $c_2$'s, and $b_{mag}$'s are not Gaussian parameters, we are left with $20 \ (25)$ EFT parameters to scan over in the numerical sampling, along with cosmological parameters and nuisances entering in corrections to additional observational systematics. 

For the point-mass contributions to galaxy-galaxy lensing, eq.~\eqref{eq:pm}, we follow the treatment described in~\cite{DES:2021zxv}. 
As computed on a fiducial cosmology, the analytical marginalisation amounts to simply redefine the data precision matrix $C^{-1}$ as~\cite{MacCrann:2019ntb}
\begin{equation}
  \label{eq:Cpointmass}
C^{-1} \rightarrow C^{-1} - U(I + U^T C^{-1} U)^{-1} U^T C^{-1} \ .
\end{equation}
Here $I$ is the identity matrix and $U$ is a $N_{\rm data} \times N_{lens}$ matrix, where the $i$-th vector column $\vec{V}$ is defined as $V_a = 0$ if $a$ does not correspond to an element of $\gamma_t^{ij}(\theta)$ or if $a$ does not correspond to lens $i$, and $V_a = \sigma_{B_i} \  \beta^{ij}  \ \theta_i^{-2}$ otherwise, where $\beta_{ij}$ is defined in eq.~\eqref{eq:beta} and $\sigma_{B_i}$ is the width of the Gaussian prior.
As ref.~\cite{2105.13545}, we use the same large prior $\sigma_{B_i} = 10^{17} \, M_{\odot}$.

\subsection{Prior choices} \label{sec:prior}

\paragraph{Cosmological parameters} 
In our analysis, we scan over the cosmological parameters $\omega_{cdm}$ (physical cold dark matter abundance), $h$ (reduced Hubble parameter), and $\ln (10^{10} A_s)$ ($\log$-amplitude of primordial fluctuations) with large flat priors.
The physical baryon abundance $\omega_b$ and the spectral tilt $n_{s}$ are either set to the truth of the simulations in sec.~\ref{sec:validation} or to the values preferred by respectively Big-Bang Nucleosynthesis (BBN) experiments~\cite{Mossa:2020gjc}, $\omega_b = 0.02235$, and Planck~\cite{Planck:2018vyg}, $n_{s} = 0.965$, when analysing DES Y3 data in sec.~\ref{sec:results}.\footnote{We expect that, given the precision of DES data, varying $\omega_b$ within the BBN prior instead of fixing it leads to practically the same results, in particular on $h$. Since $n_s$ is poorly constrained by DES Y3 data, we do not expect our results to change significantly if it is left free, though uncertainties may increase slightly.}
In the latter case, we also fix the neutrino mass to the minimal value, following the Planck prescription. 
As commonly done, we also present our results in terms of the fractional matter abundance $\Omega_m$, and clustering and lensing amplitudes $\sigma_8$ and $S_8 \equiv \sigma_8 / \sqrt{\Omega_m}$. 

\paragraph{EFTofLSS parameters}
For the EFT parameters listed in sec.~\ref{sec:likelihood}, we impose a multivariate Gaussian prior chosen as follows. 
To keep our EFTofLSS predictions within physical range, we restrict the size of the EFT parameters such that the nonlinear one-loop corrections stay within perturbation theory. 
We therefore choose a width of $2$ given that $2 \gtrsim \mathcal{O}(b_1)$, with $k_{\rm M} = 0.7 \ h/\textrm{Mpc}$, for all nonlinear EFT parameters (i.e., other than $b_1$'s). 
Note that this general `order-of-magnitude' argument was found to cover well the parameter space spanned by actual physical models based on galaxy-halo connection~\cite{Ivanov:2024xgb,Zhang:2024thl,Akitsu:2024lyt}. 
We also refine our prior expectation estimating the size of the one loop correction such that the theory error, i.e., the size of the two loop, stays under control over the scales analysed~\cite{Braganca:2023pcp,DAmico:2022gki}, as we describe in sec.~\ref{sec:pp}. 

Next, from fig.~\ref{fig:nz}, we can estimate the expected differences within sets of EFT parameters $b$'s, e.g., $b_1$'s, $c_2$'s, computed for different redshift bins.
We start by observing that the difference in the mean redshift from bins $i$ to $i+1$ is about $\Delta z \sim 0.1$. 
Assuming a redshift evolution for the biases, $b(z) \propto D(z)^p$ with $D(z)$ the growth factor and $p \approx 1$, we expect that the difference in the $b$'s between two neighbouring bins is about $10\%$. 
To account for this fact, we can impose, through the prior covariance, a correlation on parameters from two neighbouring bins in redshift (and setting accordingly the correlations with the other bins farther in redshift). 
Similarly, the difference in the $b$'s between galaxy clustering and galaxy-galaxy lensing is also small as their integral kernels are both peaking around the maximum of the lens $n(z)$'s. 
Finally, the same is to be expected for the differences between the counterterms in the two cosmic shear correlation functions, $c_{ct}^{\xi_+}$ and $c_{ct}^{\xi_-}$.
We therefore also impose a prior correlation between the parameters entering those various observables. 
To estimate the size of those correlations, we follow the procedure outlined in app.~\ref{app:correlation}. 
Therefore, we conservatively choose a correlation of $\rho = 1- \epsilon^2/2$ with $\epsilon = 20\%$, a loose expectation on how much the parameters can be different, that we impose between two parameters \textit{i)} of the same kind entering in the same observable and of two neighbouring redshift bins (for instance, $b_{gg}^i$ and $b_{gg}^{i+1}$); \textit{ii)} of the same kind entering both $w$ and $\gamma_t$ and sharing the same lens redshift bin (for instance, $b_{gg}^i$ and $b_{gs}^i$); and \textit{iii)} $c_{ct}^{\xi_+}$ and $c_{ct}^{\xi_-}$ sharing the same source redshift bin.
For $b_1$'s, we impose the correlation through a lognormal prior as described in ref.~\cite{DAmico:2022osl}. 
Overall, we find that the prior marginally penalises the $\chi^2$ while effectively allowing us to account for the redshift correlations, as discussed in sec.~\ref{sec:sims}.

For the counterterms entering in the predictions of cosmic shear, $c_{ct,(i,j)}^{\xi_\pm}$, we center the prior on the best-fits obtained fitting predictions using the Halofit model~\cite{Takahashi:2012em} for DES Y3 source redshift bins (on a fiducial cosmology). 
Here we are considering Halofit, by design, to be a good proxy of $N$-body simulations of cold dark matter, and therefore use this `UV' information in our prior. 
Halofit has some degrees of inaccuracies depending on the scales, and the counterterms in the shear correlation functions are also degenerate with the leading contributions from baryonic physics as discussed in app.~\ref{app:baryons}. 
Moreover, one can expect some level of cosmological dependence. 
We therefore do not entirely fix $c_{ct,(i,j)}^{\xi_\pm}$ but still marginalise them with a Gaussian prior of width $2$. 
For the nonlinear quadratic biases $c_2$'s and cubic biases $b_3$'s, we center their prior on their coevolution values given as a function of $b_1$'s (see \textit{e.g.}, eq.~(3.17) of ref.~\cite{Maus:2024sbb}). 
For the other nonlinear EFT parameters, as we do not use further information, we simply center their prior on $0$.
For the lens magnification biases $b_{mag}$'s, we marginalise over them with a Gaussian prior of width $2$ centered on the fixed values in the analysis of DES collaboration (see tables I and IV of ref.~\cite{DES:2021wwk}).

\paragraph{Observational systematics}
For the nuisance parameters entering in the corrections to observational systematics, we use priors according to the means and uncertainties estimated for DES Y3 listed in table I of ref.~\cite{DES:2021bwg} when analysing the \texttt{Buzzard}, and tables I and IV of ref.~\cite{DES:2021wwk} when analysing the observational data.

\paragraph{Degrees of freedom} 
When assessing the goodness-of-fit, we will compare the best-fit $\chi^2$, defined as the minimum of~\eqref{eq:likelihood} obtained when minimising~\eqref{eq:likelihood} \emph{together with the prior},  with the degrees of freedom (dof), that we count as $dof = n_{\rm data} - n_{\rm params}$, where $n_{\rm data}$ is the number of data points included in the analysis and $n_{\rm params}$ is the \emph{effective} number of parameters varied in the fit. 
Because the nuisance parameters entering in the corrections to observational systematics are mostly prior dominated, we will count none. 
Then, for each family of correlated parameters, e.g., $b_1$'s, $c_2$'s, and so on, we count one \emph{effective} parameter, as correlations amongst one family are $\gtrsim 96\%$. 
Also, we count one for the two correlated families $c_{ct}^{\xi_+}$ and $c_{ct}^{\xi_-}$, as the correlations amongst them are $\gtrsim 92\%$. 
Finally, given $3$ varied cosmological parameters, we count a total of $n_{\rm params} = 10$. 

\subsection{Perturbative convergence criteria}\label{sec:pp} 
The size of the various contributions in our EFTofLSS predictions presented in sec.~\ref{sec:eftoflss} are controlled by free Wilsonian coefficients. 
For perturbation theory to be valid, higher-order terms should not only not exceed in size the lower-order ones, but should respect some order-of-magnitude sizes that ensure the convergence of the perturbative expansion, imposing restrictions on the range of values that the EFT coefficients can take.
To fulfil this requirement, we impose a prior on the size of the one-loop contributions in our predictions when scanning over the parameter space, that we define as follow. 

Let $y$ be a data sample drawn from a normal distribution $\mathcal{N}(\bar{m}, C)$ with $\bar{m}$ the true mean and $C$ the covariance. 
In average over data samples, we have $\braket{y} = \bar{m}$ and $\braket{y y^T} - \braket{y} \braket{y}^T  = C$. 
Let $m$ be a mispecified model by a theory error $\delta m$ such that $m = \bar{m} + \delta m$. 
We define the $\chi^2$ statistics of the data $y$ as
\begin{equation}
\chi^2(m) \equiv \chi^2(y|m) = (m-y)^T C^{-1} (m-y) \ . 
\end{equation}
As on average we have $\braket{\bar{m}-y} = 0$, we find that the typical cost in $\chi^2$ from using the mispecified model $m$ is
\begin{equation}\label{eq:cost}
\braket{\delta \chi^2} \equiv \braket{\chi^2(m)} - \braket{\chi^2(\bar{m})} = \delta m^T C^{-1} \delta m \ . 
\end{equation}
Heuristically, for the inferred parameters of $m$ to not suffer from misspecification bias larger than their statistical uncertainty, we require that $\braket{\delta \chi^2}$ be smaller than the typical variance of the $\chi^2$ distribution. 
This is $2d$, where $d$ is the number of degrees of freedom.\footnote{We have checked that the variance of the $\chi^2$ for DES Y3 likelihood, estimated by simulating synthetic data with noise drawn according to the DES Y3 covariance, is in fact $2d$.}
By choosing appropriately the scales to include in the analysis as we do in sec.~\ref{sec:scalecut} so that the model misspecification remains under control, $\braket{\delta \chi^2}$ can not exceed $2d$. 
Enforcing this consistency condition then imposes a bound on the size of $\delta m$ following~\eqref{eq:cost}. 

In the EFTofLSS, $\delta m$ corresponds approximately to the first truncated higher-order term in the loop expansion not included in the model $m$ of the observable of interest. 
Furthermore, perturbative convergence of the field expansion implies scaling relations between loop contributions in subsequent orders. 
Since $\delta m$ is bounded, these relations impose a bound on the low-order loop contributions included in $m$. 
For example in our case we consider 
\begin{equation}
|P_{2l}| \sim \frac{P_{1l}^2}{P_{0l}} \ ,
\end{equation}
where $P_{nl}$ is the $n$-th loop contribution to the power spectrum. 
To find the induced bound on the size of the one-loop contribution, we now assume the following ansatz,
\begin{equation}\label{eq:oneloop}
|P_{1l}(k)| \sim A \max\left[ S_{1l}(k) , S_{\rm ct}(k) \right] \ . 
\end{equation}
Here $A$ is a normalisation factor and $S$'s are the expected scalings of contributions in the EFTofLSS at one loop, 
\begin{equation}
S_{1l}(k) = b_1^i b_1^j P_{\rm lin}(k) \left( \frac{k}{k_{\rm NL}} \right)^{3+n_s(k)} \ , \qquad S_{\rm ct}(k) = (b_1^i c_{\rm ct}^j + c_{\rm ct}^i b_1^j) P_{\rm lin}(k) \left( \frac{k}{k_{\rm NL}} \right)^{2} \ ,
\end{equation} 
where $n_s(k) \equiv d\log P_{\rm lin} / d\log k$, $P_{\rm lin}$ is the matter linear power spectrum, and $b_1^i$ corresponds to the values of $b_1$ in the $i$-th observational bin considered in this work ($b_1^i = 1$ if $i$ corresponds to cosmic shear).
We thus find that $A$ is related to the two-loop contribution as (setting $c_{\rm ct} = 1$, as we are interested in an estimate)
\begin{equation}
|P_{2l}| \sim A^2 \max \left[ b_1^i b_1^j \left( \frac{k}{k_{\rm NL}} \right)^{2(3+n_s(k))}, \ \frac{(b_1^i + b_1^j)^2}{b_1^i b_1^j} \left( \frac{k}{k_{\rm NL}} \right)^{4} \right] P_{\rm lin}(k) \ .
\end{equation} 
Let us denote $\xi[P]$ the observable of interest, that is in our case a projected angular correlation function. 
Since $\delta m \approx \xi[|P_{2l}|]$, we can solve for maximal $A$ given the bound
\begin{equation}
\delta m^T C^{-1} \delta m \lesssim 2d \ .
\end{equation}
This value of $A$ yields an expected upper bound $\xi_{\rm max}$ on the size of the one-loop contribution $\xi[P_{1l}]$ given~\eqref{eq:oneloop}. 

To ensure that the one-loop contribution $\xi[P_{1l}]$ does not exceed this expectation when scanning over the parameter space, we add to the log-posterior a prior $\pi$ defined as
\begin{equation}\label{eq:pp}
-2 \log \pi = \frac{1}{n_{\rm data}} \sum_\alpha \left( \frac{\xi^\alpha[P_{1l}]}{\xi_{\rm max}^\alpha} \right)^2 \ ,
\end{equation} 
where the index $\alpha$ runs from $1$ to the length of the data vector, $n_{\rm data}$, such that the penalty in $\chi^2$ is $1$ when the one loop saturates the bound $\xi_{\rm max}$. 
All analyses in this work are performed under this consistency requirement, that we test against simulations in app.~\ref{app:buz18}.

\subsection{Numerical tools} 
Posterior distributions are sampled using the Metropolis-Hastings algorithm as implemented in \code{MontePython 3}~\cite{Brinckmann:2018cvx}. 
Linear matter power spectra and other linear cosmology inputs are calculated using the Boltzmann code~\code{CLASS}~\cite{Blas:2011rf},  accelerated with neural networks~\cite{Gunther:2022pto} during exploratory phases of this work. 
The EFTofLSS predictions for the projected angular correlation functions at the one loop are calculated with a new theory code, \texttt{PyFowl}, that also include the additional observational modelling considered in this work. 
Its implementation is detailed in app.~\ref{app:code}.\footnote{Made publicly available with the 3$\times$2pt likelihood used in this analysis at \url{https://github.com/pierrexyz/pyfowl} }
Convergence of Monte-Carlo Markov chains is assessed using the Gelman-Rubin criterion~\cite{Gelman:1992zz}: we consider chain convergence when the $R$ parameter is $R-1 \leq 0.02$.
Triangle plots and credible intervals are obtained with \code{GetDist}~\cite{Lewis:2019xzd}. 
To find the maximum a posteriori, referred to as the best-fit, we use simulated annealing as implemented in \texttt{Procoli}~\cite{Karwal:2024qpt}. 
We make use of some code snippets borrowed from \texttt{CosmoSIS}~\cite{Zuntz:2014csq} for reading the \texttt{fits} files from the DES Y3 cosmology data release.


\begin{table}
\centering
\scriptsize
\begin{tabular}{|ccccccc|}\hline
$\Delta_{\rm sys}/\sigma_{\rm stat}$ & $\Omega_m$ 			 &  $ h$					&  $S_8 $ 					& $\omega_{cdm}$ 				& $\ln (10^{10}A_s)$ 	& $\sigma_8$	\\ \hline \hline
   Synthetic &   $-0.03$  & $-0.01$ & $-0.27$ & $-0.16$ &  $0.15$ & $-0.03$	\\ \hline \hline
     Buzzard &  -   & -  & -  &  - & -0.06  &  - \\
Buzzard + theo. sys.	 &  -  &  -  & -0.02 &   - & -0.14  & -  	\\ 
Buzzard + obs. sys.  &  - &  - &  -0.2 & - & -0.25 & -  	\\ \hline \hline
Buzzard V$\times 18$  &  - &  - &  - & 0.31 & -0.33 & -  	\\ \hline \hline
DES Y3 + theo. sys. (\texttt{MagLim}) &  -0.03 & 0.07 & 0.25 & 0.08 & 	0.09 & 0.16 \\ 
DES Y3 + theo. sys.  (\texttt{RedMagic}) &  0.04 & -0.07 & 0.23 & -0.08 & 	0.19 & 0.07 \\ 
DES Y3 + sub. baryons (\texttt{MagLim}) &  0.01 & 0.06 &  0.13 & 0.11 &  -0.04  & 0.07 \\ \hline 
\end{tabular}
\caption{\footnotesize  
Summary of systematic errors $\Delta_{\rm sys}$ on the inferred cosmological parameters, relative to their statistical uncertainties $\sigma_{\rm stat}$, in the 3$\times$2pt analysis for DES Y3 data from the EFTofLSS. 
From top to bottom, $\Delta_{\rm sys}$ are measured as follows. 
\emph{Synthetic}: shift of the mean to the truth. 
\emph{Buzzard}: if the absolute shift of the mean to the truth is less than the $1\sigma_{\rm noise}^{V \times 18}$ uncertainties from the noise in the simulations (of volume $\sim 18$ of DES Y3 data, \emph{i.e.}, $1\sigma_{\rm noise}^{V \times 18} \simeq 1\sigma_{\rm stat}/\sqrt{18} $), we report no systematic error. Otherwise, $\Delta_{\rm sys}$ is measured as the shift of the mean to the $1\sigma_{\rm noise}^{V \times 18}$ region. 
\emph{Buzzard + theo. sys.}: as Buzzard, but also marginalising over the theory error. 
\emph{Buzzard + obs. sys.}: as Buzzard, but also marginalising over observational corrections. 
\emph{Buzzard V$\times 18$}: as Buzzard, but analysed with a covariance corresponding to the total simulations volume, about $18$ times DES Y3 data volume.
\emph{DES Y3 + theo. sys.}: shift in the mean from the analysis of DES Y3 data with or without marginalising over the theory error. 
\emph{DES Y3 + sub. baryons}: shift in the mean from the analysis of DES Y3 data with or without marginalising over subleading baryonic effects. 
}
\label{tab:sys}
\end{table}

\section{Pipeline validation\label{sec:validation}}

In this section, we test our cosmological analysis pipeline, described in previous sections, in several ways. 
First, in sec.~\ref{sec:scalecut}, we determine the scales that can be included in the analysis based on estimates of the theory error in the EFTofLSS, so that the angular separations we keep are under analytical control for each observable and lead to tolerable systematic shifts in the final cosmological constraints. 
Next, we check in sec.~\ref{sec:synth} that we can consistently recover cosmological parameters from noiseless synthetic data generated with the EFTofLSS, allowing us to gauge potential prior volume projection effects on the cosmological parameter distributions, due to the marginalization over the nuisance parameters.
Finally, we test our pipeline against the \texttt{Buzzard} simulations in sec.~\ref{sec:sims} with various setups.

A summary of our findings of this section is given in table~\ref{tab:sys}. 
Overall, we find systematic errors at a level $\lesssim 0.3\sigma_{\rm stat}$ from our pipeline, thus validating our cosmological 3$\times$2pt analysis of DES Y3 within the EFTofLSS, whose results are presented in the next section. 
In passing, we check if baryons, beyond their leading effects captured in the EFTofLSS predictions given in sec.~\ref{sec:theory}, have sizeable subleading contributions from isocurvature modes as described in app.~\ref{app:baryons} that can lead to significant shifts if ignored. 
We find, by properly including them, that it is not the case, as the posteriors shifts by $\lesssim 0.1\sigma_{\rm stat}$. 
For simplicity, we therefore do not include these extra baryonic contributions in our main results. 

Note that, since our scale cuts are either more conservative, or rather close to the ones chosen in the cosmological analysis from the DES collaboration~\cite{DES:2021wwk} (see figs.~\ref{fig:bestfit_gal}~and~\ref{fig:bestfit_shear}), we assume in this work that the data vector that we are given is free from (unacceptably large) systematic errors. 
We stress that the systematic error assessment presented in this section is only on our cosmological analysis setup. 

\subsection{Scale cut from theory error}~\label{sec:scalecut}
In the EFTofLSS, the predictions for observables are organised in a perturbative expansion.
At a given perturbative order, this allows us to quantitatively estimate the theory error as the next loop order not included in the predictions. 
For the one-loop model used in this work, sec.~\ref{sec:theory}, this theory error is then given by the contribution from the two-loop terms. 
This provides a principled way to choose the scale cut in the data analysis: keeping only the angular scales over which the contributions from the next-to-next-leading order (NNLO) terms stay sufficiently small with respect to the data uncertainties, we make sure to not introduce significant systematic errors from the theoretical modelling~\cite{Zhang:2021yna,Simon:2022csv,DAmico:2022ukl}. 
As a rule of thumb, by keeping the overall signal-to-noise ratio of the NNLO terms to be less than a given tolerance fraction $X$, we expect to introduce relative systematic shifts in the parameter determination of order $X$.  
To precisely assess the impact of the residual theory error, we fit the data by marginalizing over NNLO terms. We check that this introduces shifts in the measurement of the cosmological parameters of less than our tolerance threshold $\lesssim 0.3\sigma_{\rm stat}$, with respect to the baseline analysis.
In the following, we provide more details on this overall procedure. 

\begin{figure}[ht!]
\centering
\includegraphics[width=0.99\textwidth]{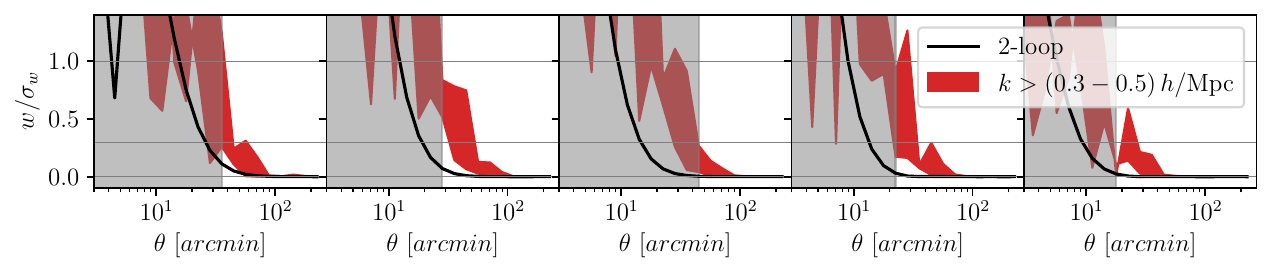}
\caption{\footnotesize  
Scale cut determination from the relative size of theory errors with respect to data uncertainties: estimate of two-loop contributions (in black) and projection from high-$k$ modes (in red). 
The angular separations kept in the analysis (not in the shaded regions) are chosen such that the size of the theory errors does not exceed, roughly, $\sim 0.3-0.5\sigma$. 
For conciseness, only $w$'s from the \texttt{Buzzard} simulations are shown, and we stress that our scale cuts determined this way are tuned for each datasets considered in this work. 
}
\label{fig:scalecut_buz}
\end{figure}

\paragraph{Two loop error} To estimate the size of the theory error, we consider the following NNLO contributions to the power spectrum~\cite{Zhang:2021yna}:
\begin{align}
P^{\rm 2,hd}_{ab} \left( k , z \right) & \propto \frac{k^2}{k_{\rm M}^2} P^{\rm 1}_{ab}(k, z) \, , \label{eq:2loop_hd} \\
P^{\rm 2,ct}_{ab} \left( k , z \right) & \propto \frac{k^4}{k_{\rm M}^4} P^{\rm 0}_{ab}(k, z) \ , \label{eq:2loop_ct}
\end{align}
where $P^L$ with $L = 0, 1, 2$, denote the contributions in the power spectrum at different loop order (linear, one-loop, or two-loop), and $ab = gg, gm, mm$. 
The two-loop counterterm~\eqref{eq:2loop_ct} is found to be the largest one when analyzing 3D clustering, as it is enhanced by the larger length scale entering the renormalization of velocity contact operators appearing in redshift-space distortion terms~\cite{DAmico:2021ymi}. 
In the case of the 2D projected correlation functions, for which redshift-space distortions are suppressed by the thickness of the redshift bins, as shown in fig.~\ref{fig:contributions}, we find that the higher-derivative term~\eqref{eq:2loop_hd} is the larger one for all observables (assuming the coefficients in front of them are both order one).
We will therefore use~\eqref{eq:2loop_hd} to estimate the size of the theory error. 
As the NNLO terms are, obviously, important only at small angular separations (and very small at large ones), we use the Limber approximation to calculate their contribution to the projected correlation functions.
Schematically, we have:
\begin{equation}\label{eq:2loop}
\xi^{\rm 2,hd}_{ab}(\theta)  = c^{\rm 2,hd}_{ab} \int \frac{\rmd l \, l}{2\pi} J_\mu(l \theta) \int \rmd \chi \, \frac{f_a(\chi) f_b(\chi)}{\chi^2} P^{\rm 2,hd}_{ab} \left( \frac{l}{\chi}, z \right) \ , 
\end{equation}
where $\mu = 0,2,4$, depending on the correlation $ab$ (see sec.~\ref{sec:eftoflss}), and $c^{\rm 2,hd}_{ab}$ are order-one coefficients controlling the size of the NNLO terms. 

In fig.~\ref{fig:scalecut_buz}, we show the scale cuts determined with the size of the theory error for the \texttt{Buzzard} simulations. 
We stress that the scale cut choice depends on the dataset considered. 
On each one of \texttt{Buzzard}, \texttt{MagLim}, or \texttt{redMaGiC}, we perform this procedure to determine the scale cuts. 
Later on, in this section, we validate them by marginalising over the theory error in the fit and checking that the induced shifts in the cosmological parameters with respect to the baseline analysis are under our tolerance threshold.

\paragraph{Projection of high-$k$ modes} The support of the line-of-sight integrals when projecting the 3D Fourier power spectra into angular correlation functions, e.g., eqs.~(\ref{eq:wiflat}, \ref{eq:Cgg}), runs up to high-$k$ modes that are not under predictive control within the EFTofLSS. 
They are down-weighted to some extent by the projection integral kernels, that, however, can be rather broad, especially for source galaxies as it can be seen in fig.~\ref{fig:scalecut_buz}. 

A conservative choice would then be to remove all angular separations in the observables for which the integrated contributions from uncontrolled high-$k$ modes become too large with respect to the data uncertainties.\footnote{An alternative approach consists in adding a small-scale theory error to the covariance~\cite{DeRose:2025myp}}
The caveat here is that we cannot precisely determine with respect to the data uncertainties which $k$ modes are described by the EFTofLSS accurately enough to be kept in the analysis.
Focusing on dark matter, one can take inspiration from lessons learnt when fitting $N$-body simulations with the EFTofLSS at one loop~\cite{DAmico:2021ymi}. 
An estimate is to consider uncontrolled high-$k$ modes as the ones starting from $k \sim 0.4 h / \textrm{Mpc}$ and then higher. 
We then compute the integrated contributions of those high-$k$ modes to the observables by summing responses to injections of $k$-band powers (of a given input power spectrum), and compare their relative size with the data error bars. 
The individual $k$-band power injections are shown in app.~\ref{app:pksensitivity} and integrated contributions are shown in fig.~\ref{fig:scalecut_buz} against the data error bars. 
In general, we find that the errors from the uncontrolled high-$k$ modes are a bit larger, yet comparable, to the two-loop errors for all observables in practically all bins. 
Note that, contrary to the scale cut determination by the theory error described earlier, the high-$k$ error estimate here cannot really take into account the size of the data uncertainties, as we do not have data errors in $k$-space.
In the end, we choose to take the most conservative scale cut between the two.

\paragraph{Theory error marginalisation} Finally, we validate the scale cuts by directly marginalising over the theory error in the fit to the data. 
We do so by comparing results, at given scale cuts, obtained from the baseline analysis, with the ones obtained upon addition of the two-loop contributions~\eqref{eq:2loop} to the one-loop theory predictions.
If the shifts of the cosmological parameters pass our tolerance threshold $\lesssim 0.3\sigma_{\rm stat}$, the scale cuts are validated. 
The results of this test for DES Y3 data are shown in table~\ref{tab:sys}. 
For the \texttt{Buzzard} simulations, for which we present results in sec.~\ref{sec:sims}, we compare the shifts between the true values and the results obtained when marginalising over the theory error. 
All these tests are passed, therefore validating the determination of the scale cuts.

\subsection{Tests on synthetic data}~\label{sec:synth}
\begin{figure}
\centering
\scriptsize
\begin{tabular}{|ccccccc|}\hline
$\Delta X / X$ & $\Omega_m$ 			 &  $ h$					&  $S_8 $ 					& $\omega_{cdm}$ 				& $\ln (10^{10}A_s)$ 	& $\sigma_8$	\\ \hline \hline
     Buzzard &  $0.01\pm 0.11$		 & $0.008^{+0.063}_{-0.94}$	& $-0.005^{+0.027}_{-0.036}$ 		& $0.019^{+0.071}_{-0.11}$ 			& $-0.011\pm 0.038$ 		& $-0.005^{+0.063}_{-0.089}$		\\
Buzzard + theo. sys.			&  $0.02^{+0.12}_{-0.11}$			 & $0.000^{+0.056}_{-0.10}$ 		& 	$-0.008^{+0.026}_{-0.036}$ 			& $0.014^{+0.068}_{-0.11}$ 			& $-0.014\pm 0.037$		& $-0.013^{+0.054}_{-0.095}$	\\ 
Buzzard + obs. sys.			&  $0.02\pm 0.11$			 & $0.005^{+0.058}_{-0.092}$ 		& 	 $-0.015^{+0.030}_{-0.036}$			& 	$0.020^{+0.065}_{-0.10}$		& $-0.020\pm 0.040$ 		& $-0.017^{+0.061}_{-0.090}$	\\ \hline\hline
Buzzard V$\times 18$			&  $0.004^{+0.035}_{-0.031}$			 & $0.025^{+0.023}_{-0.029}$ 		& 	 $-0.0098^{+0.0097}_{-0.011}$			& 	$0.063^{+0.028}_{-0.032}$		& $-0.023\pm 0.010$ 		& $0.008^{+0.022}_{-0.028}$	\\ \hline\hline
   Synthetic						&  $0.00\pm 0.10$ & $-0.001^{+0.057}_{-0.082}$ & $-0.009^{+0.026}_{-0.032}$ 		& $-0.015^{+0.066}_{-0.096}$			& $0.006\pm 0.039$ 		& $-0.003^{+0.058}_{-0.081}$	\\ \hline 
\end{tabular}\\ \vspace{0.3cm}
\includegraphics[width=0.99\textwidth]{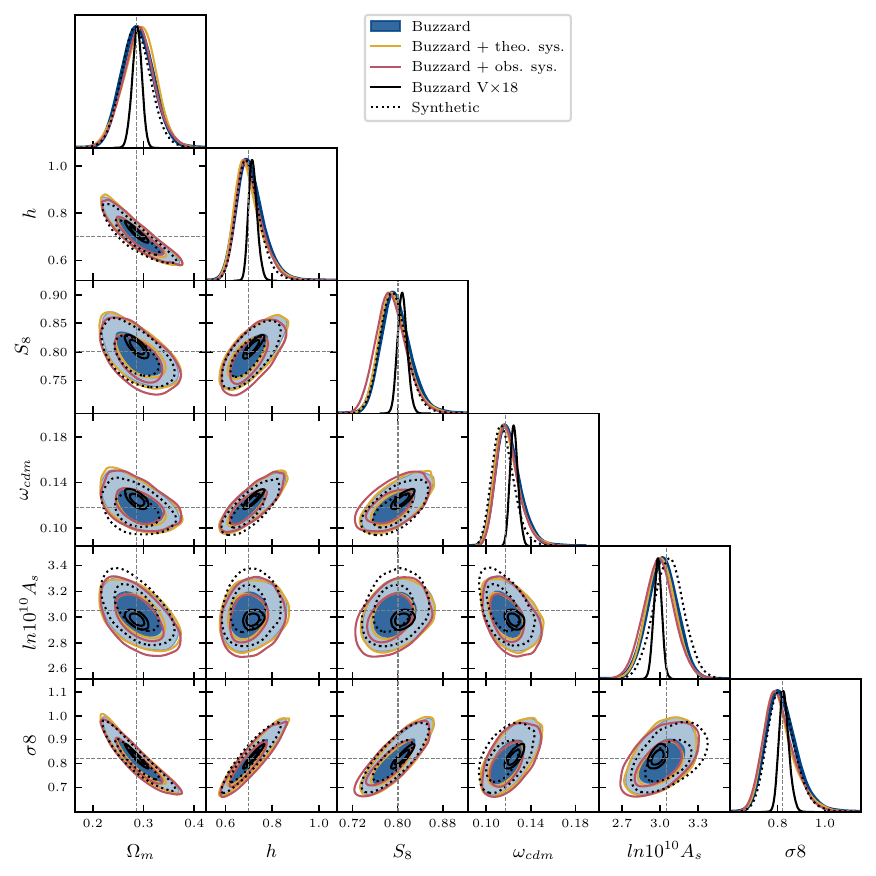}
\caption{\footnotesize  
Relative $68\%$-credible intervals and triangle plots of $\Lambda$CDM cosmological parameters from the EFTofLSS analysis of the \texttt{Buzzard} simulations, with $\omega_b$ and $n_{s}$ set to their truth. 
Also shown are results obtained using instead the covariance corresponding to the total simulation volume (about $18$ times the DES Y3 data volume) or when further marginalising over theoretical or observational systematic corrections, or on synthetic data. 
}
\label{fig:sims}
\end{figure}

We further check our pipeline by performing cosmological inference on noiseless synthetic data generated by our model. 
The model parameters are the best fit obtained on the simulations described in the next section.
This test allows us to check for prior projection effects, and to follow the procedure outlined in app.~\ref{app:CI}, in order to define statistics with a choice of integration measure such that the mean estimator is relatively unbiased. 
Adopting the linear log-measure~\eqref{eq:linM} in the analysis of synthetic data, we recover the true cosmological parameters within $\lesssim 0.3\sigma_{\rm stat}$ as shown in tab.~\ref{tab:sys} and fig.~\ref{fig:sims}. 
Thus, the prior projection effects appear to be under control, and we report our results using this measure.


\subsection{Tests on simulations}\label{sec:sims}

Next, we test our pipeline against the \texttt{Buzzard v2.0} simulations~\cite{DES:2021bwg} mentioned in sec.~\ref{sec:data}. 
The \texttt{Buzzard v2.0} consists in a suite of 18 $N$-body simulations, each designed to reproduce the lens and source samples of the DES Y3 3$\times$2pt data. 
The photometric lens redshifts are calibrated with \texttt{redMaGiC}, which is the pre-blinding main Y3 galaxy sample of DES collaboration. 
The available total simulation volume of this suite, by combining the measurements from all 18 realisations, allows us to validate our analysis setup with a sensitivity of about a quarter, $\sim 1/\sqrt{18}$, of the accuracy on the actual measured cosmological parameters. 
Therefore, the 3$\times$2pt measurements on \texttt{Buzzard} provide an independent test of the modelling described in sec.~\ref{sec:theory} of galaxy biasing, small-scales physics, redshift-space distortions, magnification, beyond-Limber integrals, together with our parametrisation and inference setup described in sec.~\ref{sec:setup}. 
More details on the \texttt{Buzzard} simulations can be found in ref.~\cite{DES:2019jmj}. 
Following ref.~\cite{DES:2021bwg}, for all analyses performed on \texttt{Buzzard} simulations except when specified otherwise, we fit the mean of measurements of the 18 independent realisations but with the covariance of DES Y3 data volume. 

\begin{figure}
\centering
\includegraphics[width=0.99\textwidth]{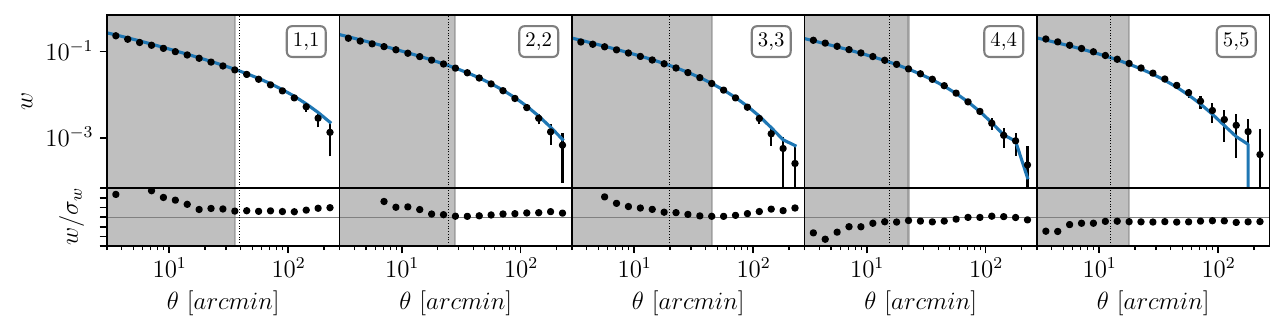}
\includegraphics[width=0.99\textwidth]{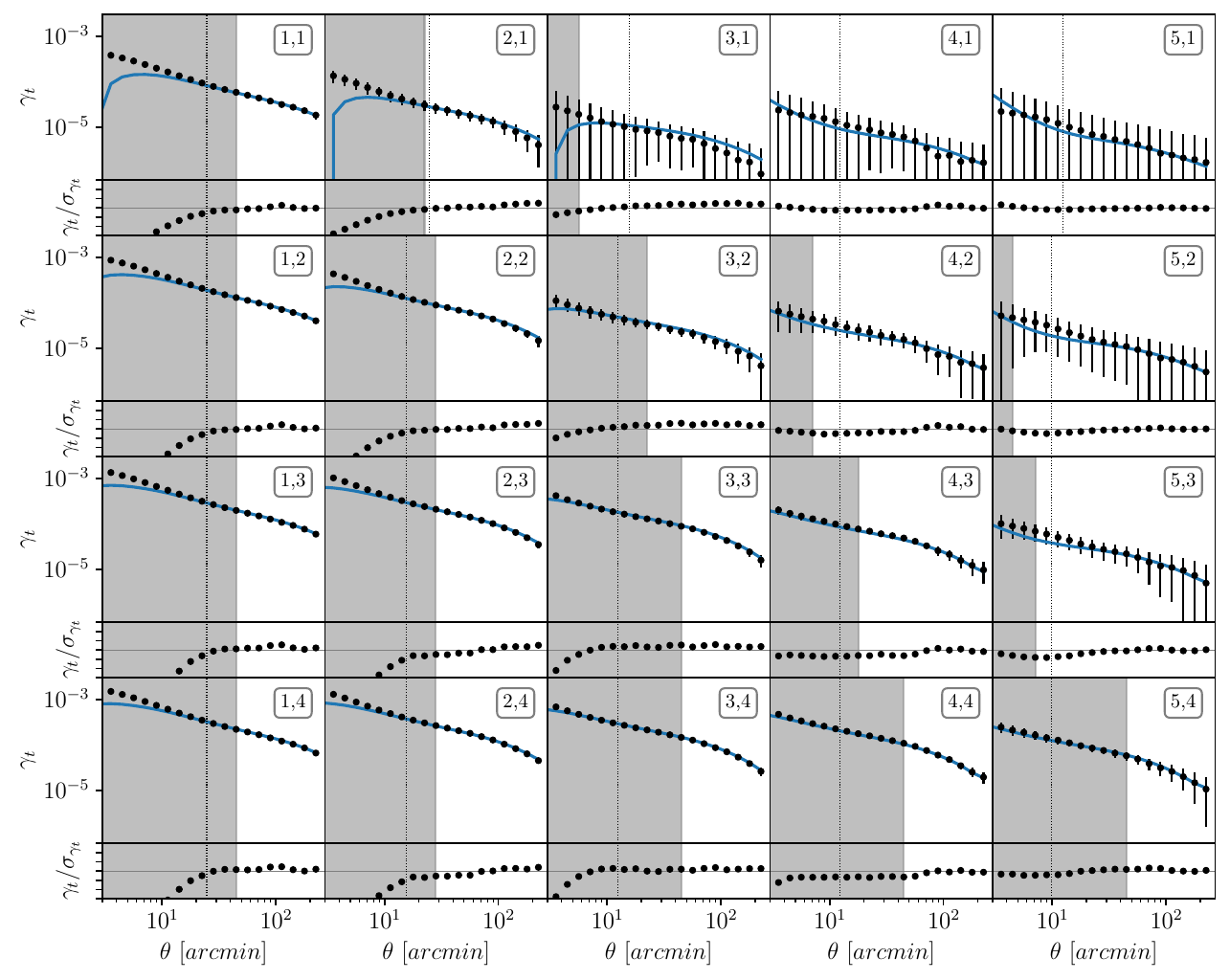}
\caption{\footnotesize  
\texttt{Buzzard} two-point angular correlation functions: galaxy clustering $w$ and galaxy-galaxy lensing $\gamma_t$.
In the upper part of each plot, the black dots are the data points with their error bars, and the blue lines are the best-fit predictions from the EFTofLSS presented in this work.
The lower part of each plot shows the residuals of the best-fit curves relative to the data diagonal errors (with $y$-axis corresponding to $\pm 3\sigma$).
The shaded regions are excluded by the scale cuts used in this analysis; for reference, the DES collaboration scale cut choice~\cite{DES:2021bwg} is shown in dotted vertical lines.
For the $\gamma_t$ correlations, the rows (first indices) correspond to the lenses, while the columns (second indices) corresponds to the sources.
We remind that the measurements are an average over 18 independent realisations whereas the error bars shown correspond to the full DES Y3 data volume.
}
\label{fig:buz_bestfit_gal}
\end{figure}

\begin{figure}
\centering
\includegraphics[width=0.99\textwidth]{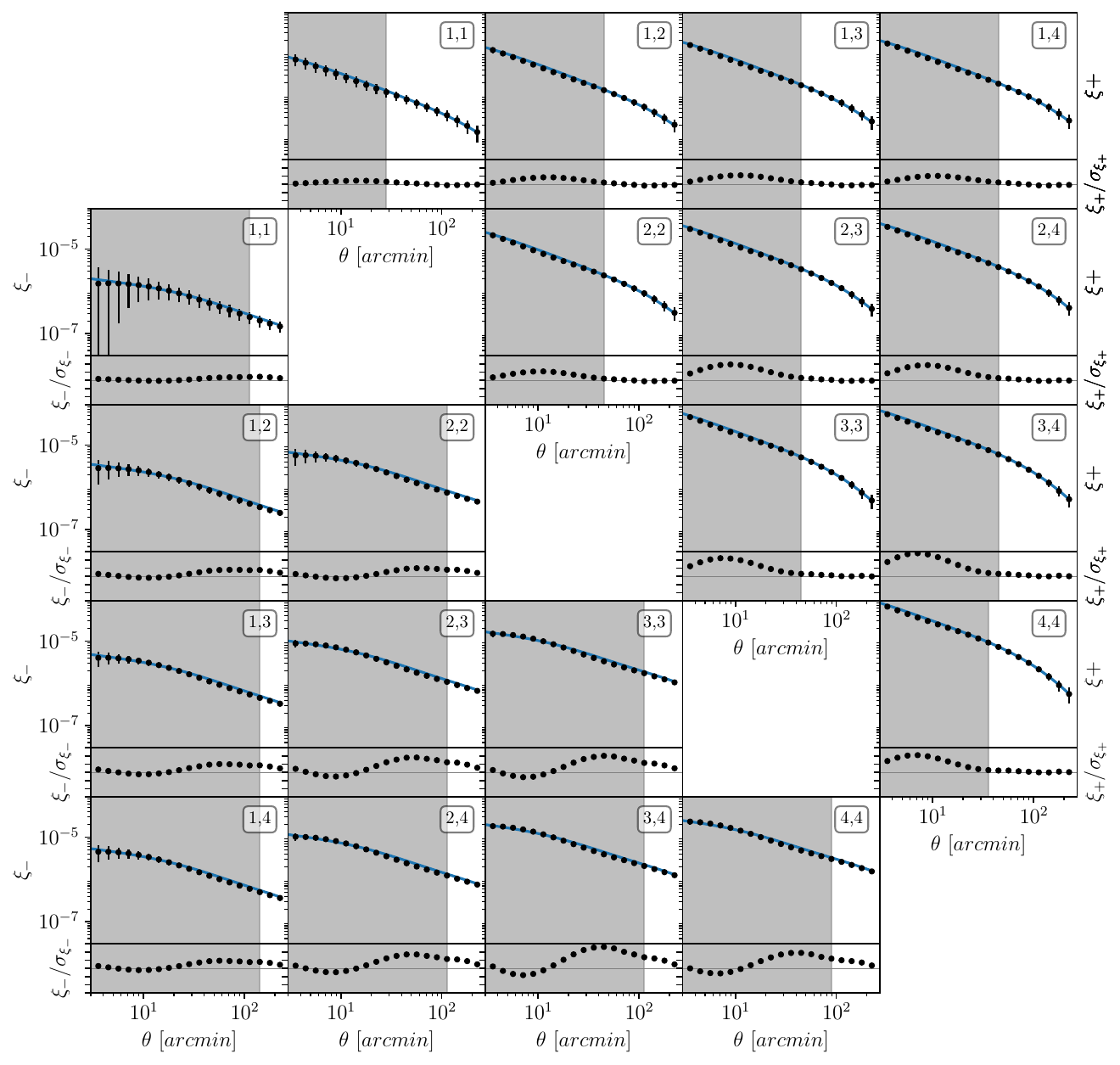}
\caption{\footnotesize  
\texttt{Buzzard}  two-point angular correlation functions: cosmic shear $\xi_\pm$. 
In the upper part of each plot, the black dots are the data points with their error bars, and the blue lines are the best-fit predictions from the EFTofLSS presented in this work.
The lower part of each plot shows the residuals of the best-fit curves relative to the data diagonal errors (with $y$-axis corresponding to $\pm 3\sigma$).
The shaded regions are excluded by the scale cuts used in this analysis; for reference, the DES collaboration scale cut choice~\cite{DES:2021bwg} is shown in dotted vertical lines.
We remind that the measurements are an average over 18 independent realisations whereas the error bars shown correspond to the full DES Y3 data volume.
}
\label{fig:buz_bestfit_shear}
\end{figure}

\paragraph{Goodness of fit} 
Before performing the full cosmological analysis of the \texttt{Buzzard} simulations, it is useful to do a fit with cosmology fixed to the truth to quickly estimate the goodness of fit. 
For the full 3$\times$2pt, 2$\times$2pt (galaxy clustering $w$ and galaxy-galaxy lensing $\gamma_t$), and cosmic shear $\xi_\pm$,  
our scale cuts make for a total number of data points of $406, 285$, and $121$, respectively, and we have correspondingly $80, 60$, and $20$ free parameters (see sec.~\ref{sec:likelihood}). 
We find best-fit $\chi^2$'s of respectively $14.4 \ (181.7), 13.8 \ (171.9)$, and $0.5 \ (9.3)$,\footnote{the $\chi^2$'s from the 2$\times$2pt and cosmic shear do not necessarily add up to the one of the 3$\times$2pt since the covariance is non-diagonal. }  where the numbers in parenthesis are obtained when fitting with a covariance rescaled to the total simulation volume.
The first set of numbers can be compared to the $\chi^2$'s obtained by DES collaboration~\cite{DES:2021bwg}, where they use the Halofit model~\cite{Takahashi:2012em} for all observables, with linear biases and magnification biases, common between the same lens redshift bin in $w$ and $\gamma_t$. 
With their scale cuts, they find respectively $15.0$ for $494$ points, $13.6$ for $285$ points, and $1.4$ for $209$ points. 
Although the data points, and their total numbers, are not the same between the two analyses, they are of the same order.\footnote{We note however that, using Halofit in the analysis, more points in the cosmic shear are included.
As the Halofit model is by design a good approximation of $N$-body simulations, it is not surprising that it can outperform the one-loop EFTofLSS for the nonlinear matter power spectrum. 
This happens also because the \texttt{Buzzard} simulations do not include baryonic effects on the matter distribution. 
Therefore, this might lead to an over-optimistic scale-cut when using Halofit, or other predictors based on $N$-body simulations without baryons.
See further discussions in ref.~\cite{DES:2021bwg,DES:2021vln,DES:2021bvc} and in sec.~\ref{app:baryons}.} 
We see that the best-fit $\chi^2$'s are comparable, indicating that, at this point, both modellings seem appropriate for the \texttt{Buzzard} simulations. 
The second set of numbers obtained with the covariance rescaled to the full simulation volume allows us to compute the $p$-value of the fit. 
We find a $p$-value of $\sim 1$ for all data combinations, indicating an excellent fit to the \texttt{Buzzard} simulations.\footnote{This is somewhat surprising, since our self-consistently determined scale cut is appropriate for the DES-Y3 volume.}
In figs.~\ref{fig:buz_bestfit_gal}~and~\ref{fig:buz_bestfit_shear}, we show the best-fit curves to the \texttt{Buzzard} 3$\times$2pt measurements. 
Similarly as in ref.~\cite{DES:2021bwg}, we note some systematically discrepant residuals in \textit{e.g.}, the first redshift bin of $w$. 
We stress that the 3$\times$2pt covariance is very non-diagonal, warning us against premature conclusions when looking solely at the residuals compared to its diagonal elements only. 

Before moving on, we assess how well the best-fit (i.e., maximum a posteriori) falls within our prior defined in sec.~\ref{sec:prior}. 
To answer this question, we look at the cost in $\chi^2$ of the prior on the best fit of the \texttt{Buzzard} simulations.
We find that the prior contributes about $10.7$ for $7$ effective EFT parameters, as counted in sec.~\ref{sec:prior}.
This corresponds to a $p$-value of $15\%$, which means that the \texttt{Buzzard} best fit falls within $\sim 1.5\sigma$ of our prior distribution. 
We conclude that our agnostic priors, informed only by naturalness of the EFTofLSS, do not affect the fit to the \texttt{Buzzard} simulations, so we validate them.

\paragraph{Pipeline validation test} 
Having validated the goodness of fit fixing the cosmology, we now turn to the full cosmological inference of the \texttt{Buzzard} simulations. 
The results are presented in table~\ref{tab:sys} and fig.~\ref{fig:sims}. 
For all test configurations considered, we measure the systematic shifts to the truth for the inferred cosmological parameters. 
However, the noise in the simulations dictate the accuracy with which we can measure a systematic shift, \textit{i.e.}, the uncertainty on the measurements of the systematic error is $1\sigma_{\rm noise} \simeq 1\sigma_{\rm stat}/\sqrt{18}$ where $\sigma_{\rm stat}$ is the standard deviation measured with a DES Y3 volume covariance (found on \texttt{Buzzard}). 
Therefore, if the shifts are within the $1\sigma_{\rm noise}$ uncertainties, we declare no detection of systematic error. 
Beyond that, we report a systematic error as the distance of the mean to the $1\sigma_{\rm noise}$ region. 
For the results obtained both with the baseline setup, or upon marginalisation over the theory error, we find that we can recover cosmological constraints at $\lesssim 0.1 \sigma_{\rm stat}$ (for the scale cuts chosen in sec.~\ref{sec:scalecut}).
Similar conclusions are obtained when fitting the \texttt{Buzzard} with a covariance rescaled to the total simulation volume, about $\sim 18$ times the one of DES Y3 data: we find systematic errors of $\lesssim 0.3 \sigma_{\rm stat}$ for all cosmological parameters, with the maximal systematic shift accumulating along the principal axis of the elliptical contour in the $\omega_{cdm} - \ln(10^{10}A_s)$ plane.
Results obtained with a diagonal covariance in app.~\ref{app:buz18} seem to suggest that this systematic shift is due to inaccuracies in the modelling of the covariance we use, maybe too approximate to analyse a data volume as large as the one of the full \texttt{Buzzard} simulation suite.
Given that overall the systematic errors that we have detected measuring cosmological parameters on the \texttt{Buzzard} simulations are tolerable, we validate our EFTofLSS modelling, pipeline, and inference setup.

\paragraph{Observational systematics}
The \texttt{Buzzard} suite that we have been offered assume perfect knowledge of photometric redshifts. 
Studies by the DES collaboration have shown that their treatment of the photo-$z$ errors described in~\ref{sec:observational} are robust in recovering unbiased cosmological constraints~\cite{DES:2021wwk}. 
We thus treat the uncertainties in photometric redshifts the same way in the present work when analysing the observational data. 
We perform an end-to-end check of the pipeline by fitting the \texttt{Buzzard} simulations also marginalising over the systematic uncertainties within their prior described in sec.~\ref{sec:prior}. 
Results are shown in fig.~\ref{fig:sims}. 
We see that marginalising over photo-$z$ errors, intrinsic alignments, and shear calibration, does not introduce further shifts in the measured cosmological parameters at a level beyond $\lesssim 0.2 \sigma_{\rm stat}$. 
This test therefore validates the treatment of observational systematics in the present analysis, assuming that the modelling, that we take to be the one considered by the DES collaboration~\cite{DES:2021wwk}, is sufficient.


\begin{figure}
\centering
\scriptsize
\begin{tabular}{|ccccccc|}\hline
& $\Omega_m$ 			 &  $ h$					&  $S_8 $ 					& $\omega_{cdm}$ 				& $\ln (10^{10}A_s)$ 	& $\sigma_8$	\\ \hline \hline
     EFT-DES 3$\times$2pt (\texttt{MagLim}) &  \makecell{0.252 \\ $0.272^{+0.019}_{-0.025}$}			 & \makecell{0.808 \\ $0.773\pm 0.049$} 	& \makecell{0.873 \\ $0.833 \pm 0.032$} 		& \makecell{0.142 \\ $0.139^{+0.011}_{-0.013}$} 			& \makecell{3.07 \\ $2.96\pm 0.13$} 		& \makecell{0.952 \\ $0.879\pm 0.060$}		\\ \hline \hline
          EFT-DES 3$\times$2pt (\texttt{RedMagic}) &  \makecell{0.204 \\ $0.250^{+0.028}_{-0.025}$}			 & \makecell{0.905 \\ $0.769^{+0.042}_{-0.071}$} 	& \makecell{0.857 \\ $0.803^{+0.027}_{-0.033}$} 		& \makecell{0.144 \\ $0.123^{+0.008}_{-0.012}$} 			& \makecell{3.18 \\ $3.13\pm 0.12$} 		& \makecell{1.039 \\ $0.885^{+0.053}_{-0.083}$}		\\ \hline \hline
      DES collab. 3$\times$2pt		&  $0.314\pm 0.027$			 & $0.684^{+0.037}_{-0.046}$ 		& 	$0.790^{+0.013}_{-0.015}$ 			& $0.1236\pm 0.0074$ 			& $2.921^{+0.088}_{-0.078}$ 		& $0.775\pm 0.046$	\\ 
    EFT-BOSS 2+3pt			&  $0.311\pm 0.010$			 & $0.692\pm 0.011$ 		& 	$0.794\pm 0.037$ 			& $0.1255\pm 0.0057$ 			& $2.94\pm 0.11$ 		& $0.808\pm 0.041$	\\ 
    Planck 							&  $0.3191^{+0.0085}_{-0.016}$ & $0.671^{+0.012}_{-0.0067}$ & $0.807^{+0.018}_{-0.0079}$ 		& $0.1201\pm 0.0013$ 			& $3.046\pm 0.015$ 		& $0.832\pm 0.013$	\\ \hline 
\end{tabular}\\
\vspace{0.3cm}
\includegraphics[width=0.99\textwidth]{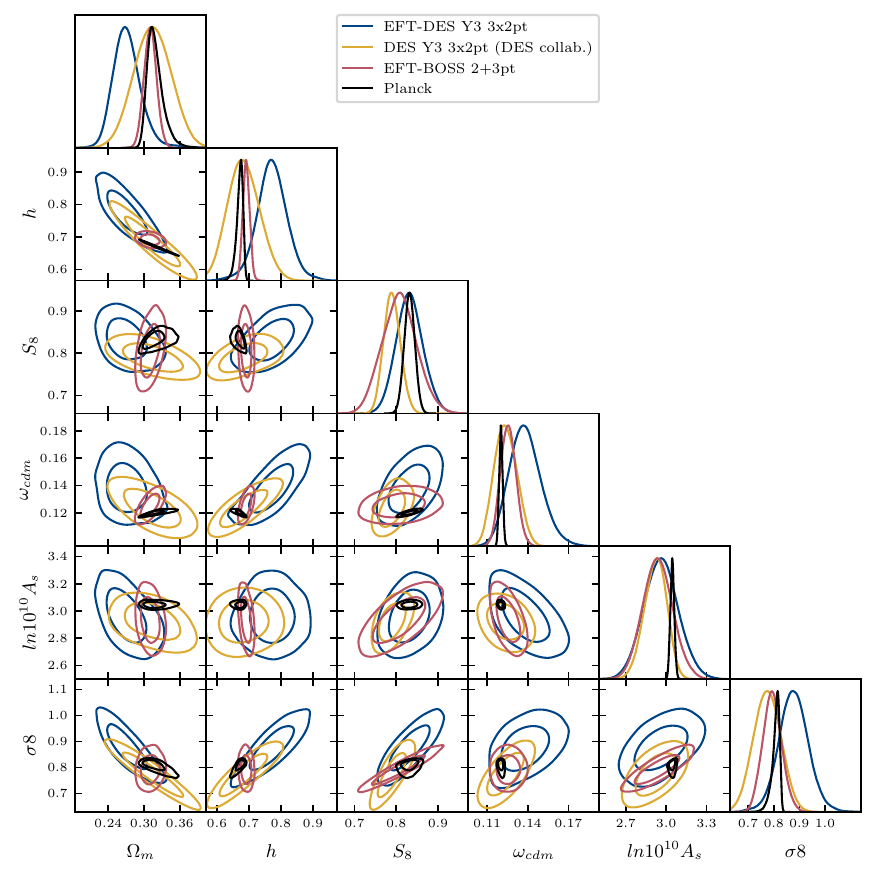}
\caption{\footnotesize  
Best-fits, 68\%-credible intervals, and triangle plots of $\Lambda$CDM cosmological parameters from the EFTofLSS analysis of DES Y3 3$\times$2pt data, with $\omega_b$ set to BBN and Planck preferred values, respectively. 
As a comparison, we show results of DES Y3 3$\times$2pt analysis from \textit{i)} DES Y3 3$\times$2pt analysis by the DES collaboration~\cite{DES:2021wwk} where the publicly-released MCMC chain products are post-processed with a BBN prior on $\omega_b$ and a Planck prior on $n_s$; \textit{ii)} BOSS galaxy clustering power spectrum and bispectrum analysis from the EFTofLSS at one loop~\cite{DAmico:2022osl}; and \textit{iii)} CMB data from Planck with free neutrino mass~\cite{Planck:2018vyg}. }
\label{fig:desy3_boss_planck}
\end{figure}

\section{Results} \label{sec:results}

Having validated (sec.~\ref{sec:validation}) the analysis pipeline explained in sec.~\ref{sec:setup}, based on the EFTofLSS predictions worked out in sec.~\ref{sec:theory}, we now apply it on the DES Y3 3$\times$2pt data described in sec.~\ref{sec:data}.
The best-fit $\chi^2$'s are given in table~\ref{tab:bestfit}, the best-fit values in figs.~\ref{fig:desy3_boss_planck}, and the best-fit curves in~\ref{fig:bestfit_gal},~\ref{fig:bestfit_shear}~and~\ref{fig:redmag_gal}. 
The inferred cosmological parameters are presented in fig.~\ref{fig:desy3_boss_planck}, where they are compared with constraints obtained from other experiments (or analyses). 
In the following we first review the internal consistencies of the results from our EFTofLSS analysis of DES Y3 data before putting them into perspective, comparing to results from other probes.

\subsection{Base $\Lambda$CDM results}

\begin{table}\scriptsize
\centering
\begin{tabular}{|cccccc|}\hline
              & 3$\times$2pt (\texttt{MagLim}) & 3$\times$2pt (\texttt{redMaGic}) & $\xi_{\pm}$ & $w + \gamma_t$ (\texttt{MagLim}) & $w + \gamma_t$ (\texttt{redMaGiC}) \\ \hline \hline
$\min \chi^2$ & 402                            & 526                              & 154         & 256                              & 368                                \\
dof           & 372                            & 484                              & 163         & 208                              & 318                                \\
$p$-value     & 0.14                           & 0.09                             & 0.68        & 0.013                            & 0.029 \\ \hline                      
\end{tabular}
\caption{\footnotesize 
Best-fit $\chi^2$'s, degrees of freedom (dof), and $p$-values obtained in the fit to DES Y3 3$\times$2pt data with our EFTofLSS pipeline. 
The individual $\chi^2$-contributions from cosmic shear $\xi_{\pm}$ and 2$\times$2pt $w + \gamma_t$ are also shown. 
 }
\label{tab:bestfit}
\end{table}

\paragraph{Goodness-of-fit} 
Before investigating the cosmological constraints, we first evaluate the quality of our fit to DES Y3 $3\times$2pt data with our EFTofLSS pipeline under $\Lambda$CDM. 
The best-fit $\chi^2$ values and corresponding $p$-values are listed in table~\ref{tab:bestfit}. 
We adopt $p>1\%$ as a threshold, consistent with DES Y3 analysis~\cite{DES:2021wwk}, with the difference that their $p$-value is derived from posterior predictive distributions (PPD)~\cite{DES:2020lei}, which accounts for parameter uncertainty. 
Our best-fit estimates of the $p$-value constitutes a lower bound of the one from PPD, and is therefore conservative. 
For both analyses of the \texttt{MagLim} or \texttt{RedMagic} galaxy lens samples, we find the goodness-of-fit to be acceptable. 
This contrasts with DES Y3 analysis~\cite{DES:2021wwk}, where \texttt{RedMagic} sample yielded $p < 1\%$,  leading the collaboration to reject that fit after unblinding. 
Differences between their analysis with ours are discussed below. 
We also examine the $p$-values for each component of the likelihood at the best-fit point of the 3$\times$2pt analysis. 
While the lensing correlations $\xi_{+/-}$ are generally well fit, the $p$-values for the 2$\times$2pt are somewhat low for both \texttt{MagLim} or \texttt{RedMagic}, though still acceptable.
Figs.~\ref{fig:bestfit_gal},~\ref{fig:bestfit_shear}, and~\ref{fig:redmag_gal} display the best-fit curves compared to the data, as well as the residuals normalised to the diagonal error bars. 
We do not see significant discrepancies in the fit to $\xi_{+}$, but in $\xi_-$, $w$, or $\gamma_t$, we find several regions where neighbouring best-fit points lie systematically above or below the data. 
Keeping in mind that the non-diagonal nature of the covariance matrix can complicate the interpretation, these deviations might indicate mild inconsistencies. 
For instance, the best-fit curve to bin 2 of $w$ for both \texttt{MagLim} or \texttt{RedMagic} is consistently lower than the data points, which indicates some levels of unaccounted systematics such as unmitigated redshift uncertainties.
As for $\xi_-$ and $\gamma_t$, we note these observables are sensitive to intrinsic alignments, which in our analysis are only modelled at linear level (see sec.~\ref{sec:observational}). 
Reassessing the fit to DES data with a more refined treatments of systematics is thus well motivated and will be pursued in future work, especially in light of the upcoming DES Y6 data release. 

\paragraph{Baseline $\Lambda$CDM constraints}
Following DES Y3 analysis~\cite{DES:2021wwk}, we use the \texttt{MagLim} sample (with vetoed bins 5 and 6 as explained in sec.~\ref{sec:data}) in our baseline analysis, while the \texttt{redMaGiC} sample is used for cross-check purpose only (see below). 
Results are presented in fig.~\ref{fig:desy3_boss_planck}. 
By fixing the baryons abundance $\omega_b$ to the preferred value by BBN experiments, we obtain a measurements from DES Y3 3$\times$2pt not only on $\Omega_m$ and $S_8$, but also on $h$. 
We determine them, at $68\%$CL, to a precision of $8.1\%, 3.8\%,$ and $6.3\%$, respectively, yielding $\Omega_m = 0.272 \pm 0.022$, $S_8 = 0.833 \pm 0.032$, and $h = 0.773 \pm 0.049$. 
In sec.~\ref{sec:comparison}, we compare these results to constraints from other cosmological probes, and in sec.~\ref{sec:H0}, we discuss how $H_0$ is measured in the 3$\times$2pt analysis. 

\paragraph{\texttt{MagLim} vs. \texttt{redMaGiC} sample} 
Using the \texttt{redMaGiC} sample instead of the main \texttt{MagLim} sample (see sec.~\ref{sec:data}), we obtain the results shown in figs.~\ref{fig:main}~and~\ref{fig:desy3_boss_planck}. 
The constraints on $\Omega_m$ and $h$ are about $15\%$ looser than the ones obtained from the \texttt{MagLim} sample, while the one on $S_8$ are comparable. 
Overall, the 1D posteriors are compatible at $\lesssim 1\sigma$. 
Assuming that most of the objects in those two samples are the same but bearing in mind that there are nonetheless different selections in brightness, color, redshift, scale cut, etc., this provides a crude upper bound on the systematic error in the DES Y3 data at the catalog level. 
Taken at face value, this seems to indicate that the cosmological results are not dominated by systematics, although some effects could shift constraints from both samples in the same direction.

\subsection{Comparison to other probes}\label{sec:comparison}

In this section, we compare the posterior distributions of cosmological parameters obtained in this work with those from various datasets and analysis configurations.
To quantify the statistical consistency amongst them, we use the standard Gaussian tension metric $T_1$ in units of standard deviation ($\sigma$) on $1D$ posteriors, 
\begin{equation}
T_1 = \frac{|\mu_1 - \mu_2|}{\sqrt{\sigma_1^2 + \sigma_2^2}} \ ,
\end{equation}
where $\mu_1$ ($\sigma_1$) and $\mu_2$ ($\sigma_2$) are the means (standard deviations) of the two $1D$ posteriors we want to compare.  
We also make use of a generalisation to $ND$ posteriors: given two $ND$ Gaussian distributions with means $\pmb{\mu_1}$ and $\pmb{\mu_2}$, and covariance matrices $\Sigma_1$ and $\Sigma_2$, we can compute their distance as
\begin{equation}\label{eq:TN}
T_N = \sqrt{(\pmb{\mu_1}-\pmb{\mu_2})^T (\Sigma_1 + \Sigma_2)^{-1} (\pmb{\mu_1}-\pmb{\mu_2})} \ .
\end{equation}

\paragraph{Comparison with DES collaboration Y3 results}
We first compare in fig.~\ref{fig:desy3_boss_planck} our constraints from DES Y3 3$\times$2pt with results obtained by DES collaboration in ref.~\cite{DES:2021wwk} (hereafter \emph{DEScollab}) for the same dataset. 
Specifically, for the comparison to be meaningful, we take the publicly-released MCMC chains\footnote{made available here: \url{https://desdr-server.ncsa.illinois.edu/despublic/y3a2_files/chains/}} from~\cite{DES:2021wwk} that we post-process with a BBN prior on $\omega_b$ and a Planck prior on $n_s$ via Gaussians of mean and width $\omega_b=0.02235 \pm 0.00036$ and $n_s = 0.9649 \pm 0.0044$. 
We find that the 1D posteriors of $\Omega_m$, $h$, and $S_8$ agree within $\lesssim 1.3\sigma$, with our reconstructed $S_8$ slightly above the one from \emph{DEScollab}. 
Using eq.~\eqref{eq:TN}, we find that the marginalised 2D posteriors of $(\Omega_m, h)$ are consistent at the $1.3\sigma$ level under the Gaussian approximation, and the overall posteriors of $(\Omega_m, h, S_8)$ are consistent at the $ 1.4\sigma$ level. 
Given that we are comparing two analyses of the same dataset, we now comment on the possible origin of these differences.  
To start, we note that the \emph{DEScollab} pipeline has been validated for measurements of $(\Omega_m, S_8)$ with $\omega_b$ and $n_s$ treated as a free parameter, notably against the \texttt{Buzzard} mocks (see sec.~\ref{sec:sims}), while we have validated our pipeline for $(\Omega_m, h, S_8)$ for fixed $\omega_b$ and $n_s$.

Let us now compare the analysis choices.
\emph{DEScollab} uses a linear bias prescription with a Halofit power spectrum to model the galaxy clustering signal instead of the EFTofLSS treatment of nonlinearities that we use. 
Moreover, their analysis retains more angular scales for $w$ and $\gamma_t$ as shown in fig.~\ref{fig:bestfit_gal} compared to ours. This difference originates mainly in that our scale cuts are subject to the self-consistent theoretical criterion described in sec.~\ref{sec:scalecut}, before being validated on simulations. 
\emph{DEScollab} also presents an alternative analysis based on a modelling including additional one-loop corrections~\cite{DES:2020yyz,DES:2021rex}, for which the nonlinear bias values are fixed to $b_1$ through relations from co-evolution model or calibrated on mock catalogs of \texttt{RedMaGiC}-like galaxies, leaving essentially $b_1$ as a free parameter. 
For this alternative analysis, \emph{DEScollab} retains scales down to $4 \, \hinvMpc$ instead of $8 \, \hinvMpc$, validated on the \texttt{Buzzard} simulations~\cite{DES:2021bwg}. 
The resulting constraints are close to their baseline results, and we find a similar level of agreement with our results as previously at the $1.5\sigma$ level.\footnote{Note that in this analysis, \emph{DEScollab} leaves the total neutrino mass free. }   
Leaving all biases free as in our analysis allows in principle to capture more complex, potentially relevant physics at play in galaxy formation (in particular, in regimes beyond simplifying assumptions used to construct simulations), however at the expense of introducing additional degeneracies. 
We also note that the counterterms described in sec.~\ref{sec:eftoflss}, which absorb unphysical contributions from the loop integrals, are absent in \emph{DEScollab} nonlinear modelling. 

Similarly, there are large differences in the scale cut choice for the shear correlation functions as shown in fig.~\ref{fig:bestfit_shear}. 
\emph{DEScollab} relies on the Halofit power spectrum that allows in principle to go beyond the regime of validity of perturbation theory, within a certain degree of precision.
Their scale cuts were chosen to be safe from baryonic effects, based on estimates from hydrodynamic simulations (see however discussions in \textit{e.g.},~\cite{DES:2021vln,DES:2021bvc,Amon:2022azi}). 
Our one-loop EFTofLSS analysis does not allow us to retain nearly as many angular scales, however, baryonic effects are under control, as detailed in app.~\ref{app:baryons}. 
There are also differences in the modeling of the intrinsic alignments. 
\emph{DEScollab} uses a model for the intrinsic alignments that extends beyond the linear one that we use in this work, the Tidal
Alignment and Tidal Torquing (TATT) model~\cite{Blazek:2017wbz}. 
TATT relies on the expansion of tensor fluctuations projected along the line-of-sight up to quadratic order. 
One-loop diagrams of the type $13$ in principle contributes equally as the one-loop $22$-diagrams predicted in TATT, and subtle IR-cancellations are expected between the $22$ and $13$ diagrams~\cite{Lewandowski:2017kes}. 
Whether using a partial set of one-loop contributions improves the modeling of intrinsic alignments may deserve further investigations. 
For additional insight, see the full one-loop calculation presented in ref.~\cite{Chen:2023yyb}.

Finally, a last difference lies in the choice of prior. 
As presented in sec.~\ref{sec:setup}, in our analysis $w$ and $\gamma_t$ for the \emph{same} lens galaxies are \emph{not} described by the same model parameters, effectively capturing the difference in the time evolution of the EFT parameters that are projected onto different redshift kernels. 
Similar considerations were applied for the counterterms appearing in $\xi_+$ and $\xi_-$. 

To conclude, our comparison highlights the impact on cosmological constraints from the choice of modeling, scale cut, and prior in the DES 3$\times$2pt analysis. 
As we retain less angular scales and marginalise over more nuisance parameters, parameter uncertainties from our analysis end up to be larger than the ones from \emph{DEScollab}.

\paragraph{Comparison to \emph{Planck} and BOSS} 
In fig.~\ref{fig:desy3_boss_planck}, we compare our results from DES Y3 3$\times$2pt with independent cosmological probes, namely CMB data from \textit{Planck}~\cite{Planck:2018vyg} and galaxy clustering data from an EFTofLSS analysis of BOSS~\cite{DAmico:2022osl}. 
The 1D posteriors of $\Omega_m$, $h$, and $S_8$, are consistent between all these datasets at the $\lesssim 2\sigma$ level, with the farthest deviation visible in $h$ between DES and \textit{Planck}. 
Our inferred value for $S_8$ is slightly above the ones from \textit{Planck}, in stark contrast with results from other analyses of projected LSS data (see \textit{e.g.},~\cite{Reeves:2025axp} for a recent summary). 
Our DES constraint on $S_8$ is comparable in precision to the BOSS one, and about a factor $3$ looser than the \textit{Planck} one, while our DES constraint on $\Omega_m$ is a factor $2$ looser than the \textit{Planck} one. 
Our constraint on $H_0$ is not competitive, yet insightful since the measurement comes from a different ruler than the one in galaxy clustering or CMB data, as we explain below. 
Under the Gaussian approximation, the full posterior of $(\Omega_m, h, S_8)$ that we infer from DES data is in mild tension at the $1.7\sigma$ and $2.3\sigma$ level with those from BOSS and \emph{Planck}, respectively, mainly driven by the discrepancy visible in the $\Omega_m-h$ plane. 
In comparison, we find that \emph{DEScollab} + BBN + $n_s$-prior is consistent with \emph{Planck} at the $2.3\sigma$ level in the 3D $(\Omega_m, h, S_8)$-space, with the discrepancy mainly pulled by a low $S_8$ from \emph{DEScollab} at $1.9\sigma$. 
We now turn to investigate what drives the 3$\times$2pt fit in the $\Omega_m$–$h$ plane.

\subsection{A new measurement of $H_0$}\label{sec:H0}
It is a common lore that weak lensing is hardly sensitive to $H_0$. 
Here we derive an analytic approximation for the main degeneracy between $\Omega_m$ and $h$ visible in fig.~\ref{fig:main} or fig.~\ref{fig:desy3_boss_planck}, and explain how it gets partially broken through the 3$\times$2pt combined analysis.

\paragraph{Matter-radiation equality scale} 
Since the BAO peak lies at large angular separations and is further smeared by the projection along the line of sight, projected correlation functions are almost independent of the sound horizon (see \textit{e.g.},~\cite{Reeves:2025axp} for a more quantitative discussion).
The key physical scale that enables measurement of $h$ is then the matter–radiation equality scale, $\sim k_{\rm eq}^{-1}$, which governs the overall shape of the projected correlation function.
$k_{\rm eq}$ is the comoving wavenumber that enters the horizon at matter-radiation equality $a_{\rm eq}$ (\textit{i.e.}, is equal to the Hubble radius back then), yielding
\begin{equation}
k_{\rm eq} = H_0 \frac{\sqrt{2} \Omega_m}{\sqrt{\Omega_r} c} = 100 h \times \frac{\sqrt{2} \Omega_m}{\sqrt{\Omega_r}} \frac{ \rm km/s}{c \, \rm Mpc} \ ,
\end{equation}
which is about $k_{\rm eq} \approx 0.073 \, \Omega_m h^2\, \textrm{Mpc}^{-1}$, given that $\Omega_r h^2 \approx 4.15 \times 10^{-5}$. 
The associated length scale is typically too large to be measured directly, however it enters in the amplitude of the power spectrum. 
For $k \gg k_{\rm eq}$, the overall broadband shape of the matter linear power spectrum scales approximately as~\cite{Eisenstein:1997ik,Dodelson:2003ft}
\begin{equation}
P_{ lin}(k, z) \approx D(z)^2 \, \frac{T(k)^2}{k^3} A_s \left(\frac{k}{k_{*}}\right)^{n_s -1}   
\propto D(z)^2 \, A_s \left(\frac{k}{k_{*}}\right)^{n_s -1} k (\Omega_m H_0^2)^{-2} \left(\frac{k_{\rm eq}}{k}\right)^{4}  \log \left( \frac{k}{k_{\rm eq}} \right)^2  \ ,
\end{equation}
where $P_{lin}$ is given in units of Mpc${}^3$ and where we have approximated the transfer function as $T(k) \approx -(\frac{3}{2} \Omega_m H_0^2)^{-1} k^2 T_\Phi(k)$, with $T_\Phi(k) \propto \left(\frac{k_{\rm eq}}{k}\right)^{2}  \log \left( \frac{k}{k_{\rm eq}} \right)$.  
For $n_s \approx 1$, we consider the following scaling solution,
\begin{equation}\label{eq:scalingsol}
P_{ lin}(k, z) \propto D(z)^2 \, A_s (\Omega_m h^2)^2 \, k^\nu  \ , 
\end{equation}
where $\nu \approx -2$ for the scales of interest, $k \sim 0.1 \hinvMpc$, that are not yet in the regime $k \gg k_{\rm eq}$ as assumed above.

\paragraph{Scaling of projected angular correlation functions} 
How does $k_{\rm eq}$ projects in angular space?
The angular two-point function $\xi_{ij}(\theta)$ is given, in the Limber approximation, by
\begin{equation}\label{eq:limber_}
\xi_{ij}(\theta) = \int \frac{\rmd \ell \, \ell}{2\pi} J_\mu(\ell \theta) \int \rmd \chi \, \frac{f_i(\chi) f_j(\chi)}{\chi^2} P_{ij} \left( \frac{\ell}{\chi}, z(\chi) \right) \, ,
\end{equation}
where $\mu = 0,2,4,$ depending on the correlation $ij$, with $f_i$ and $f_j$ the respective line-of-sight kernels. 
Here we take $P_{ij}$ to be proportional to the scaling solution~\eqref{eq:scalingsol}, further multiplied by biases $b_i b_j$ where $b_i \equiv 1$ if $i$ corresponds to matter. 
To solve for the scaling solution~\eqref{eq:scalingsol}, we first evaluate the Bessel transform as in app.~\ref{app:code} by performing the change of variables $\ell \rightarrow k \chi$ and $\theta \rightarrow r / \chi$ and using the master integral~\eqref{eq:masterint}. 
Here we use the fact that most of the signal-to-noise ratio in 3$\times$2pt is coming from small angular separations $\theta \sim \mu/\ell$ for which most of the support of the line-of-sight integral in~\eqref{eq:limber_} corresponds to Fourier modes $k \sim \ell/\chi \gg k_{\rm eq}$, where $\chi$ is taken as the comoving distance where the kernel peaks. 
We find
\begin{equation}\label{eq:xi_scaling_solution}
\xi_{ij}(\theta) \propto b_i b_j \, A_s \Omega_m^2 h^4 \, \mathcal{K}_{ij} B_\mu \times \theta^{-2-\nu} \ ,
\end{equation}
where
\begin{equation}\label{eq:SB}
\mathcal{K}_{ij}  \propto \int d\chi \, \chi^{-\nu} \frac{f_i(\chi) f_j(\chi)}{\chi^2} D(\chi)^2 \, ,  \quad B_\mu  = \frac{2^\nu}{\pi} \frac{\Gamma(1+\nu/2 +\mu/2)}{\Gamma(-\nu/2 + \mu/2)}  \, . 
\end{equation}
Neglecting the mild cosmological dependence in the growth function $D$, $\mathcal{K}_{ij}$ is almost cosmology-independent for $\xi_{gg}(\theta) = w(\theta)$. 
For correlations involving shear, however, the lensing efficiency~\eqref{eq:lensing_efficiency} provides an extra dependence on $\Omega_m$.\footnote{Since \eqref{eq:lensing_efficiency} is defined in comoving Mpc, explicit factors of $h$ appear in the prefactor. However, by expressing comoving distances in $\Mpcinvh$, correspondingly comoving wavenumbers $k$ in $\hinvMpc$ and the power spectrum $P$ in units of $(\Mpcinvh)^3$, all explicit dependencies of $h$ cancel out in~\eqref{eq:limber_}. The remaining dependence on $h$ is thus entirely encoded in the shape and amplitude of $P$.  } 
Therefore, the main degeneracy in the $\Omega_m - h$ plane is partially broken. 
We find that the projected angular correlation functions scale roughly as $\xi_{ss} \sim A_s \Omega_m^4 h^4$, $\xi_{gs} \sim b_1 A_s \Omega_m^3 h^4$, and $\xi_{gg} \sim b_1^2 A_s \Omega_m^2 h^4$,  which implies that the 3$\times$2pt combination allows to break degeneracies between $b_1$, $A_s$, $\Omega_m$, and $h$. 
Note that our trivial scaling estimate for shear matches roughly the scaling of~\cite{Jain:1996st}.

\begin{figure}
\centering
\includegraphics[width=0.51\textwidth]{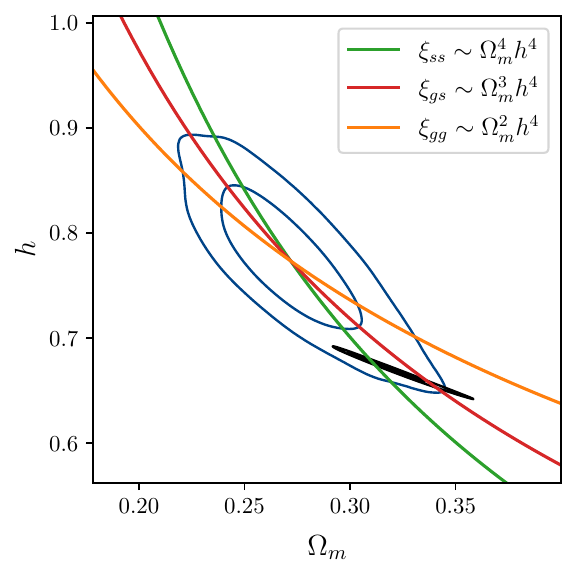}
\caption{\footnotesize  
2D posterior distribution in the $\Omega_m-h$ plane from the EFTofLSS analysis of DES Y3 data, plotted against the main degeneracy lines estimated from scaling solutions. 
The \emph{Planck} contour is shown in black for comparison. 
}
\label{fig:desy3_deg}
\end{figure}

\paragraph{The $\Omega_m-h$ degeneracy} 
To understand what drives the fit in the $\Omega_m-h$ plane, we show in fig.~\ref{fig:desy3_deg} the DES posterior alongside the main degeneracy lines implied by our scaling approximations.  
We find good agreement between our estimates and the dominant degeneracy in the DES posterior.
Notably, the largest deviation from the \emph{Planck} contour occurs along the degeneracy direction associated with correlations involving shear. 
This suggests that weak lensing favours a lower value of $\Omega_m$ measured through its sensitivity to the gravitational potential (see eqs.~\eqref{eq:psilm}~and~\eqref{eq:lensing_efficiency}), which in turn implies a higher value of $H_0$. 
See \textit{e.g.},~\cite{Garcia-Garcia:2024gzy} for similar findings in a different analysis setup of weak lensing data.

\section{Conclusions}\label{sec:conclusion}

In this work, we have conducted a 3$\times$2pt analysis of projected angular correlation functions of DES Y3 data using a newly-developed pipeline constructed around predictions from the EFTofLSS at one loop. 
Our findings are summarised as follows, together with a few research directions worth pursuing: 
\begin{itemize}
\item By fixing $\omega_b$ to the value preferred by BBN and $n_s$ to the value preferred by \emph{Planck}, we are able to measure not only $S_8$ and $\Omega_m$, but also $H_0$ as explained in sec.~\ref{sec:H0}. 
As shown in fig.~\ref{fig:desy3_boss_planck}, our inferred value of $S_8$ is fully consistent with \emph{Planck}, while $\Omega_m$ measured primarily by shear appears low, which in turn implies a high value of $H_0$ given their relatively strong anticorrelation. 
Overall, the consistency with \emph{Planck} in the $(\Omega_m, h, S_8)$-space is $ \sim 2.3\sigma$. 
Our result can also be compared (see fig.~\ref{fig:main}) to DES collaboration analysis~\cite{DES:2021wwk} in combination with a BBN prior on $\omega_b$ and a \emph{Planck} prior on $n_s$, revealing some discrepancies up to $\sim 1.4 \sigma$.
Our work suggests that general flexible predictions of projected statistics enabled by the EFTofLSS together with the self-consistent analysis choices described in sec.~\ref{sec:scalecut} already impact current 3$\times$2pt analyses, and are likely to become increasingly relevant for incoming precision measurements such as Euclid photometric data or LSST. 
\item A number of incremental, yet important improvements of our analysis pipeline can be immediately foreseen. 
Although our goodness-of-fit analysis suggests the presence of unaccounted observational systematics --- warranting further investigation on that front --- we list here a few potential improvements on the modelling side.
To begin with, a reassessment of the cosmological results using a consistent one-loop modelling of intrinsic alignments~\cite{Chen:2023yyb}, rather than the linear model currently adopted, is necessary. 
Next, we have used angular projected statistics, that tend to spread out Fourier modes over all angular separations (to various degree), in particular short ones not captured by the EFTofLSS at one loop as estimated in sec.~\ref{sec:scalecut}. 
It would be interesting to perform the same analysis but in harmonic spaces, for which DES Y3 $C_\ell$'s have been recently measured and analysed~\cite{DES:2022qpf,DES:2024oud}. 
We foresee that $C_\ell$'s can have advantages when it comes to defining scale cuts, such that a potentially larger overall fraction of Fourier modes (and thus controlled information) can be kept as being less spread out. 
Finally, projected LSS statistics would benefit greatly from higher loops extending the analytical control of the theory over higher Fourier modes. 
In particular, the two-loop matter power spectrum in the EFTofLSS~\cite{Carrasco:2013mua,Anastasiou:2025jsy,Bakx:2025jwa} with a consistent accounting of the baryonic effects at this order~\cite{Braganca:2020nhv} is readily available to model shear correlation functions.
\item It would be interesting to investigate how the EFTofLSS likelihood of projected LSS data can shed light on the recent hints of evolving dark energy in galaxy-clustering data combined with CMB and supernova data~\cite{DESI:2024mwx,DESI:2024hhd,Lu:2025gki,DESI:2025zgx}, neutrino masses~\cite{Racco:2024lbu,Reeves:2025axp}, or primordial non-Gaussianities~\cite{DAmico:2022gki,Cabass:2022ymb}, and in combination with other probes. 
With the new publicly released pipeline with this article, we hope that our work can encourage fruitful developments in analyses of incoming photometric data such as Euclid~\cite{Euclid:2024yrr} or LSST~\cite{LSSTDarkEnergyScience:2012kar}, alone or in combination with other datasets.   
\end{itemize}
We leave these promising avenues for future work.



\section*{Acknowledgements}

\noindent  We thank Marco Bonici, Ross Cawthon, and Alexander Reeves for discussions. 
We thank Joe DeRose for providing the Buzzard simulation suites. 
We thank Jessie Muir and Daniel Gruen for support in handling the DES data. 
We are in debt to Jessie Muir, Dhayaa Anbajagane, and other people in DES for pointing to us a mistake in our treatment of the DES publicly-released cosmology chains in the first arXiv version of this paper.
We also thank Laura Reymond for letting us compare our implementation against hers. 
Preliminary results were presented at the workshop ``Theoretical modeling of Large-Scale Structure of the Universe'' at the Higgs Centre for Theoretical Physics in Edinburgh on June 3-5 2024, for which we thank the hosts and organisers. 
We acknowledge the support of Euler Cluster by High Performance Computing Group from ETHZ Scientific IT Services, where parts of the analysis were carried, and the use and support of the HPC (High Performance Computing) facility of the University of Parma, Italy, where parts of the analysis were carried.
This work made use of the \texttt{Python} libraries \href{https://numpy.org/}{\texttt{NumPy}}, \href{https://scipy.org/}{\texttt{SciPy}}, \href{https://matplotlib.org/}{\texttt{Matplotlib}}, \href{https://www.astropy.org/}{\texttt{Astropy}}, and \href{https://pyyaml.org/}{\texttt{PyYAML}}, as well as the software \href{https://www.wolfram.com/mathematica/}{\texttt{Mathematica}}. 
Our color schemes are taken from \href{https://personal.sron.nl/~pault/}{Paul Tol's notes}. 
PZ acknowledges support from Fondazione Cariplo under the grant No 2023-1205. 

This project used public archival data from the Dark Energy Survey (DES). Funding for the DES Projects has been provided by the U.S. Department of Energy, the U.S. National Science Foundation, the Ministry of Science and Education of Spain, the Science and Technology FacilitiesCouncil of the United Kingdom, the Higher Education Funding Council for England, the National Center for Supercomputing Applications at the University of Illinois at Urbana-Champaign, the Kavli Institute of Cosmological Physics at the University of Chicago, the Center for Cosmology and Astro-Particle Physics at the Ohio State University, the Mitchell Institute for Fundamental Physics and Astronomy at Texas A\&M University, Financiadora de Estudos e Projetos, Funda{\c c}{\~a}o Carlos Chagas Filho de Amparo {\`a} Pesquisa do Estado do Rio de Janeiro, Conselho Nacional de Desenvolvimento Cient{\'i}fico e Tecnol{\'o}gico and the Minist{\'e}rio da Ci{\^e}ncia, Tecnologia e Inova{\c c}{\~a}o, the Deutsche Forschungsgemeinschaft, and the Collaborating Institutions in the Dark Energy Survey.
The Collaborating Institutions are Argonne National Laboratory, the University of California at Santa Cruz, the University of Cambridge, Centro de Investigaciones Energ{\'e}ticas, Medioambientales y Tecnol{\'o}gicas-Madrid, the University of Chicago, University College London, the DES-Brazil Consortium, the University of Edinburgh, the Eidgen{\"o}ssische Technische Hochschule (ETH) Z{\"u}rich,  Fermi National Accelerator Laboratory, the University of Illinois at Urbana-Champaign, the Institut de Ci{\`e}ncies de l'Espai (IEEC/CSIC), the Institut de F{\'i}sica d'Altes Energies, Lawrence Berkeley National Laboratory, the Ludwig-Maximilians Universit{\"a}t M{\"u}nchen and the associated Excellence Cluster Universe, the University of Michigan, the National Optical Astronomy Observatory, the University of Nottingham, The Ohio State University, the OzDES Membership Consortium, the University of Pennsylvania, the University of Portsmouth, SLAC National Accelerator Laboratory, Stanford University, the University of Sussex, and Texas A\&M University.
Based in part on observations at Cerro Tololo Inter-American Observatory, National Optical Astronomy Observatory, which is operated by the Association of Universities for Research in Astronomy (AURA) under a cooperative agreement with the National Science Foundation.

%

\appendix

\section{Magnification bias}
\label{app:mag}

The angular distorsion induced by lensing will change the apparent magnitude of objects.
Since any galaxy survey has a selection in magnitude, this effect causes a modification in the observed number density of objects, which is a systematic dubbed ``magnification bias''.
It can be calculated and added to the galaxy density; here we follow the derivation presented in~\cite{Hui:2007cu}.

To derive the expression for the observed number density $n$, let us start by defining the true number density of galaxies $\Phi$ at redshift $z$, angular position $\hn$, and emitting flux in the range $\rmd f$ around $f$ as
\begin{equation}
  n(z, \hn) = \Phi(z, \hn, f) \rmd f \, .
\end{equation}
The \emph{observed} number density of galaxies is lensed by matter in the foreground.
Lensing induces the map $\hn \to \hn_L$, $f \to f_L$, with $z$ unaffected:
\begin{equation}
  \hn = \hn_L + \del \hn \, , \qquad f = A^{-1} f_L \, , \qquad A^{-1} = \det \frac{\de \hn_i}{\de \hn_{L,j}}  \, ,
\end{equation}
where $\del \hn$ is the angular displacement induced by lensing, and $A^{-1}$ the magnification.
Since the map conserves the galaxy number, we can write
\begin{equation}
  \Phi_L(z, \hn_L, f_L) \rmd f_L \rmd^2 \hn_L = \Phi(z, \hn, f) \rmd f \rmd^2 \hn \, .
\end{equation}
In addition to lensing, we have to take into account that the observed number density of galaxies is related to $\Phi$ by an efficiency function $\eps(f)$, which in the simplest case is just a step-function, meaning that we only observe galaxies above a minimum flux: $\eps(f) = \Theta_H(f - f_{\rm min})$.
Thus, introducing $\eps(f)$ and using the conservation equation, we get 
\begin{multline}
  n_L(z, \hn_L) = \int_0^\infty \eps(f_L) \Phi_L(z, \hn_L, f_L) \rmd f_L 
  = \int_0^\infty \eps(A f) \Phi(z, \hn, f) \rmd f \det \frac{\de \hn_i}{\de \hn_{L,j}} \\
  = \frac{1}{A(z, \hn)} \int_0^\infty \eps(A f) \Phi(z, \hn_L + \del \hn, f) \rmd f \, .
\end{multline}

At this point, we make the weak lensing approximation, substituting
\begin{equation}
  A \simeq 1 + 2 \kappa \, , \quad |\kappa| \ll 1 \, ,
\end{equation}
where $\kappa$ is the convergence.
Expanding in $\kappa$ and working at first order, we get
\begin{equation}
  n_L(z, \hn_L) = (1 - 2 \kappa(z, \hn_L)) \int_0^\infty  \(\eps(f) + 2 \kappa f \frac{\rmd \eps}{\rmd f} \) \Phi(z, \hn_L + \del \hn, f) \rmd f \, .
\end{equation}
This can be rewritten in the usual form
\begin{equation}
  n_L(z, \hn_L) = n(z, \hn_L + \delta \hn) \left[ 1 + (5 s - 2) \kappa \right] \, ,
\end{equation}
where
\begin{equation}
  s \equiv \frac{2}{5 n(z, \hn_L + \del \hn)} \int \frac{\rmd \eps}{\rmd \ln f} \Phi(z, \hn_L + \del \hn, f) \rmd f \to \frac{\rmd \log_{10} n}{\rmd m_{\rm min}}  \, .
\end{equation}
The last expression is derived in the limit of sharp filter $\eps(f) = \Theta_H(f - f_{\rm min})$, using the definition of the (apparent) magnitude $m_{\rm min} \equiv - 2.5 \log_{10} f_{\rm min}$.\footnote{
Explicitly, given $n = \int \epsilon(f, f_{\rm min}) \Phi(f) df$, 
\begin{equation}
\frac{\rmd \log_{10} n}{\rmd m_{\rm min}} = -\frac{2}{5} \frac{\rmd \ln n}{\rmd \ln f_{\rm min}} = -\frac{2}{5 n} \frac{\rmd n}{\rmd \ln f_{\rm min}} = -\frac{2}{5 n} \int \frac{\rmd \epsilon(f, f_{\rm min})}{\rmd \ln f_{\rm min}} \Phi(f) df = \frac{2}{5 n} \int \frac{\rmd \epsilon(f, f_{\rm min})}{\rmd \ln f} \Phi(f) df  \ , 
\end{equation}
where in the last step we have used the fact that $\frac{\rmd \eps}{\rmd \ln f_{\rm min}} = - \frac{\rmd \eps}{\rmd \ln f}$ when $\eps = \Theta_H(f - f_{\rm min})$. 
}

The last step is to derive the fractional overdensity $\delta$, which will be corrected for the magnification bias.
To do so, we first notice that, on average, $\braket{n_L} = \braket{n}$, since $\braket{\kappa} = 0$.
Therefore we get, neglecting second-order terms (in both $\kappa$ and $\delta$), 
\begin{equation}
  1 + \del_L(z, \hn_L) = 1 + \del(z, \hn_L + \del \hn) + (5 s - 2) \kappa  \, .
\end{equation}
We finally find that the observed $\delta$ is
\begin{equation}
  \label{eq:splitmag}
  \del_L(z, \hn) = \del(z, \hn) + \del_\mu(z, \hn) \, , \quad
  \del_\mu = (5 s - 2) \kappa = b_{\rm mag} \kappa \, ,
\end{equation}
which is the definition of the magnification bias $b_{\rm mag}$.
In terms of the gravitational potential $\phi$, the convergence is given by
\begin{equation}\label{eq:kappa}
  \kappa(z(\chi), \hn) = \int_0^\chi \rmd \tchi \frac{\chi - \tchi}{\chi \tchi} \nabla_{\hn}^2 \phi(z(\tchi), \tchi \hn)
  = \int_0^\chi \rmd \tchi \frac{\chi - \tchi}{\chi \tchi} \int_{\vk} \phi(z(\tchi), \vk) \nabla_{\hn}^2 e^{i \tchi \hn \cdot \vk} \, .
\end{equation}
Finally, one expands in spherical harmonics $\kappa(z(\chi), \hn) = \sum_{\ell m} \kappa_{\ell m}(z(\chi)) Y_{\ell m}(\hn)$:
\begin{equation}\label{eq:kappa_ylm}
  \begin{split} 
  \kappa_{\ell m}(z(\chi)) =& - 4 \pi i^\ell \ell (\ell+1) \int_{\vk} Y^*_{\ell m}(\hk) \int_0^\chi \rmd \tchi \frac{\chi - \tchi}{\chi \tchi} j_\ell(\tchi k) \phi(z(\tchi), \vk) \\
  =& 4 \pi i^\ell \ell (\ell+1) \int_{\vk} Y^*_{\ell m}(\hk) \int_0^\chi \rmd \tchi \frac{\chi - \tchi}{\chi \tchi} j_\ell(\tchi k) \frac{3}{2} \Om_{m,0} H_0^2 \frac{1+z(\tchi)}{k^2} \del_m(z(\tchi), \vk) \, .
  \end{split}
\end{equation}
In the last line we used the Poisson equation $k^2 \phi(z, \vk) = - \frac{3}{2} \Om_{m,0} H_0^2 (1+z) \del_m(z, \vk)$.

\section{Limber approximation}
\label{app:limber}
In this appendix, we discuss the Limber approximation used to evaluate the lensing correlation functions.
We start from Eq.~\eqref{eq:Cpsi}:
\begin{equation}
  C^{ij}_{\psi \psi}(\ell) = \frac{8}{\pi} \int_0^\infty \rmd \chi \int_0^\infty \rmd \tchi \int_0^\infty \frac{\rmd k}{k^2} \,  j_\ell(k \chi) j_\ell(k \tchi) \, \frac{f^i_\kappa(\chi)}{\chi^2} \frac{f^j_\kappa(\tchi)}{\tchi^2} \, P_{mm}\(k, z, \tilde{z} \) \, ,
\end{equation}
where for simplicity of notation we use $z=z(\chi)$, $\tilde{z} = z(\tchi)$.

A commonly used approximation is the Limber approximation, which amounts to replacing the Bessel functions with
\begin{equation}
  j_\ell(k \chi) \simeq \sqrt{\frac{\pi}{2 (l + \frac{1}{2})}} \del_D\(\ell + \frac{1}{2} - k \chi \) \, .
\end{equation}
This is valid for $\ell \gg 1$, since the Bessel sharply rises to a peak at $k \chi \simeq \ell + 1/2$ and oscillates rapidly afterwards; the width of the peak scales as $\ell^{1/3}$.
Corrections to the Limber approximation have been evaluated in~\cite{LoVerde:2008re}.
Here we will mostly follow their arguments, taking into account that we have the unequal-time power spectrum in the integral.
First, we will substitute the spherical Bessel functions with Bessel functions, which have simpler analytical properties, using the formula $j_\ell(x) = \sqrt{\pi / (2 x)} J_{\nu}(x)$, with $\nu = \ell+\frac{1}{2}$.
For large $\nu$, $\int_0^\infty \rmd x J_\nu(x) = 1$ and $\int_0^\infty \rmd x J_\nu(x) f(x) \simeq f(\nu)$ for $f(x)$ slowly varying around the peak of $J_\nu(x)$.
We are led to the expression: 
\begin{equation}
  C^{ij}_{\psi}(\ell) = 4 \int_0^\infty \frac{\rmd k}{k^3}
  \int_0^\infty \rmd \chi \, J_\nu(k \chi) \frac{f^i_\kappa(\chi)}{\chi^{5/2}}
  \int_0^\infty \rmd \tchi \,  J_\nu(k \tchi)   \frac{f^j_\kappa(\tchi)}{\tchi^{5/2}} \, P_{mm}\(k, z, \tilde{z} \) \, .
\end{equation}
To simplify this, we note that we can express the power spectrum as a sum of terms in which the dependence on $z$, $\tilde{z}$ is separable:
\begin{equation}
  P_{mm}(k, z, \tilde{z}) = \sum_{n_1, n_2} D(z)^{n_1} D(\tilde{z})^{n_2} P_{n_1,n_2}(k) \, .
\end{equation}
Using this decomposition, the integral simplifies as
\begin{equation}\label{eq:Cpsi}
  C_{\psi}(\ell) = 4 \sum_{n_1, n_2} \int_0^\infty \frac{\rmd k}{k^3} P_{n_1, n_2}(k) I_{\nu}^{n_1, i}(k) I_{\nu}^{n_2, j}(k) \, ,
\end{equation}
where $I_\nu^{n, i}(k)$ is the following 1D integral:
\begin{equation}
  I_\nu^{n, i}(k) = \int_0^\infty \rmd \chi \,  J_\nu(k \chi) D(z)^n \frac{f^i_\kappa(\chi)}{\chi^{5/2}}
  \equiv \int_0^\infty \rmd \chi \,  J_\nu(k \chi) F_{n,i}(\chi) \, .
\end{equation}
For the Limber approximation to be accurate, we expect $F^{n,i}(\chi)$ to be slowly varying around $\chi = \nu/k$.
To quantify the deviations, we can Taylor expand the $F$ around $k \chi = \nu$:
\begin{equation}
  I_\nu^{n, i}(k) = \sum_{m=1}^\infty \frac{1}{m!} \left. \frac{\rmd^m F_{n,i}(r)}{\rmd \, r^m} \right|_{r = \frac{\nu}{k}} \int_0^\infty \frac{\rmd x}{k} J_\nu(x) \frac{(x - \nu)^m}{k^m} \, .
\end{equation}
The integral of powers are only convergent for $-1 < m <1/2$, but can be extended to all natural $m$ by analytic continuation.
An equivalent procedure is followed in~\cite{LoVerde:2008re}.
The first few terms read:
\begin{equation}
  \begin{split}\label{eq:II}
    k I_\nu^{n, i}(k) &= F_{n,i}(r)|_{r=\frac{\nu}{k}} -  \frac{F''_{n,i}(r)|_{r=\frac{\nu}{k}}}{2 k^2} - \nu \frac{F^{(3)}_{n,i}(r)|_{r=\frac{\nu}{k}}}{6 k^3} + \mathcal{O}(\nu^{-4} ) \\
    &= F_{n,i}(r)|_{r=\frac{\nu}{k}} - \frac{1}{\nu^2} \[ \left. \frac{r^2}{2} F''_{n,i}(r)\right|_{r=\frac{\nu}{k}} + \left. \frac{r^3}{6} F^{(3)}_{n,i}(r)\right|_{r=\frac{\nu}{k}} \] + \mathcal{O}(\nu^{-4} ) \, .  
  \end{split}  
\end{equation}
Thus, inserting~\eqref{eq:II} in~\eqref{eq:Cpsi}, the power spectrum is
\begin{equation}
  \begin{split}
  C^{ij}_{\psi}(\ell) =& \, 4 \sum_{n_1, n_2} \int_0^\infty \frac{\rmd k}{k} \frac{P_{n_1, n_2}(k)}{k^4} F_{n_1,i}(r) F_{n_2,j}(r) |_{r=\frac{\nu}{k}}\\
  \times & \left\{ 1 -  \frac{1}{6 \nu^2} \left. \[ 3 r^2 \( \frac{F_{n_1,i}''(r)}{F_{n_1,i}(r)}  + \frac{F_{n_2,j}''(r)}{F_{n_2,j}(r)} \) + r^3 \( \frac{F_{n_1,i}^{(3)}(r)}{F_{n_1,i}(r)} + \frac{F_{n_2,j}^{(3)}(r)}{F_{n_2,j}(r)} \) \] \right|_{r=\frac{\nu}{k}} \right\} \, .
  \end{split}
  \label{eq:fullCpsi}
\end{equation}
At lowest order, we get the standard expression for the Limber approximation~\cite{LoVerde:2008re}, 
\begin{equation}
  C^{ij}_{\psi}(\ell) = \, \frac{4}{(\ell+\frac{1}{2})^5} \int_0^\infty \rmd k P_{mm}(k) f_{\kappa}^i(r) f_{\kappa}^j(r) \, .
  \label{eq:limber}
\end{equation}

Let us go back to the corrections.
After a few integration by parts, Eq.~\eqref{eq:fullCpsi} can be rewritten in the form:
\begin{equation}
  \begin{split}
  C^{ij}_{\psi}(\ell) =& \, 4 \sum_{n_1, n_2} \int_0^\infty \frac{\rmd k}{k} \frac{P_{n_1, n_2}(k)}{k^4} F_{n_1,i}(r) F_{n_2,j}(r) \\
  \times & \[ 1 + \frac{1}{2 \nu^2} \( \frac{\rmd \ln F_{n_1,i}(r)}{\rmd \ln r} \frac{\rmd F_{n_2,j}(r)}{\rmd \ln r} \frac{\rmd \ln \tilde{P}_{n_1, n_2}(k)}{\rmd \ln k} - s(k) \) \] \, ,
  \end{split}
\end{equation}
where
\begin{equation}
  s(k) = \frac{k^2}{3 \tilde{P}_{n_1, n_2}(k)} \( 3 \tilde{P}_{n_1, n_2}''(k) + k \tilde{P}_{n_1, n_2}^{(3)}(k) \) \, ,
\end{equation}
and we used the notation $\tilde{P}_{n_1, n_2}(k) = P_{n_1, n_2}(k) / k^4$.
The first term is the usual Limber approximation.
The corrections depend on $\nu = \ell + 1/2$, and on the kernels $F_{n,i}(r)$.
To get a rough estimation of the validity of the Limber approximation, following~\cite{LoVerde:2008re}, we can approximate $F_{n,i}(r)$ by a Gaussian with peak at $\mu$ and width $\sigma$. 
Then the log derivative is $r (\mu-r)/\sigma^2$.
Suppose that both kernels are peaked at the same comoving distance $\bar{r}$.
In this case, we can estimate that the Limber approximation is valid for $\nu \gtrsim \bar{r}/\sigma$, where $\sigma^2$ is the sum in quadrature of the widths of $F_{n,i}(r)$, $F_{n,j}(r)$.
If the kernels are peaked at different distances $\mu_i$ and $\mu_j$, then the approximation becomes worse, requiring $\nu \gtrsim |\mu_i^2 - \mu_j^2|/\sigma^2$. 
As a typical example, taking $\sigma \sim 500$ Mpc and $|\mu_i - \mu_j| \sim 2000$ Mpc, we find $\ell_{\rm min} \sim 16$. 
We therefore use only the Limber approximation for the loop contributions that start contributing significantly only at high $\ell$'s.

\section{Baryonic effects}
\label{app:baryons}
The EFTofLSS provides a consistent framework to predict the clustering of dark matter and baryons on large scales, where the leading corrections in perturbations have a simple and calculable functional form even after the onset of baryonic processes~\cite{Lewandowski:2014rca,Braganca:2020nhv}. 
We here provide a derivation of two-point correlators of biased tracers in the EFTofLSS at the one-loop in the presence of baryons, following~\cite{Lewandowski:2014rca}.

After integrating out the short modes, a universe with two fluids interacting gravitationally and through momentum exchange (cold dark matter and baryons) can be described by the following set of equations for the long-wavelength modes: 
\begin{align}
& \nabla^2 \phi = \frac{3}{2} H_0^2 \frac{a_0^3}{a}(\Omega_c \delta_c + \Omega_b \delta_b) \, , \\
& \dot{\delta}_\sigma = -\frac{1}{a} \partial ((1+\delta_\sigma) v_\sigma^i) \, , \\
& \partial_i \dot{v}_\sigma^i + H \partial_i v^i_\sigma + \frac{1}{a} \partial_i (v_\sigma^j \partial_j v_\sigma^i) + \frac{1}{a} \partial^2 \phi  = - \frac{1}{a} \partial_i (\rho_\sigma^{-1}\partial_j \tau^{ij}) \pm \frac{1}{a} \partial_i (\rho_\sigma^{-1}V^i) \, ,
\end{align}
where the indices $\sigma = b, c$ refer to baryons and cold dark matter, respectively. 
The first equation is the Poisson equation relating the Laplacian of the gravitational potential to both the dark matter and baryon density contrasts. 
The second and third lines are respectively the continuity and Euler equations for each species. 
The effect of the short-distance physics is encoded in the effective stress energy tensors $\rho_\sigma^{-1} \tau^{ij}_\sigma$ and the momentum-exchange interaction terms $\pm \rho_\sigma^{-1} V^i$, where the signs are opposite since in the sum the total momentum is conserved. 
Due to the equivalence principle, at lowest order, the r.h.s. of the Euler equations can be re-expressed as terms proportional to $\partial_i \partial_j \phi$ and $\partial_i v_\sigma^j$, that in turn can be expressed as function of the density contrasts through the Poisson equation~\cite{Baumann:2010tm,Lewandowski:2014rca,Braganca:2020nhv}: 
\begin{equation}
- \frac{1}{a} \partial_i (\rho_\sigma^{-1}\partial_j \tau^{ij}) \pm \frac{1}{a} \partial_i (\rho_\sigma^{-1}V^i) = c_{\sigma,b}(a) \partial^2 \delta_b + c_{\sigma,c}(a) \partial^2 \delta_c  + \dots \, ,
\end{equation}
where $c_{\sigma, b}(a), c_{\sigma, c}(a)$ are free time coefficients. 
Here, we neglect the counterterm proportional to the relative velocity, which as argued in~\cite{Braganca:2020nhv} is negligibly small.
At lowest order, these give rise to four counterterms. 
It is convenient to transform to the basis of adiabatic and isocurvature modes, defined by
\begin{equation}
\Omega_m \delta_A = \Omega_c \delta_c + \Omega_b \delta_b \, , \qquad \delta_I = \delta_c - \delta_b \, ,
\end{equation}
and similar definitions for the velocities.
The adiabatic (isocurvature) mode represents the total (relative) density fluctuation. 
The isocurvature mode becomes more and more suppressed with time. At leading order, the current ratio is about
\begin{equation}
\frac{\delta_I(k, a_0)}{\delta_A(k, a_0)} \sim \frac{D(a_i)}{D(a_0)} \frac{\delta_I(k, a_i)}{\delta_A(k, a_i)} \sim 10^{-2} \, .
\end{equation}
Thus, for practical purposes, we will be interested only in the adiabatic auto-correlation, which is what is considered in the baseline model, and eventually the adiabatic isocurvature cross-correlation. 

For a biased tracer `g', we can write an EFT expansion in biased operators and spatial derivatives $b_{\alpha,\sigma} \mathcal{O}_{\alpha,\sigma}$: 
\begin{align}
\delta_g & = \sum_\alpha b_{\alpha,c} \mathcal{O}_{\alpha,c} + \sum_\alpha b_{\alpha,b} \mathcal{O}_{\alpha,b} \,  \\
	& = \sum_\alpha b_{\alpha,A} \mathcal{O}_{\alpha,A} + \sum_\alpha b_{\alpha,I} \mathcal{O}_{\alpha,I} \, ,
\end{align}
where in the second line, we have recast the expansion in the basis of adiabatic and isocurvature modes. 
For example, the EFT expansion contains the linear bias terms and the counterterms: 
\begin{equation}
\sum_\alpha b_{\alpha,\sigma} \mathcal{O}_{\alpha,\sigma} \supset \lbrace b_{1,\sigma} \delta_{\sigma}^{\rm lin}, c_{\sigma,b} (k^2/k_{\rm M}^2) \delta_{b}^{\rm lin}, c_{\sigma,c} (k^2/k_{\rm M}^2) \delta_{c}^{\rm lin} \rbrace \, .
\end{equation}
Such expansion up to the order relevant to a one-loop calculation has been derived elsewhere for one species, see e.g.~\cite{Senatore:2014eva,Angulo:2015eqa,Fujita:2016dne}. 
Here, the generalization is straightforward, keeping in mind that there are now two expansions, one for each species, of the same functional form but with different EFT parameters, and there are two counterterms in each expansion instead of one, that are proportional to the density contrast of each species. 

Given the size of the isocurvature modes, we will only consider the leading linear correction from the isocurvature modes. 
Thus, in the isocurvature EFT expansion, we only keep the linear isocurvature bias term and the adiabatic counterterm, while in the adiabatic EFT expansion, we keep the linear and nonlinear adiabatic terms and the adiabatic counterterm, but drop the isocurvature counterterm:
\begin{align}
\delta_{g,I} & \equiv \sum_\alpha b_{\alpha,I} \mathcal{O}_{\alpha,I}  = b_{1,I} \delta_{I}^{\rm lin} + c_{I,A} (k^2/k_{\rm M}^2) \delta_{A}^{\rm lin} + \dots \, , \label{eq:delta_g_I} \\
\delta_{g,A} & \equiv \sum_\alpha b_{\alpha,A} \mathcal{O}_{\alpha,A} = b_{1,A} \delta_{A}^{\rm lin} + \sum b_{{\rm NL}, A} \mathcal{O}_{{\rm NL}, A} + c_{A,A} (k^2/k_{\rm M}^2) \delta_{A}^{\rm lin} + \dots \, ,
\end{align}       
where $\sum b_{{\rm NL}, A} \mathcal{O}_{{\rm NL}, A}$ denotes the nonlinear adiabatic terms. 

The galaxy-galaxy two-point correlator follows (the derivation is similar for the galaxy-matter and matter-matter correlators): 
\begin{align}
\braket{\delta_g \delta_g} & = \braket{\delta_{g,A} \delta_{g,A}} + \braket{\delta_{g,A} \delta_{g,I}} + \braket{\delta_{g,I} \delta_{g,I}} \, \\
	& \simeq \braket{\delta_{g,A} \delta_{g,A}}' + b_{1,A} b_{1,I} \braket{\delta^{\rm lin}_{A} \delta^{\rm lin}_{I}} \, , \label{eq:corr_adia_iso}
\end{align}
where at the second line we have dropped the isocurvature auto-correlation which is negligible.
$\braket{\delta_{g,A} \delta_{g,A}}'$ is given by the baseline model presented in sec.~\ref{sec:eftoflss} with a redefinition of the counterterm coefficient $2 b_{1,A} c_{\rm gg}' \equiv 2 b_{1,A} c_{A,A} + b_{1,A} c_{I,A} + b_{1,I} c_{A,A}$, such that counterterms from the adiabatic isocurvature cross-correlation are absorbed. 
Therefore, the baseline model automatically incorporates the leading short-scale baryonic effects through the counterterm. 
We find worth emphasizing that such EFT correction can accommodate, at least in principle, any baryonic processes, from star formations to active galactic nuclei feedback, supernovae, black hole accretion, etc. 
For tests against simulations of the responses of the EFTofLSS to various baryonic processes, see~\cite{Lewandowski:2014rca,Foreman:2015uva,Braganca:2020nhv}. 

In sec.~\ref{sec:validation}, we check that further marginalising over the subleading contributions from the linear adiabatic isocurvature cross power spectrum with $b_{ia}$'s drawn from a Gaussian prior centered on $0$ of width $2$ (one for each lens) leads to negligible shifts in the inferred cosmological parameters, therefore justifying to neglect all baryonic effects beyond the leading one automatically captured by the EFTofLSS counterterm. 
We stress that this occurs in our analysis since we restrict the scales only to the one accessible by the EFTofLSS. 
In particular, we do not make use of knowledge from $N$-body simulations to fit the cosmic shear, which would allow one, in principle, to access scales beyond perturbation theory, as long as baryonic feedbacks are under control. 
Given current debates around the size of baryonic effects at small scales, in that sense, our results are more robust. 


\section{Credible intervals}
\label{app:CI}
We now specify the way we report our cosmological results by defining the statistics we use. 
Given a posterior distribution $\mathcal{P}$ of inferred parameters $\pmb \lambda \in \mathbb{R}^k$, we can define moments as expectation values over $n \leq k$ parameters $\lambda_{\alpha_{1\dots n}}\equiv \lambda_{\alpha_1} \times \dots \times \lambda_{\alpha_n}$ as
\begin{equation}\label{eq:estimator}
\mathbb{E}_\mathcal{P}[\lambda_{\alpha_{1\dots n}}] = \frac{1}{Z} \int  \mathcal{M(\pmb \lambda)} d^k\pmb \lambda \ \lambda_{\alpha_{1\dots n}} \ \mathcal{P}(\pmb\lambda) \ , 
\end{equation}
where $\mathcal{M(\pmb \lambda)} d^k\pmb \lambda$ is the integration measure and $Z \equiv \mathbb{E}_\mathcal{P}[1]$ is some normalisation. 
The credible interval for parameter $\lambda_\alpha$ read $\mu_\alpha \pm \sigma_\alpha$, where $\mu_\alpha$ is the mean and $\sigma_\alpha$ is the standard deviation defined as
 \begin{equation}
\mu_\alpha = \mathbb{E}_\mathcal{P}[\lambda_\alpha] \ , \qquad \sigma_\alpha^2 = \mathbb{E}_\mathcal{P}[\lambda_\alpha^2] - \mu_\alpha^2 \ .
 \end{equation}
The standard choice for the measure is Lebesgue ($\mathcal{M} \equiv 1$), and then for Gaussian distributions the mean is unbiased with minimal variance.
This property is desirable when it comes to report our credence on physical quantities, and it can be shown that the Jeffreys measure, $\mathcal{M_F(\pmb \lambda)} = \sqrt{\det |\mathcal{F}(\pmb \lambda)|}$, where $\mathcal{F}$ is the Fisher matrix, is usually a good choice for the kind of posterior distributions we are dealing with here~\cite{Reeves:2025bxc}. 
While in the limit where the data volume becomes infinite, the posteriors are expected to become Gaussian, for a finite data size, however, they can be highly non-Gaussian, especially when looking at the projection on the space of interest, i.e., the cosmological parameters, upon marginalising over a large subspace of nuisance parameters (see e.g.,~\cite{DAmico:2022osl,Simon:2022csv,Reeves:2025bxc}). 
As shown in ref.~\cite{Reeves:2025bxc}, $\mathcal{M_F(\pmb \lambda)} d^k \pmb \lambda$ is, under certain conditions, a reasonable measure choice such that the marginalised posteriors appear unbiased. 

We implement the Jeffreys measure for the linear parameters that we analytically marginalise over by simply dropping the log-determinant in eq.~\eqref{eq:marg}.
The rest of the measure over the scanned parameters is not practical to compute over an entire sampling, as it involves the computation of derivatives of the theory model, which for us have to be numerical --- although this could be feasible with automatic differentiation.
To proceed, we therefore expand the measure around the best fit $\pmb \lambda^*$ such as
\begin{equation}\label{eq:expandM}
\log \mathcal{M_F(\pmb \lambda)} = \log \mathcal{M_F(\pmb \lambda^*)} + \partial_\alpha \log \mathcal{M_F(\pmb \lambda)} \Big|_{\pmb \lambda = \pmb \lambda^*} (\lambda_\alpha -\lambda_\alpha^*) +  \dots \ , 
\end{equation}
where $\partial_\alpha \equiv \frac{\partial}{\partial \lambda_\alpha}$. 
The terms multiplying $\lambda_\alpha^*$ are constants and therefore cancel out in the definition of the expectation values.
It follows that the integration measure $\mathcal{M_F(\pmb \lambda)} d^k \pmb{\lambda}$ is equivalent, up to higher-order terms, to adding to $\log \mathcal{P}$ the following piece, linear in $\pmb \lambda$, 
\begin{equation}
\log \mathcal{M}_\mathcal{F}'(\pmb \lambda) =  \frac{1}{2} \lambda_\alpha \, \partial_\alpha \log \det |\mathcal{F}(\pmb \lambda)| \Big|_{\pmb \lambda = \pmb \lambda^*}     \ ,
\end{equation}
where 
\begin{equation}\label{eq:dlogF}
\frac{1}{2}\partial_\alpha \log \det |\mathcal{F}| = \frac{1}{2} \mathcal{F}^{-1}_{\mu \nu} \ \partial_\alpha \mathcal{F}_{\mu \nu} 
\end{equation}
is simply a number that can be computed once and for all credible intervals. 

Yet in practice, instead of computing eq.~\eqref{eq:dlogF}, we find more convenient to perform the equivalent procedure of generating (noiseless) synthetic data on the best fit and `experimentally' calibrate the linear log-measure, such that the recovered parameters are unbiased~\cite{DAmico:2022osl}. 
Our procedure on synthetic data adds to the log-likelihood a linear term that is more precise, in principle, than the Jeffreys prior since the latter is based on the Fisher, although the Jeffreys has the advantage of a being a prior correction that can be implemented without any use of synthetic data. 
This construction thus ensures to achieve the desirable properties in the statistics we use to report our results. 
Based on our findings in sec.~\ref{sec:synth}, the linear log-measure we choose to report all cosmological results obtained from our DES Y3 likelihoods is
\begin{equation}
  \label{eq:linM}
\log \mathcal{M}(\pmb \lambda) = - 55\, \omega_{cdm} - 9\,  b_1 \ ,
\end{equation}
where $b_1$ is the linear bias of $w_{1,1}$.  
Note that in principle, our statistics could be defined for each different datasets considered: \texttt{Buzzard}, \texttt{MagLim}, or \texttt{redMaGiC}. 
However, their respective best-fits are found to lie close, and the measure expansion~\eqref{eq:expandM} can be performed around arbitrary points as long as they are reasonably well within the $1\sigma$-region. 
Therefore for simplicity we always use eq.~\eqref{eq:linM} for reporting the cosmological results.


\section{Parameter correlation}
\label{app:correlation}
In this appendix we discuss how one can estimate the correlation between two parameters that are known to be close in value, such as a galaxy bias measured in two neighbouring redshift bins. 
These estimates are used to construct the prior we impose on the EFT parameters in our analysis presented in sec.~\ref{sec:prior}. 

Let $c_1$ and $c_2$ be two parameters such that $c_1 \equiv c_* + \delta/2 \ , c_2 \equiv c_* - \delta/2$. 
Assuming that $c_1$ and $c_2$ are distributed according a bivariate Normal distribution with mean $(c_1, c_2)$ and covariance $C_{ij} \equiv \sigma (\delta^K_{ij} + (1-\delta^K_{ij}) \rho)$, where $\rho$ is the correlation coefficient between $c_1$ and $c_2$, we have $\sigma_{c_*}^2 = \sigma^2 (1 + \rho)/2$ and $\sigma_\delta^2 = 2\sigma^2 (1 - \rho) \equiv \sigma^2 \epsilon^2$. 
As we are interested in finding $\rho$, we can simply solve for the variance ratio of the difference to the mean: 
\begin{equation}
\frac{\sigma_\delta^2}{\sigma_{c_*}^2} \equiv  \frac{\epsilon^2}{(1+\rho)/2} \simeq \epsilon^2 \ , 
\end{equation}
where the second equality is true as long as $\rho$ is close to one (and so $\epsilon^2 \ll 1)$. 
Therefore, to find $\rho \equiv 1- \epsilon^2/2$, we simply need to estimate $\epsilon \equiv \sigma_\delta / \sigma$. 
Intuitively, 
\begin{itemize}
\item $\sigma$ is the our prior on how much we expect $c_1$ or $c_2$ can vary. An order-of-magnitude estimate is to take $\sigma$ as the natural size of the parameters so $\sigma \sim \mathcal{O}(c_*)$. 
\item $\sigma_\delta$ is our prior on how much we expect $c_1$ and $c_2$ can be different. Similarly, we can estimate its size as $\sigma_\delta \sim \mathcal{O}(\delta)$. 
\end{itemize}
Thus, a rough estimate of the size of $\epsilon$ is $\sim \mathcal{O}(\delta/c_*)$.  

We now focus on the case of two neighbouring redshift bins. 
For definiteness we show the formulae for galaxy clustering, although the method described here is general. 
To find an estimate of $\epsilon$, we realise that $c_1$ and $c_2$ are \emph{effective} bin-integrated parameters associated to redshift bins 1 and 2 respectively, such that
\begin{equation}
c_{i} c_{j} \ I_{ij}[1,1]  \approx c_*^2 \ I_{ij}[D_i,D_j] \ ,
\end{equation}
where 
\begin{equation}\footnotesize
I_{ij}[X_i,X_j] := \int d\ell \ J_\alpha (\ell \theta) \int dk \ k^2 \int d\chi_1 \ \int d\chi_2 \ X_i(\chi_1) X_j(\chi_2) W_i(\chi_1) W_j(\chi_2) j_\ell(\chi_1 k) j_\ell(\chi_2 k) P(k, \chi_1, \chi_2) \ .
\end{equation}
Here we assume that the time dependence of the \emph{intrinsic} parameters $c_i(z)$ is $D_i(\chi) \simeq D(\chi_i)/D(\chi_*)$, where $i = 1, 2$. 
Therefore, 
\begin{equation}\label{eq:I}
\epsilon^2 \sim \frac{(c_1-c_2)^2}{c_*^2} \approx \frac{I_{11}[D_1,D_1]}{I_{11}[1,1]} + \frac{I_{22}[D_2,D_2]}{I_{22}[1,1]} - 2 \frac{I_{12}[D_1,D_2]}{I_{12}[1,1]} \ .
\end{equation}
In the limit of $W_i(\chi) \rightarrow \delta_D(\chi - \chi_i)$, we recover the naive estimate discussed in sec.~\ref{sec:prior}, 
\begin{equation}
\epsilon \sim \frac{|D(\chi_1) - D(\chi_2)|}{D(\chi_*)} . 
\end{equation}

In practice, for $b_1$'s, we estimate $I_{ij}$ simply with $P \equiv P_{\rm lin}$ in eq.~\eqref{eq:I}. 
For the nonlinear parameters, we do the same but with the replacement $P_{\rm lin} \rightarrow (k/k_{\rm NL})^2/(1+(k/k_{\rm NL})^2) P_{\rm lin}$, which is a good proxy for the size of the loop, given that $\sim k^2/k_{\rm NL}^2$ is the parameter controlling the loop expansion in the EFTofLSS. 
For the correlations that we are interested in discussed in sec.~\ref{sec:prior}, we find $\epsilon$ around $5-15\%$. 
To be conservative, we choose for our prior $\epsilon = 0.2$.

\section{Additional tests on simulations}\label{app:buz18}

\begin{figure}[ht!]
\centering
\scriptsize
\begin{tabular}{|ccccccc|}\hline
$\Delta X / X$ & $\Omega_m$ 			 &  $ h$					&  $S_8 $ 					& $\omega_{cdm}$ 				& $\ln (10^{10}A_s)$ 	& $\sigma_8$	\\ \hline \hline
Buzzard V$\times 18$			&  $0.004^{+0.035}_{-0.031}$			 & $0.025^{+0.023}_{-0.029}$ 		& 	 $0.0098^{+0.0097}_{-0.011}$			& 	$0.063^{+0.028}_{-0.032}$		& $-0.023\pm 0.010$ 		& $0.008^{+0.022}_{-0.028}$	\\
Buzzard V$\times 18$, diag. cov.			&  $0.0095\pm 0.032$			 & $0.005\pm 0.023$ 		& 	 $-0.004^{+0.009}_{-0.010}$			& 	$0.023^{+0.019}_{-0.022}$		& $-0.0155\pm 0.0083$ 		& $-0.008 \pm 0.025$	\\ \hline 
\end{tabular}\\ \vspace{0.3cm}
\includegraphics[width=0.99\textwidth]{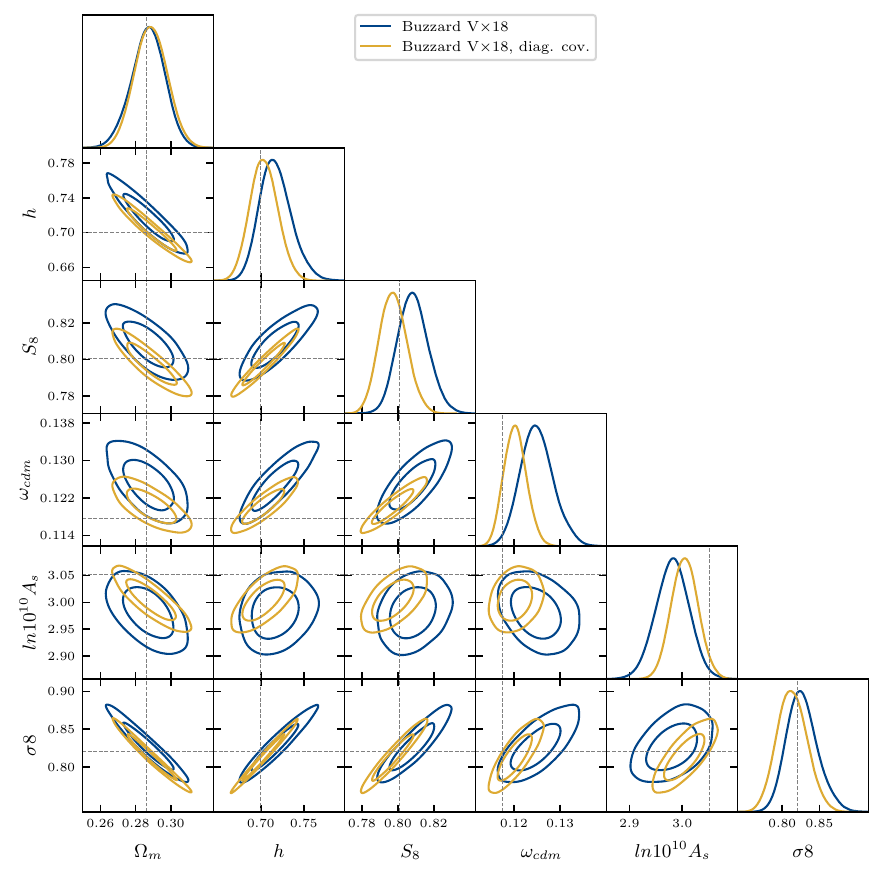}
\caption{\footnotesize  
Relative $68\%$-credible intervals and triangle plots of $\Lambda$CDM cosmological parameters from the EFTofLSS analysis of the \texttt{Buzzard} simulations, with $\omega_b$ and $n_{s}$ set to their truth. 
Contrary to most results presented in the main text, the results here are obtained fitting with a covariance for the total simulation volume, about $18$ times the one of DES Y3 data. 
}
\label{fig:buz18}
\end{figure}

In this appendix, we present additional tests on simulations. 
First, we show results fitting the \texttt{Buzzard} simulations with a covariance for the total volume of the simulation suite, about $18$ times the one of DES Y3 data.
Posteriors and credible intervals are shown in fig.~\ref{fig:buz18}. 
In the main text, we have already discussed the results obtained fitting with the covariance obtained from ref.~\cite{DES:2020ypx}. 
We have found that, although all systematic shifts measured on the 1D posteriors of the cosmological parameters are tolerably small ($\lesssim 0.3	\sigma$), a theory error seems to appear in the direction orthogonal to the principal axis of the contour ellipse in the $\omega_{cdm} - \ln(10^{10}A_s)$ plane.
To understand this fact, we run the same fit but using only the diagonal entries of the covariance. 
For that case, we see that we recover the truth better, with the largest systematic shifts measured to be $-0.19\sigma_{\rm stat}$ on $\ln(10^{10}A_s)$ where $\sigma_{\rm stat}$ is the statistical uncertainty corresponding to the DES Y3 data volume. 
Our result seems to indicate that, at that high-precision level, inaccuracies or approximations in the modelling of the covariance, particularly in the the cross-correlations, can lead to appreciable shifts in the measured cosmological parameters, though acceptable for the analysis of DES Y3 data.
It would be interesting to model the covariance of 3$\times$2pt analyses within the EFTofLSS. 
We leave this to future work. 

\begin{figure}[ht!]
\centering
\includegraphics[width=0.99\textwidth]{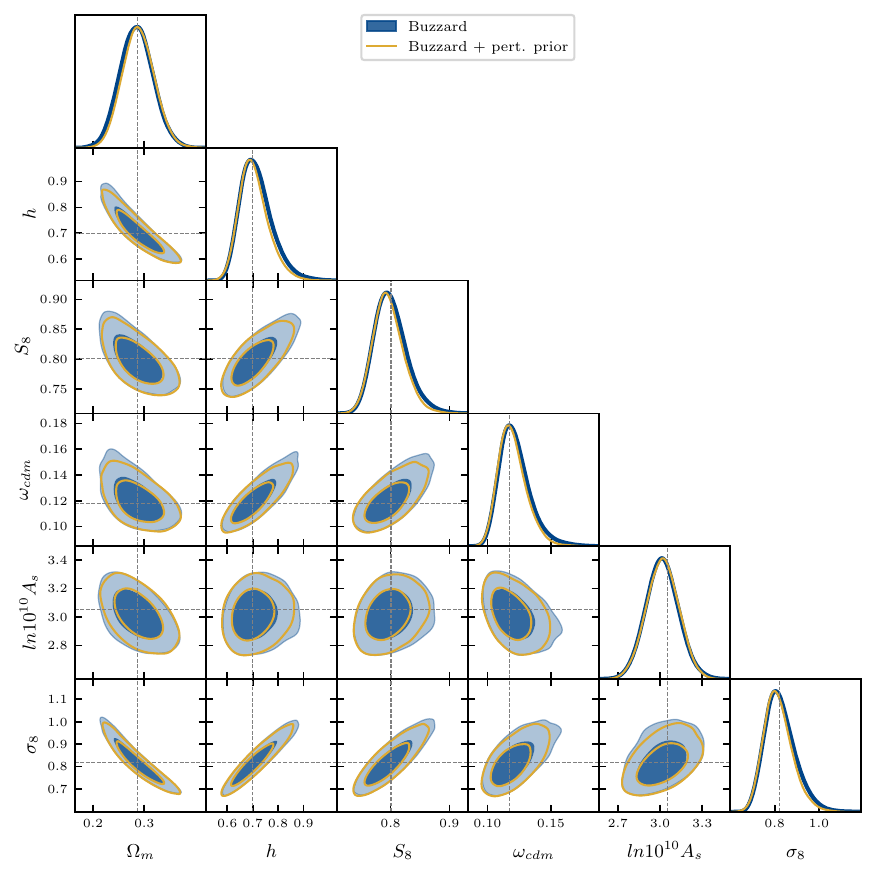}
\caption{\footnotesize  
Triangle plots of $\Lambda$CDM cosmological parameters from the EFTofLSS analysis of the \texttt{Buzzard} simulations, with $\omega_b$ and $n_{s}$ set to their truth, imposing or not a perturbativity prior. 
}
\label{fig:pp}
\end{figure}

Next, we compare in fig.~\ref{fig:pp} results on the \texttt{Buzzard} simulations with covariance corresponding to DES Y3 data volume fit with or without the perturbativity prior presented in sec.~\ref{sec:pp}. 
We see good consistency in the results, showing that most of the posterior distribution corresponds to consistent EFTofLSS predictions. 
Some fraction gets penalised, leading to visible reductions of the credible regions. 
On the best fit, we find the $\chi^2$ of the data is practically unchanged ($\delta \chi^2 = 0.01$) and the penalty from the perturbativity prior~\eqref{eq:pp} is $\sim 1$. 
This shows that the EFTofLSS fits 3$\times$2pt data on the retained scales is in good consistency with our expectation on the size of the one-loop contributions. 
As for the prior volume projection effects described in sec.~\ref{app:CI}, we find that the perturbativity prior does not play a significant role in reducing them. 


\section{Theory code implementation}
\label{app:code}

In this appendix, we describe some details about the numerical evaluation of the observables described in Section~\ref{sec:theory} implemented in the code \texttt{PyFowl} released with this paper. 
In the analysis, we evaluate the (flat-sky) projected angular correlation functions~\eqref{eq:xi+}, \eqref{eq:xi-}, \eqref{eq:gammat}, \eqref{eq:wiflat}. 
The code implements also the curved-sky formulae, as a check and for future surveys.

\paragraph{Linear terms beyond Limber}
For the terms that are linear and important are large angular scales, we first evaluate the projected correlation functions in harmonic space $C(\ell)$ on their full, beyond-Limber, expressions, and then obtain the correlation function in angular space $\xi(\theta)$ with a Bessel transform. 
Before describing the first step, we first describe the second one, to introduce the FFTLog algorithm~\cite{Hamilton:1999uv}. 
For a quick evaluation of the projection from harmonic to angular space,
\begin{equation}\label{eq:ximu}
  \xi_{\mu}(\theta) = \int \frac{\rmd \ell}{2 \pi} \ell J_{\mu}(\ell \theta) C(\ell) \, .
\end{equation} 
we can decompose the $C(\ell)$ in FFTLog as
\begin{equation}
  C(\ell) = \sum_{m=-N/2}^{N/2} c_m \ell^{-2 \nu_m} \, .
\end{equation}
At this point, we make use of the following master integral:
\begin{equation}
  \int \frac{\rmd x}{2 \pi}  J_{\mu}(x) x^{1-2 a} = \frac{2^{-2 a}}{\pi} \frac{\Gamma(1-a+\mu/2)}{\Gamma(a+\mu/2)} \, ,
\label{eq:masterint}
\end{equation}
such that
\begin{equation}
  \xi_{\mu}(\theta) = \sum_{m=-N/2}^{N/2} c_m \frac{(\theta/2)^{2 \nu_m}}{\pi \theta^2} \frac{\Gamma(1-\nu_m+\mu/2)}{\Gamma(\mu_m+\mu/2)} \, .
\end{equation}
The evaluation of $C(\ell)$'s consists in the $3D$ integrals \eqref{eq:Cpsi}, \eqref{eq:Cgpsi}, \eqref{eq:Cgg}, involving two oscillating spherical Bessel functions.
For these integrals, we employ the efficient numerical scheme described by~\cite{Fang:2019xat} (see also~\cite{Reymond:2025ixl}), that relies on the fact that the time dependence can be factorised for each line-of-sight integrals. 
The problem consists of evaluating integrals like $C_{gg}(\ell)$ at linear level:
\begin{equation}\label{eq:Cgg}
  C_{gg}(\ell) = \frac{2}{\pi} \int_0^{\infty} \rmd \chi_1 \int_0^{\infty} \rmd \chi_2 \int_0^{\infty} \rmd k \, k^2 j_{\ell}(k \chi_1) j_{\ell}(k \chi_2) f(\chi_1) f(\chi_2) P_{11}(k) \, ,
\end{equation}
where $f(\chi) = n^i_g(\chi) b(z(\chi)) D(z(\chi))$ for redshift bin $i$. 
The two $\chi$ integrals, involving the spherical Bessel and a function of $\chi$, can be easily done by FFTLog for each $\ell$'s.
Finally, the $k$ integral is a smooth integral that can be computed numerically.
The RSD and magnification integrals, that involve only the linear power spectrum, are also computed using this procedure.
We note that the RSD integral involves the second derivative $j''_{\ell}(k \chi)$, so we use a slightly modified master integral in order to avoid using the recursion relations for the derivatives of Bessel functions~\cite{Fang:2019xat}.

\paragraph{Loops in Limber} 
Since the loop contributions are relevant only at small scales, we can resort to the Limber approximation described in app.~\ref{app:limber}. 
To evaluate the power spectrum loop integrals, we use the FFTLog algorithm~\cite{Simonovic:2017mhp}, as already done in~\cite{DAmico:2019fhj,DAmico:2020kxu}. 
We start from the following decomposition of the linear power spectrum:
\begin{equation}
  P_{11}(k_n) = \sum_{m=-N/2}^{N/2} c_m k^{-2 \nu_m} \, .
\label{eq:P11dec}
\end{equation}
This allows to perform the loop integrals analytically, and get the loop power spectra as matrix multiplications with the cosmology-dependent coefficients $c_m$:
\begin{equation}
  P_{\sigma} = k^3 \sum_{m_1, m_2} c_{m_1} k^{-2 \nu_1} M_{\sigma}(\nu_1, \nu_2) k^{-2 \nu_2} c_{m_2} \, ,
\label{eq:PloopFFTLog}
\end{equation}
where $\sigma$ stands for $13$ or $22$ (see~\cite{Zhang:2021yna} for explicit expressions for $M_\sigma$).
Now, inserting eq.~\eqref{eq:Cgg} in eq.~\eqref{eq:ximu}, we get
\begin{equation}
\xi^{ij}_\mu(\theta) = \int \frac{\rmd \ell}{2\pi}\ell J_\mu(\ell \theta) \frac{2}{\pi} \int \rmd \chi_1 \int \rmd \chi_2 \int \rmd k\  k^2 \ j_\ell(k\chi_1) j_\ell(k\chi_2) f^i(\chi_1) f^j(\chi_2) P(k) \ , 
\end{equation}
where $f^i$ ($f^j$) is the line-of-sight kernel associated to the bin $i$ ($j$), and $P(k)$ denotes (a term in) the power spectrum of matter-matter, galaxy-matter, or galaxy-galaxy, depending on the angular correlation function $\xi_\mu$ under consideration.
Note that here, we have factorised put the time dependence of the power spectrum within the kernels $f$'s. 
Using the Limber approximation $j_\ell(k\chi) \rightarrow \sqrt{\frac{\pi}{2\ell}}\delta_D(\ell - k\chi)$ and performing the integral in $\ell$ and $\chi_2$, we get
 \begin{equation}\label{eq:xiefficient}
\xi_\mu^{ij}(\theta) = \int \rmd \chi \,  f^i(\chi) f^j(\chi) \zeta_\mu(r)|_{r = \chi\theta} \ ,
\end{equation}
where 
\begin{equation}\label{eq:zeta}
 \zeta_{\mu}(r) = \int \frac{\rmd k}{2 \pi} k J_{\mu}(k r) P(k) \, .
\end{equation}

Given that~\eqref{eq:P11dec} or~\eqref{eq:PloopFFTLog} are power laws in $k$, we can integrate~\eqref{eq:zeta} analytically using the master integral
\begin{equation}
  \int \rmd x  J_{\mu}(x) x^{1-2 a} = 2^{1-2 a} \frac{\Gamma(1-a+\mu/2)}{\Gamma(a+\mu/2)} \, .
\label{eq:masterint}
\end{equation}
For reference, we provide the formula at leading order,
\begin{equation}
  \zeta^{LO}_{\mu}(r) = \sum_{m=-N/2}^{N/2} c_m \frac{(r/2)^{2 \nu_m}}{\pi r^2} \frac{\Gamma(1 - \nu_m + \mu/2)}{\Gamma(\nu_m + \mu/2)} \, ,
\label{eq:XiLO}
\end{equation}
Counterterms have a similar expression to $\zeta^{LO}_{\mu}(r)$, as they involve one $P_{11}$.
For loop integral contributions, we have
\begin{equation}
  \zeta^{NLO}_{\mu}(r) = \sum_{m_1, m_2} c_{m1} c_{m2} M_{\sigma}(\nu_1, \nu_2) \frac{8 (r/2)^{2 \nu_{12}}}{\pi r^5} \frac{\Gamma(5/2 - \nu_{12} + \mu/2)}{\Gamma(\nu_{12} + \mu/2 - 3/2)} \, ,
\label{eq:XiNLO}
\end{equation}
where $\nu_{12} = \nu_1 + \nu_2$.
The evaluation of $\zeta$ can be done on a single redshift $\bar z$ for the whole survey, as long as the power spectrum time dependence can be factorised, which can be done exactly for the loop contributions, neglecting scale dependence from free-streaming species. 
As the cosmology-independent parts in~\eqref{eq:XiNLO} can be first pre-computed, we get~\eqref{eq:zeta} with one matrix multiplication per loop term, without having to first evaluate~\eqref{eq:PloopFFTLog}, nor doing any explicit integration. 
We are then left with simple 1D projections along the line-of-sight, eq.~\eqref{eq:xiefficient}. 
The evaluation of the loop contributions is thus very efficient.


\begin{figure}[ht!]
\centering
\includegraphics[width=0.99\textwidth]{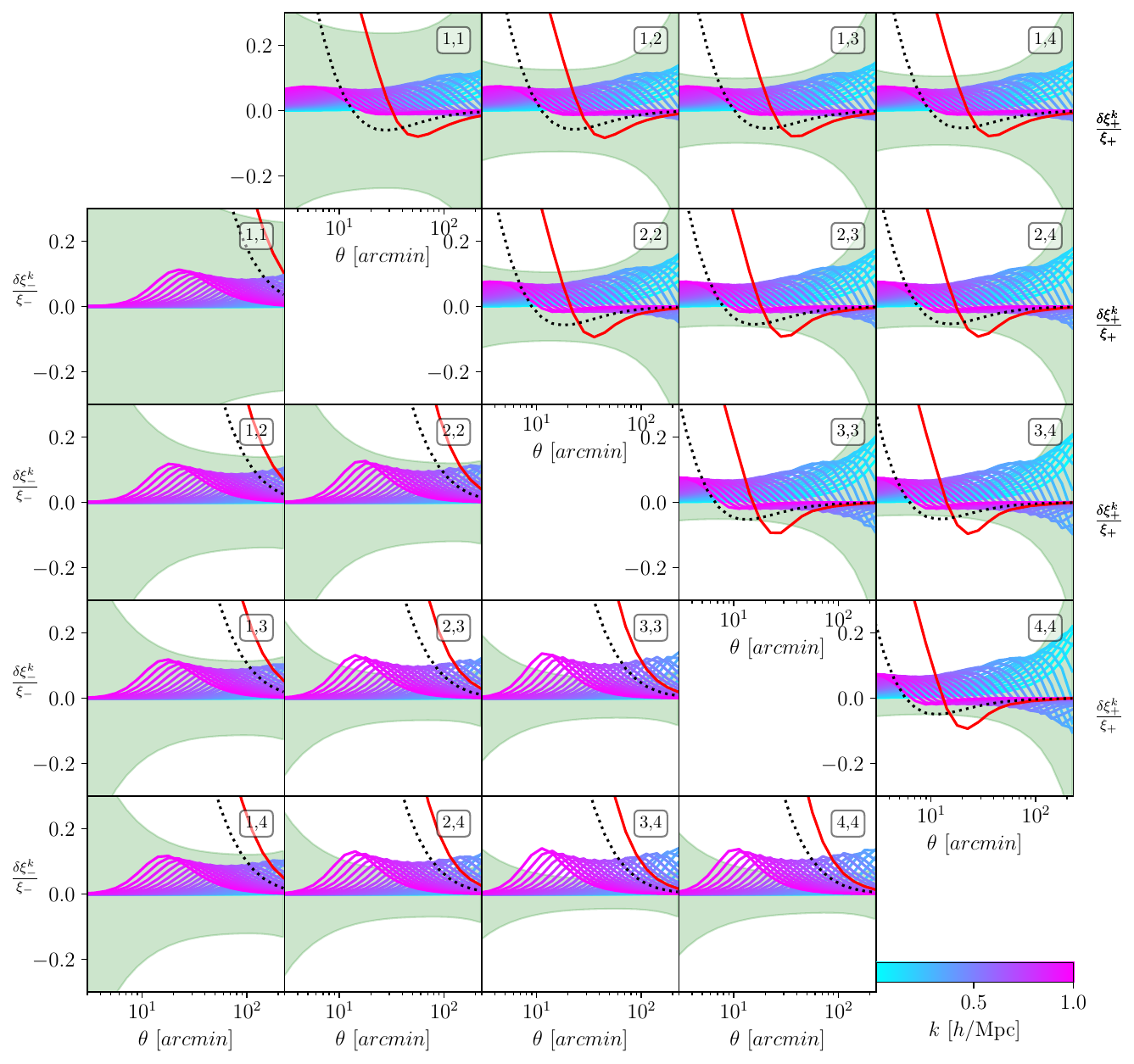}
\caption{\footnotesize  
\texttt{Buzzard} cosmic shear $\xi_\pm$ relative sensitivity on $k$-band power injections from a Halofit input power spectrum. 
The red curves represent the relative integrated contributions from $k > 0.4 \ h/\textrm{Mpc}$, that are plotted against the relative $68\%$CL data uncertainties shown as the green shaded region. 
For reference the dotted black lines represent the theory error from the two-loop contributions. 
}
\label{fig:pksensitivity}
\end{figure}

\section{Input power sensitivity}\label{app:pksensitivity}
In this appendix we study the sensitivity of our predictions for the angular correlation functions on $k$-bands of the input power spectrum. 
As the EFTofLSS predicts the power spectrum only accurately up to some $k_{\rm max}$, this allows us in sec.~\ref{sec:scalecut} to estimate the integrated contributions beyond $k_{\rm max}$ that spur our predictions as a function of angular scales.

\begin{figure}[ht!]
\centering
\includegraphics[width=0.99\textwidth]{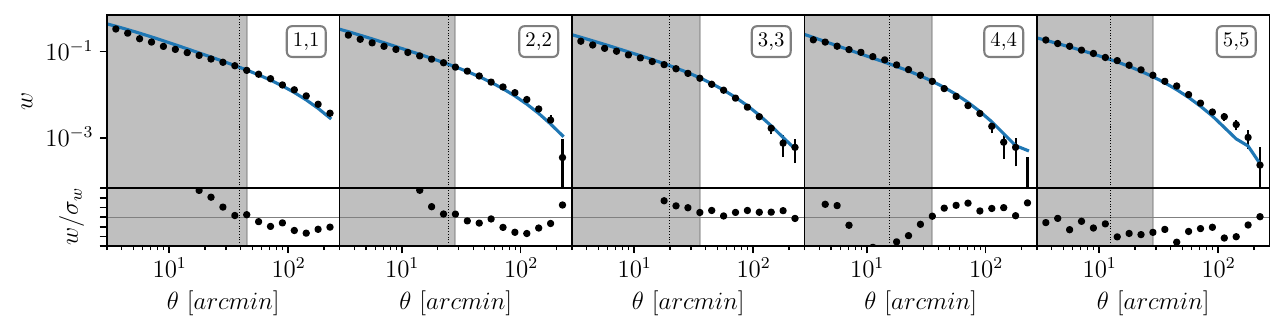}
\includegraphics[width=0.99\textwidth]{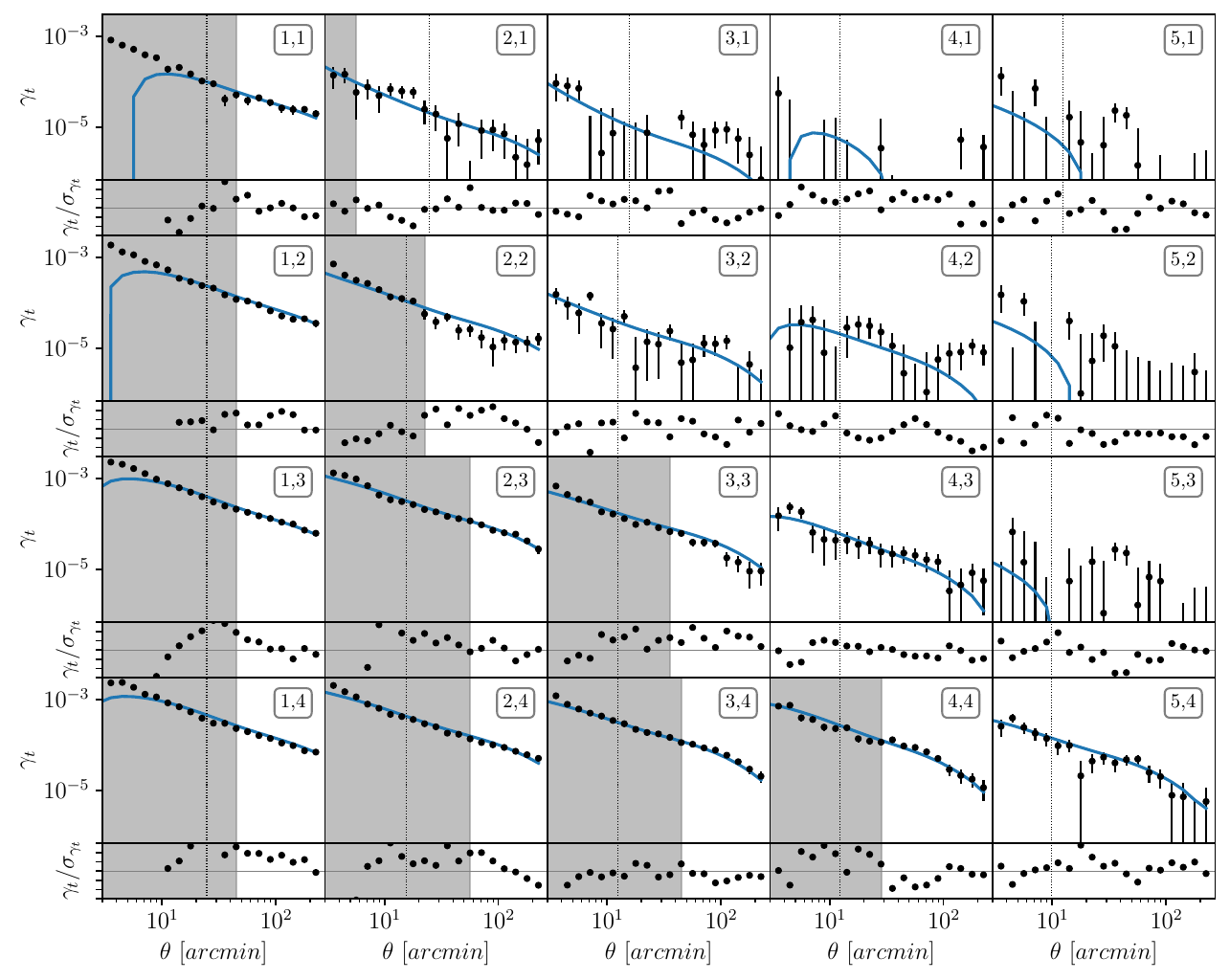}
\caption{\footnotesize  
\texttt{redMaGiC} two-point angular correlation functions: galaxy clustering $w$ and galaxy-galaxy-lensing $\gamma_t$.
two-point angular correlation functions: galaxy clustering $w$ and galaxy-galaxy lensing $\gamma_t$.
In the upper part of each plot, the black dots are the data points with their error bars, and the blue lines are the best-fit predictions from the EFTofLSS presented in this work.
The lower part of each plot shows the residuals of the best-fit curves relative to the data diagonal errors (with $y$-axis corresponding to $\pm 3\sigma$).
The shaded regions are excluded by the scale cuts used in this analysis; for reference, the DES collaboration scale cut choice~\cite{DES:2021wwk} is shown in dotted vertical lines.
For the $\gamma_t$ correlations, the rows (first indices) correspond to the lenses, while the columns (second indices) corresponds to the sources.}
\label{fig:redmag_gal}
\end{figure}

To proceed, we first decompose the input power spectrum $P(k)$ into $k$-band powers. 
These can be taken as simple top-hats, but to avoid discontinuities we choose to decompose $P(k)$ into a weighted sum of (relatively smooth) Gaussian $k$-bands of width $\Delta k \sim 0.01$, such that the \texttt{FFTLog} decomposition described in app.~\ref{app:code} occurring in \texttt{PyFowl} works well. 
By plugging each weighted Gaussian into our pipeline, we obtain the responses to our predictions, as a function of the input $k$-band powers. 
For conciseness, we only show an example of the results on the cosmic shear $\xi_\pm$ of \texttt{Buzzard} at fig.~\ref{fig:pksensitivity}.
There we use the Halofit power spectrum, while for $w$ and $\gamma_t$, we keep our EFTofLSS predictions. 
Declaring modes beyond $k > 0.4 \ h / \textrm{Mpc}$ described too inaccurately by the EFTofLSS to be included, we see that the integrated contributions from those high-$k$ modes result into an error in our predictions which can dominate the theory error estimated from the two-loop contributions in sec.~\ref{sec:scalecut}. 
In the end, we conservatively choose our scale cuts based on the maximum of the two errors, as described in sec.~\ref{sec:scalecut}. 


\section{redMaGiC best-fit}\label{app:redmag}

In fig~\ref{fig:redmag_gal}, we show the 2$\times$2pt measurements from the \texttt{redMaGiC} sample~\cite{DES:2021zxv} together with the best-fits from our EFTofLSS pipeline.

%
%

\bibliographystyle{JHEP}
\small
\bibliography{newbib}

\end{document}